\documentclass[]{aastex63}
\usepackage{multirow}

\usepackage[utf8x]{inputenc}
\usepackage{graphics,times,subfigure}
\usepackage{booktabs}
\usepackage{graphicx}
\usepackage{longtable}
\usepackage{threeparttable}
\usepackage{bm}

\received{June 18, 2023}
\revised{July 31, 2023}
\accepted{August 01, 2023}
\submitjournal{ApJ}

\shorttitle{In search of infall motion in molecular clumps IV}
\shortauthors{Yang et al.}
\begin{document}

%\title{In Search of Infall Motion in Molecular Clumps. IV. Mapping Observations toward the Second Batch of Infall Candidates} 
%\title{In Search of Infall Motion in Molecular Clumps. IV. Mapping Observations toward Confirmed Infall Sources}
\title{In Search of Infall Motion in Molecular Clumps. IV. Mapping of the Global Infall Sources}

\correspondingauthor{Xi Chen}
\email{chenxi@gzhu.edu.cn}

%\author[0000-0002-0786-7307]{Greg J. Schwarz}
%\affiliation{American Astronomical Society \\
%1667 K Street NW, Suite 800 \\

\author[0000-0002-2895-7707]{Yang Yang}
\affiliation{Center for Astrophysics, Guangzhou University, Guangzhou 510006, China}

\author[0000-0002-5435-925X]{Xi Chen}
\affiliation{Center for Astrophysics, Guangzhou University, Guangzhou 510006, China}

\author[0000-0002-5920-031X]{Zhibo Jiang}
\affiliation{Purple Mountain Observatory, Chinese Academy of Sciences, Nanjing 210023, China}
\affiliation{Center astronomy and Space Sciences, China Three Gorges University, Yichang 443002, China}

\author[0000-0003-0849-0692]{Zhiwei Chen}
\affiliation{Purple Mountain Observatory, Chinese Academy of Sciences, Nanjing 210023, China}

\author[0000-0002-8617-3521]{Shuling Yu}
\affiliation{Purple Mountain Observatory, Chinese Academy of Sciences, Nanjing 210023, China}
\affiliation{University of Science and Technology of China, Chinese Academy of Sciences, Hefei 230026, China}

\author[0000-0001-9328-4302]{Jun Li}
\affiliation{Center for Astrophysics, Guangzhou University, Guangzhou 510006, China}



\begin{abstract}
We have used the IRAM 30-m telescope to map some targets with HCO$^+$ (1-0) and H$^{13}$CO$^+$ (1-0) lines in order to search for gas infall evidence in the clumps. In this paper, we report the mapping results for 13 targets. All of these targets show HCO$^+$ emissions, while H$^{13}$CO$^+$ emissions are observed in ten of them. The HCO$^+$ integrated intensity maps of ten targets show clear clumpy structures, and nine targets show clumpy structures in the H$^{13}$CO$^+$ maps. Using the RADEX radiative transfer code, we estimate the column density of H$^{13}$CO$^+$, and determine the abundance ratio [H$^{13}$CO$^+$]/[H$_2$] to be approximately 10$^{-12}$ to 10$^{-10}$. Based on the asymmetry of the HCO$^+$ line profiles, we identify 11 targets show blue profiles, while six clumps have global infall evidence. We use the RATRAN and two-layer models to fit the HCO$^+$ line profiles of these infall sources, and analyze their spatial distribution of the infall velocity. The average infall velocities estimated by these two models are 0.24 -- 1.85 km s$^{-1}$ and 0.28 -- 1.45 km s$^{-1}$, respectively. The mass infall rate ranges from approximately 10$^{-5}$ to 10$^{-2}$ M$_{\odot}$ yr$^{-1}$, which suggests that intermediate- or high-mass stars may be forming in the target regions.
\end{abstract}



\section{Introduction} \label{sec:intro}

In the early stages of star formation, the gas in the molecular clump collapses inward the center objects under gravity \citep[e.g.,][]{Bachiller+1996}. The gas infall motion accumulates material for the mass growth of the central objects. However, the dominant physical processes driving the mass growth for central objects, particularly for massive stars, remain unclear \citep[e.g.,][]{Peretto+etal+2013}. Some theories propose that molecular clouds gather matter from large scales to their centers though global collapse, where cores and protostars experience simultaneous mass growth \citep[e.g.,][]{Peretto+etal+2006,Peretto+etal+2007,Smith+etal+2009}. Other theories suggest that the primordial fragmentation of stable molecular clouds may form compact gas reservoirs (approximately 0.1 pc), and the forming stars accumulate mass from the reservoirs \citep[e.g.,][]{McKee+Tan+2003}. Therefore, the study of gas collapse will help us better understand this physical process. In recent years, numerous studies have focused on it, to find the evidence of the gas infall motions in the clumps. However, it is difficult to observe the gas infall motion directly, mainly because the gas flows into the core cannot be easily identified.

In previous studies, the blue asymmetrical profile of optically thick lines has been commonly used as an indirect tracer of gas collapse \citep[e.g.,][]{Mardones+etal+1997}. The collapse model provides a explanation for the generation of this special line profile \citep[e.g.,][]{Leung+Brown+1977}: the envelope material falls toward the central hot objects (i.e. moving away from us along the line of sight), resulting in a redshifted self-absorption profile for the optically thick line. This will cause the self-absorption of the optically thick line to shift slightly toward the red side. The optically thick line will produce a double-peaked profile, with the blue peak stronger than the red one. However, in some cases, multiple velocity components may also produce a similar line profile. Thus, optically thin line is also required to trace the central radial velocity of the clumps. The line profile of optically thin line can be used to distinguish whether the blue profile is caused by the gas infall motion or multiple velocity components. If the optically thin line shows a single-peaked profile, with the peak located between the two peaks of the optically thick line, it suggests that the double-peaked profile of the optically thick line is more likely due to self-absorption rather than multiple velocity components. Such a combination of optically thick and thin lines is usually used to identify the infall sources.

The optically thick lines we generally used to trace gas infall motions are HCO$^+$ (1-0, 3-2, 4-3), $^{12}$CO (2-1, 3-2, 4-3), HNC (1-0, 4-3), HCN (1-0, 3-2), CS (2-1), and H$_2$CO (2$_{12}$-1$_{11}$), etc. And N$_2$H$^+$ (1-0), C$^{18}$O (1-0, 2-1, 3-2), H$^{13}$CO$^+$ (1-0), and C$^{17}$O (3-2) are usually used as optically thin lines to assist identification. At present, studies on gas infall motions mainly focus on searching for infall candidates in some certain kinds of targets, such as infrared dark clouds and massive protostars \citep[e.g.,][]{Lee+etal+2001, Klaassen+etal+2007, Klaassen+etal+2008, Chen+etal+2010, He+etal+2015, He+etal+2016, Calahan+etal+2018}. In recent years, with the continuous development of the interferometric arrays, more and more studies begin to use interferometric arrays with higher spatial resolution to observe the infall sources \citep[e.g.][]{Qin+etal+2016, Su+etal+2019, Liu+etal+2020}. However, the number of available samples remains insufficient, with only approximately four hundred identified infall candidates \citep{Yu+etal+2022}. This hinders us to have a more comprehensive understanding of the gas collapse stage. Therefore, we aim to obtain a large-scale unbiased sample of infall candidates by searching and identifying the molecular clumps with infall characteristics in the Milky Way. This will help us to analyze the physical properties of collapsing sources, and further reveal the material accumulation process in the star-forming initial stage.

In a previous study, we used the Milky Way Imaging Scroll Painting (MWISP) project made with Purple Mountain Observatory (PMO) 13.7-m telescope, which provides $^{12}$CO (1-0) and its isotopes $^{13}$CO (1-0) and C$^{18}$O (1-0) data \citep[e.g.,][]{Su+etal+2019}, to search for infall candidates along the Galactic plane. The combinations of $^{12}$CO/$^{13}$CO and $^{13}$CO/C$^{18}$O can be used as two pairs of optically thick and thin lines, to trace gas infall motions. Firstly, we carried out a blind search in all available MWISP CO data to find the sources with blue profile in the optically thick lines. About 3,500 sources were initially identified and checked \citep[][referred to as Paper I]{Jiang+etal+2023}. Subsequently, we selected 133 of them with significant blue profile in the $^{12}$CO line and the intensity of C$^{18}$O line $\textgreater$ 1 K for further confirmation. We then used the PMO 13.7-m telescope to observe HCO$^+$ (1-0) and HCN (1-0) lines, which are more effective tracers for tracing gas infall motions \citep[e.g.,][]{Peretto+etal+2013, Yuan+etal+2018, Zhang+etal+2018}. We finally identified 56 infall candidates through single-pointed observations \citep[][hereafter referred to as Paper II]{Yang+etal+2020}. However, only single-pointed observations cannot provide the spatial structure of the sources. In some cases, outflow and rotation can also cause spectral lines to show a similar blue profile \citep[e.g.,][]{Mardones+etal+1997, He+etal+2015}. But the spatial distribution map of the emission in different velocity ranges can distinguish between infall and outflow \citep[e.g.,][]{Gregersen+etal+1997}. Therefore, it is necessary to conduct mapping observations to confirm these infall candidates.

As a follow-up study, we conducted mapping observations of the selected infall candidates using HCO$^+$ (1-0) and H$^{13}$CO$^+$ (1-0) lines with the IRAM 30-m telescope. The mapping results of 24 targets have been previously reported \citep[][hereafter referred to as Paper III]{Yang+etal+2021}. Based on the mapping observation results, we confirmed that 19 of the targets have evidence of gas infall motion. The remaining five targets show red profiles (i.e. red peak stronger than the blue one), symmetrical profiles, or multiple peaks in HCO$^+$ lines, which may be caused by multiple velocity components or other kinds of gas motions. In this paper, we present the observation results of an additional 13 targets. Section \ref{sec:obs} provides a brief introduction to the mapping observations. In Section \ref{sec:results}, the mapping results of 13 targets based on HCO$^+$(1-0) and H$^{13}$CO$^+$ (1-0) lines are given. In Section \ref{sec:analysis}, we analyze the clump properties of these sources, and identify the infall candidates with an blue profile out of our sources. Meanwhile, we use two models to estimate the infall velocities and mass infall rates of the infall sources, and analyze the spatial distribution of the infall velocity of the global infall source. Finally, in Section \ref{sec:summary}, we summarize the results and analysis of this work.

\section{Observations} \label{sec:obs}

%________________________________________ Table 1: Basic parameters

\begin{table}
\begin{center}
  \caption{Source list.}\label{Tab:src-catalog}
%Please Capitalize the First Letter of Each Notional Word in table's caption
 \setlength{\tabcolsep}{1mm}{
\begin{tabular}{ccccccc}
  \hline\noalign{\smallskip}
%\cmidrule(lr){2-3}  \cmidrule(lr){4-5}
Source  &  R.A.  &  Decl.  & V$\rm{_{LSR}}$ &  Distance &  Extent of map & Association  \\ 
Name &	(J2000)	&	(J2000)	&	(km s$^{-1}$)	&	(kpc)	&	&	\\
  \hline\noalign{\smallskip}
G012.79-0.20  &  18:14:11.5  &  -17:56:15  &  35.8  &  2.40$_{-0.15}^{+0.17}$ $^1$ &  $4.5\arcmin\times4.5\arcmin$  & W33 Main, Class I, FS, Class II$^4$,  \\
 & & & & & & 95 GHz and 6.7 GHz methanol masers, H$_2$O maser$^5$ \\
G012.87-0.22  &  18:14:26.4  &  -17:52:43  &  35.4  &  2.40$_{-0.15}^{+0.17}$ $^1$ &  $4.5\arcmin\times4.5\arcmin$  & W33 Main (northeast), Class I, FS, Class II \\
G012.96-0.23  &  18:14:39.0  &  -17:48:25  &  35.2  &  2.40$_{-0.15}^{+0.17}$ $^1$ &  $2.5\arcmin\times2.5\arcmin$  & W33 A (north), FS, Class II \\
G014.00-0.17  &  18:16:29.8  &  -16:52:07  &  40.0  &  3.10 $^2$ &  $5\arcmin\times5\arcmin$  & Class I, Class II \\
G014.25-0.17  &  18:17:02.2  &  -16:38:40  &  38.1  &  $3.34_{-0.50}^{+0.42}$ $^3$ &  $3\arcmin\times3\arcmin$  & FS, Class II \\
G017.09+0.82  &  18:18:59.0  &  -13:40:19  &  22.3  &  $2.01_{-0.58}^{+0.50}$ &  $3\arcmin\times3\arcmin$  & Class I, FS, Class II \\
G025.82-0.18  &  18:39:04.2  &  -06:24:29  &  93.7  &  $5.01_{-0.28}^{+0.28}$ &  $3\arcmin\times3\arcmin$  & Class I, Class II, 95 GHz and 6.7 GHz methanol masers \\
G036.02-1.36  &  19:01:57.5  &  +02:07:55  &  31.7  &  $1.99_{-0.41}^{+0.40}$ &  $4\arcmin\times4\arcmin$  &  \\
G037.05-0.03  &  18:59:06.3  &  +03:38:41  &  81.4  &  $4.89_{-0.47}^{+0.47}$ &  $4\arcmin\times4\arcmin$  & Class I, FS, 6.7 GHz methanol masers \\
G049.07-0.33  &  19:22:42.3  &  +14:10:00  &  60.7  &  $4.68_{-0.78}^{+0.80}$ $^3$ &  $3\arcmin\times3\arcmin$  & Class I, FS, Class II, 95 GHz methanol maser \\
G081.72+1.29  &  20:35:55.6  &  +42:48:53  &   3.8  &  $1.20(\textless0.01)$ &  $3\arcmin\times3\arcmin$  &  \\
G081.90+1.43  &  20:35:51.9  &  +43:02:36  &  11.1  &  $1.18(\textless0.01)$  & $2.5\arcmin\times2.5\arcmin$  &  \\
G133.42+0.00  &  02:19:49.4  &  +61:03:32  &  -15.2  &  $0.86_{-0.51}^{+0.53}$ &  $3\arcmin\times3\arcmin$  & Class I \\
  \hline\noalign{\smallskip}
\end{tabular}}
\end{center}
\tablecomments{The distance of the sources are obtained from: $^1$ \citet{Immer+etal+2013}, $^2$ \citet{Urquhart+etal+2018}, $^3$ \citet{Ellsworth+etal+2015}, where $^1$ is the parallax distance. $^4$ Class I, FS, and Class II represent Class I YSOs, Flat-spectrum (between Class I and Class II) YSOs, and Class II YSOs, respectively \citep{Kuhn+etal+2021}. $^5$ The maser catalogs used here are as follows: the 95 GHz methanol maser catalog by \citet{Yang+etal+2017}, the 6.7 GHz methanol maser catalog by \citet{Yang+etal+2019}, the H$_2$O maser catalog by \citet{Anglada+etal+1996} and \citet{Valdettaro+etal+2001}, and the OH maser catalog by \citet{Qiao+etal+2016,Qiao+etal+2018,Qiao+etal+2020}.
}
\end{table}

%________________________________________ Table 2: Observation parameters

\begin{table}
\begin{center}
  \caption{Observation parameters}\label{Tab:obs}  
%Please Capitalize the First Letter of Each Notional Word in table's caption
 \setlength{\tabcolsep}{1mm}{
\begin{tabular}{cccccccc}
  \hline\noalign{\smallskip}
Line & Frequency & Receiver & Backend & Bandwidth & Frequency resolution & Velocity resolution & RMS  \\
  & (MHz) &  &  & (GHz) & (kHz) & (km s$^{-1}$) & (K) \\
  \hline\noalign{\smallskip}
HCO$^+$ (1-0) & 89188.5 & EMIR & VESPA & 0.2 & 40 & 0.13 & 0.02 -- 0.05  \\
H$^{13}$CO$^+$ (1-0) & 86754.3 & EMIR & FTS & 8 & 195 & 0.67 & 0.01 -- 0.03 \\ 
  \hline\noalign{\smallskip}
\end{tabular}}
\end{center}
%\tablecomments{The values in columns 8 are the rms range corresponding to the rms values. }
\end{table}

We used the IRAM 30-m telescope to observe the infall candidates identified from previous PMO single-pointed observations. The optically thick line HCO$^+$ (1-0) and optically thin line H$^{13}$CO$^+$ (1-0) at the 3 mm band were used to trace the infalling gas. Currently, we have completed mapping observations for 45 infall candidates (i.e. 37 targets, some of which cover two or more infall candidates). In the previous studies, we presented the observation results of 24 targets and conducted an analysis of their physical and chemical properties \citep[Paper III,][]{Yang+etal+2023}. In this paper, we report the observation results of an additional 13 targets (listed in Table \ref{Tab:src-catalog}). The local standard of rest (LSR) velocity values for these targets were derived from the center radial velocities traced by the MWISP C$^{18}$O data (Paper II). For the distance values, we prioritize adopting those provided in other literature, particularly parallax distance values. If the source distance cannot be found in the literature, we adopt the kinematic distance given in Paper II.

The observations were conducted with the Eight Mixer Receiver (EMIR) at the 3 mm band of the IRAM 30-m telescope from November 11 to 16, 2020 (project code: 017-20) and from January 27 to April 13, 2021 (project code: 097-20). The typical system temperature ($T_{sys}$) ranges from 90 to 135 K. We used the VESPA and FTS back-ends with bandwidths of 0.2 and 8 GHz, respectively. The frequency resolutions are 40 and 195 kHz, corresponding to velocity resolutions of approximately 0.13 and 0.67 km s$^{-1}$ in the 3 mm band, respectively (as shown in Table \ref{Tab:obs}). The angular resolution at 3 mm band is about $28\arcsec$, and the main beam efficiency is about 0.81. We used the On-the-Fly position switch observing mode to cover each observation area, and the scale of the observation area is determined through the MWISP $^{13}$CO (1-0) and C$^{18}$O (1-0) mapping data, covering the area of approximately 6 to 20 square minutes (listed in Table \ref{Tab:src-catalog}). The on-source integration time varies depending on the sizes. Including the pointing, focusing, calibration and instrumental deadtimes, it takes approximately 1 to 1.5 hours to observe one target. The total observation time for all targets amounts to about 18 hours. The CLASS software of the GILDAS package\footnote{\url{http://www.iram.fr/IRAMFR/GILDAS/}} is used for data reduction. The RMS values of average spectrum are 0.01 -- 0.03 K for H$^{13}$CO$^+$ lines, and 0.02 -- 0.05 K for HCO$^+$ lines, respectively. Then, based on the H$^{13}$CO$^+$ and HCO$^+$ mapping data, we will analyze the physical properties of these targets, and identify infall sources according to the HCO$^+$ line profile.

\section{HCO$^+$ (1-0) and H$^{13}$CO$^+$ (1-0) Maps} \label{sec:results}

\begin{figure*}[h]
  \begin{minipage}[t]{0.19\linewidth}
  \centering
   \includegraphics[width=39mm]{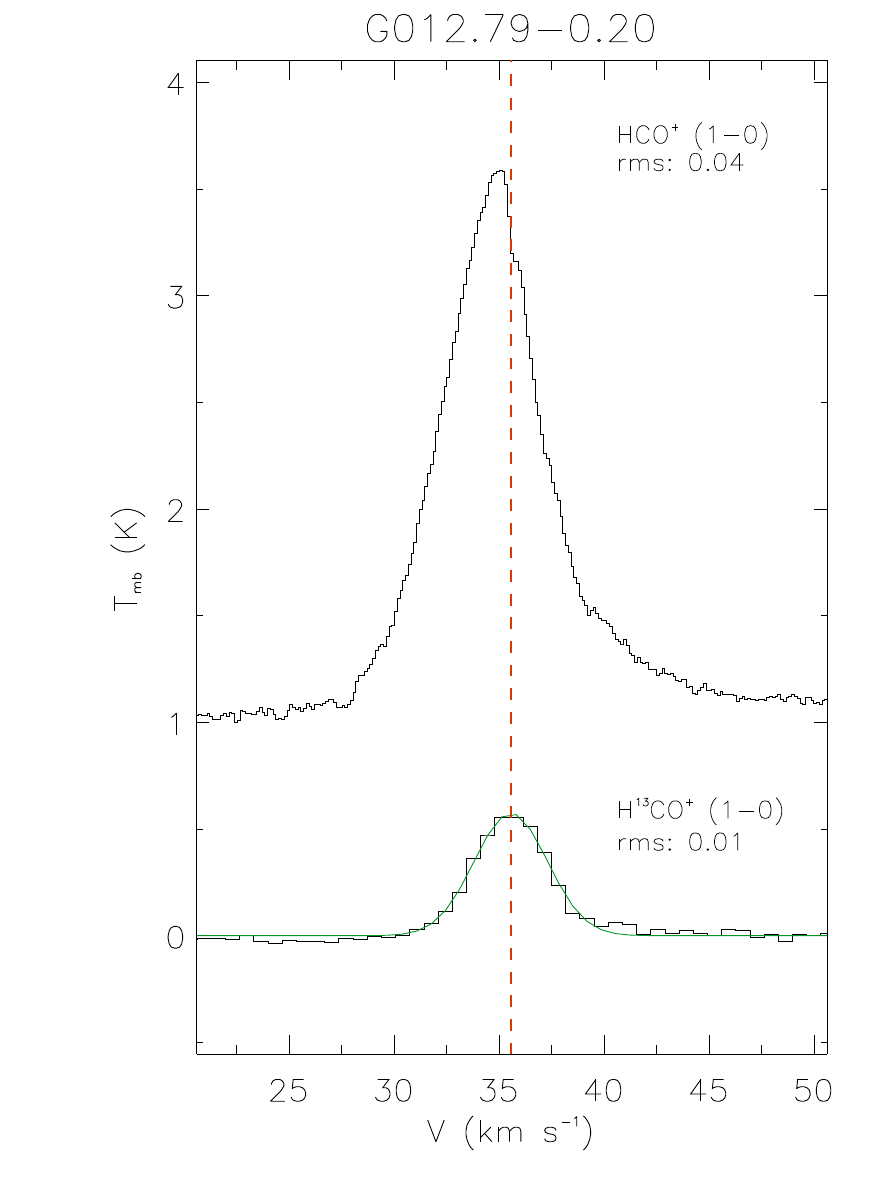}
  \end{minipage}%
  \begin{minipage}[t]{0.19\linewidth}
  \centering
   \includegraphics[width=39mm]{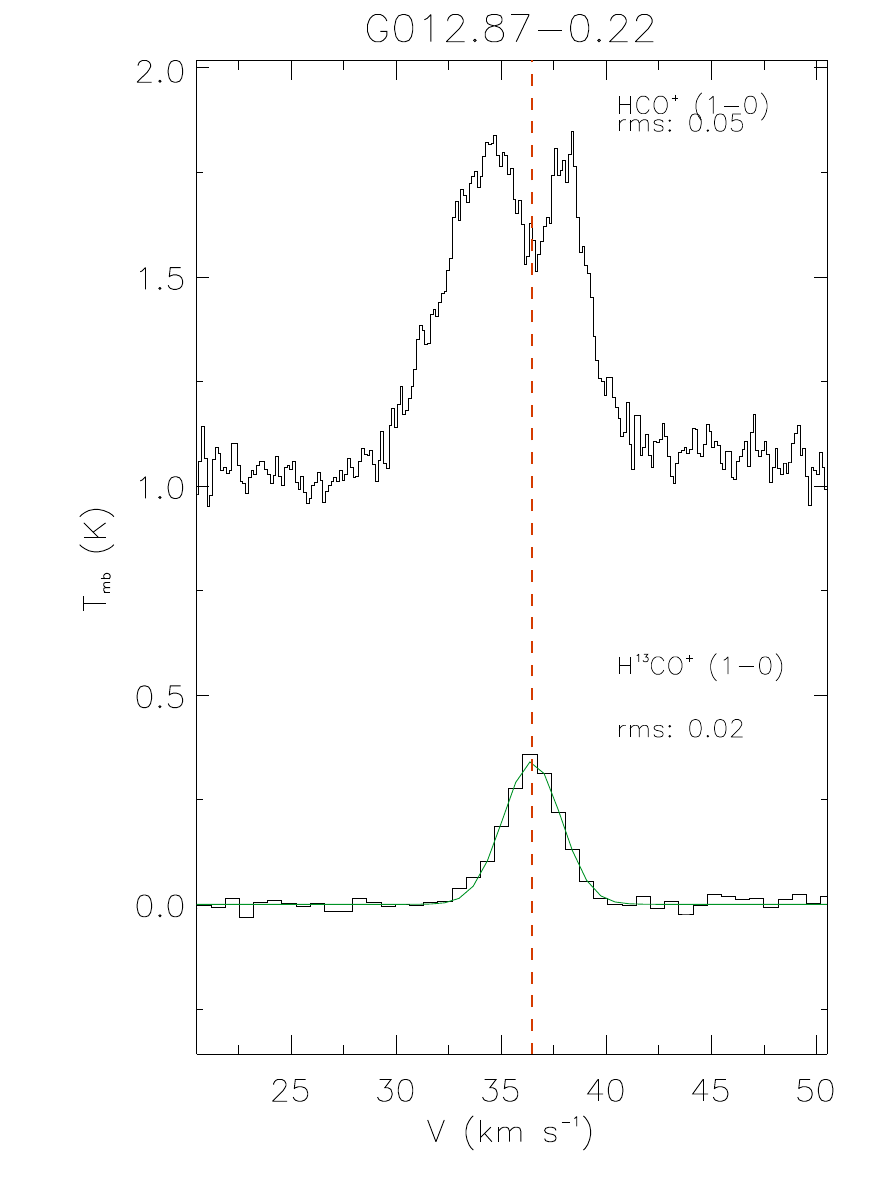}
  \end{minipage}%
  \begin{minipage}[t]{0.19\linewidth}
  \centering
   \includegraphics[width=39mm]{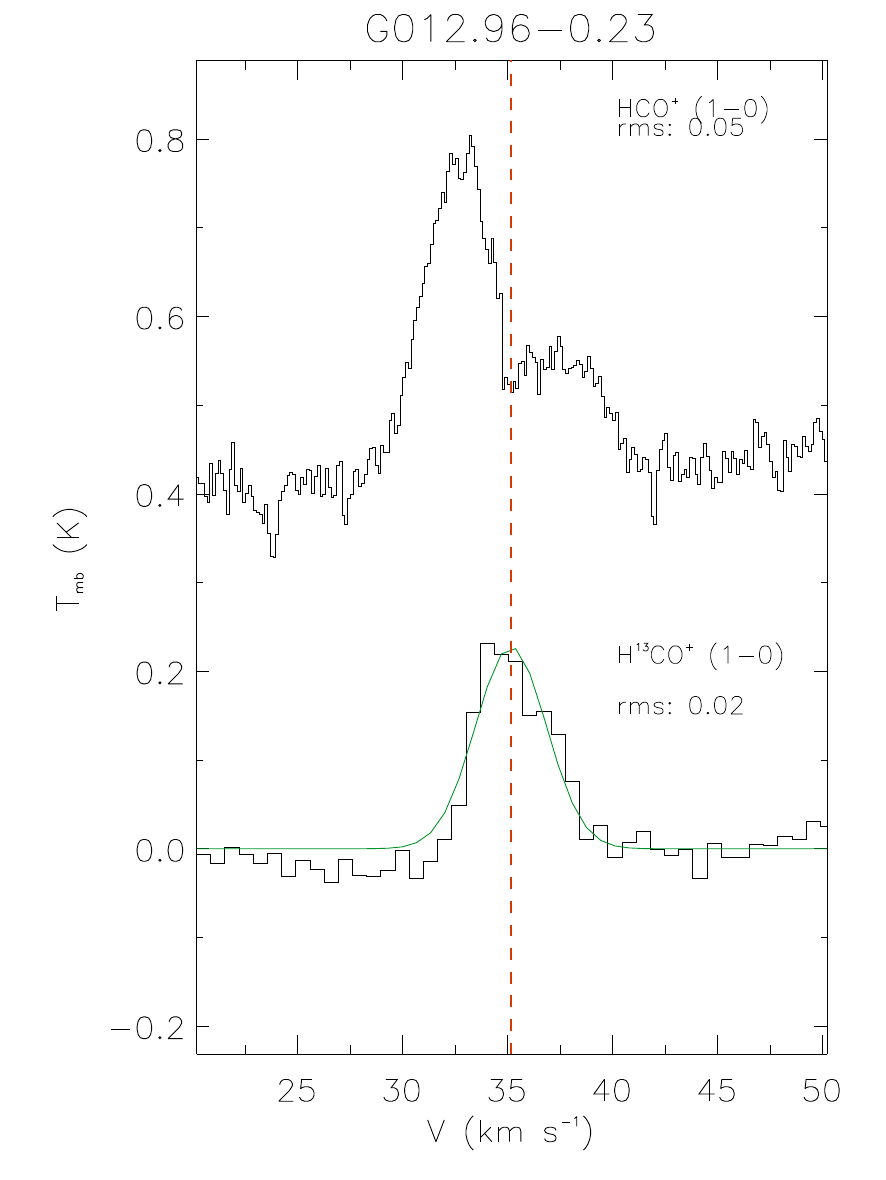}%*
  \end{minipage}%
  \begin{minipage}[t]{0.19\linewidth}
  \centering
   \includegraphics[width=39mm]{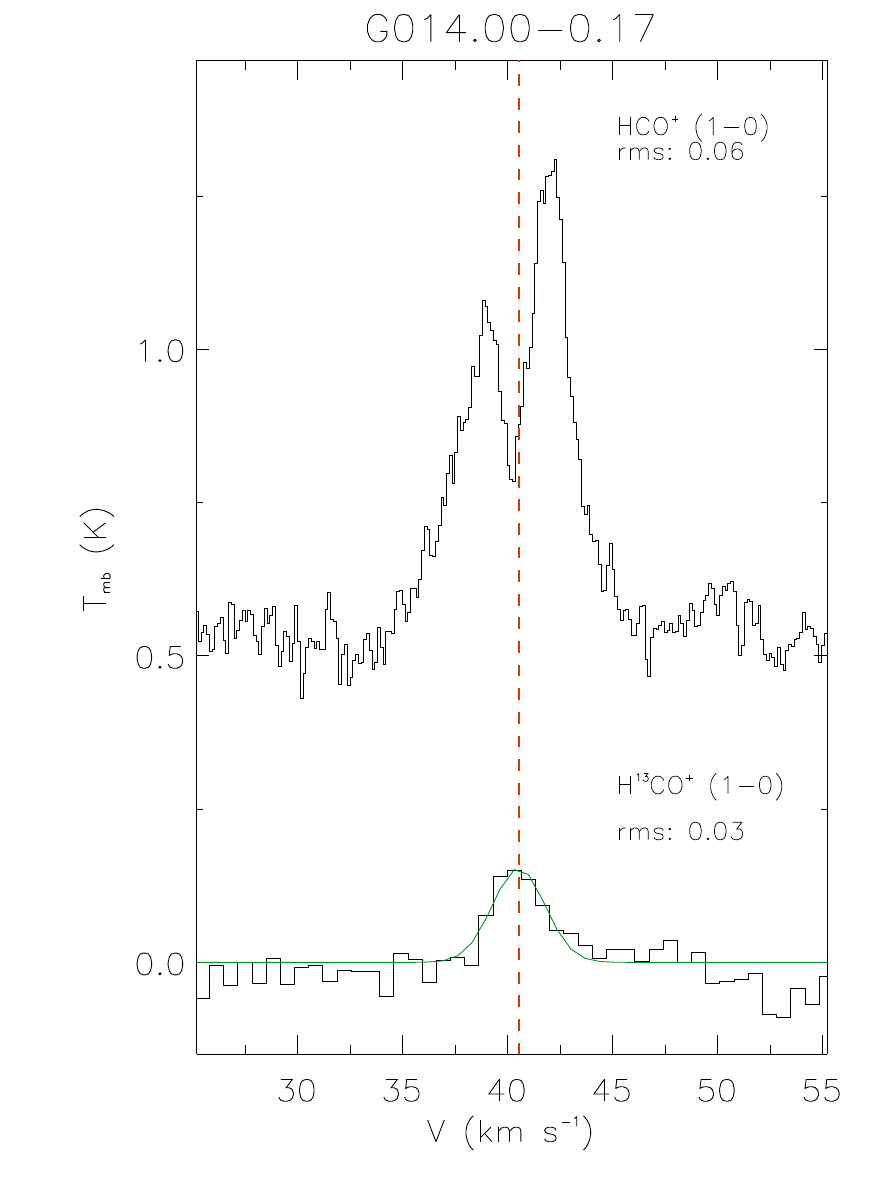}
  \end{minipage}%
  \begin{minipage}[t]{0.19\linewidth}
  \centering
   \includegraphics[width=39mm]{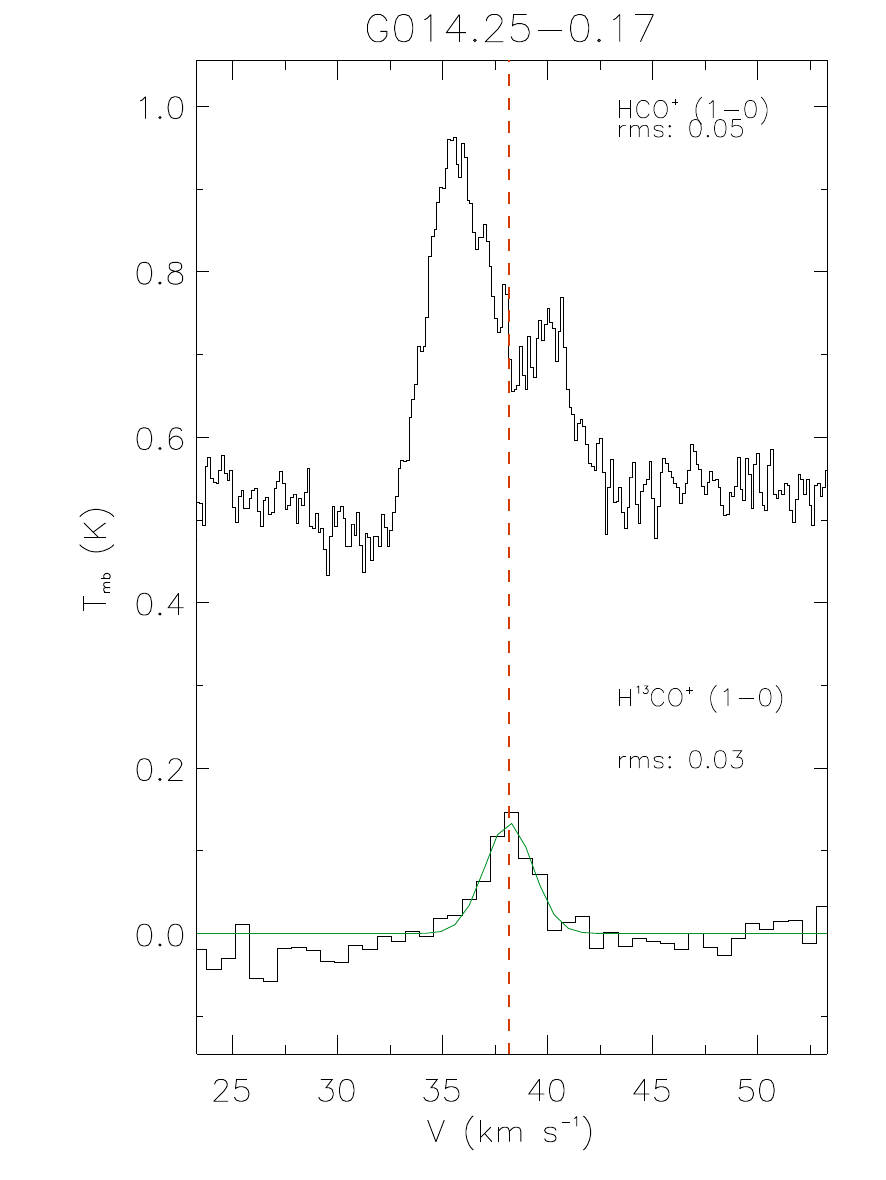}
  \end{minipage}%   
  
  \begin{minipage}[t]{0.19\linewidth}
  \centering
   \includegraphics[width=39mm]{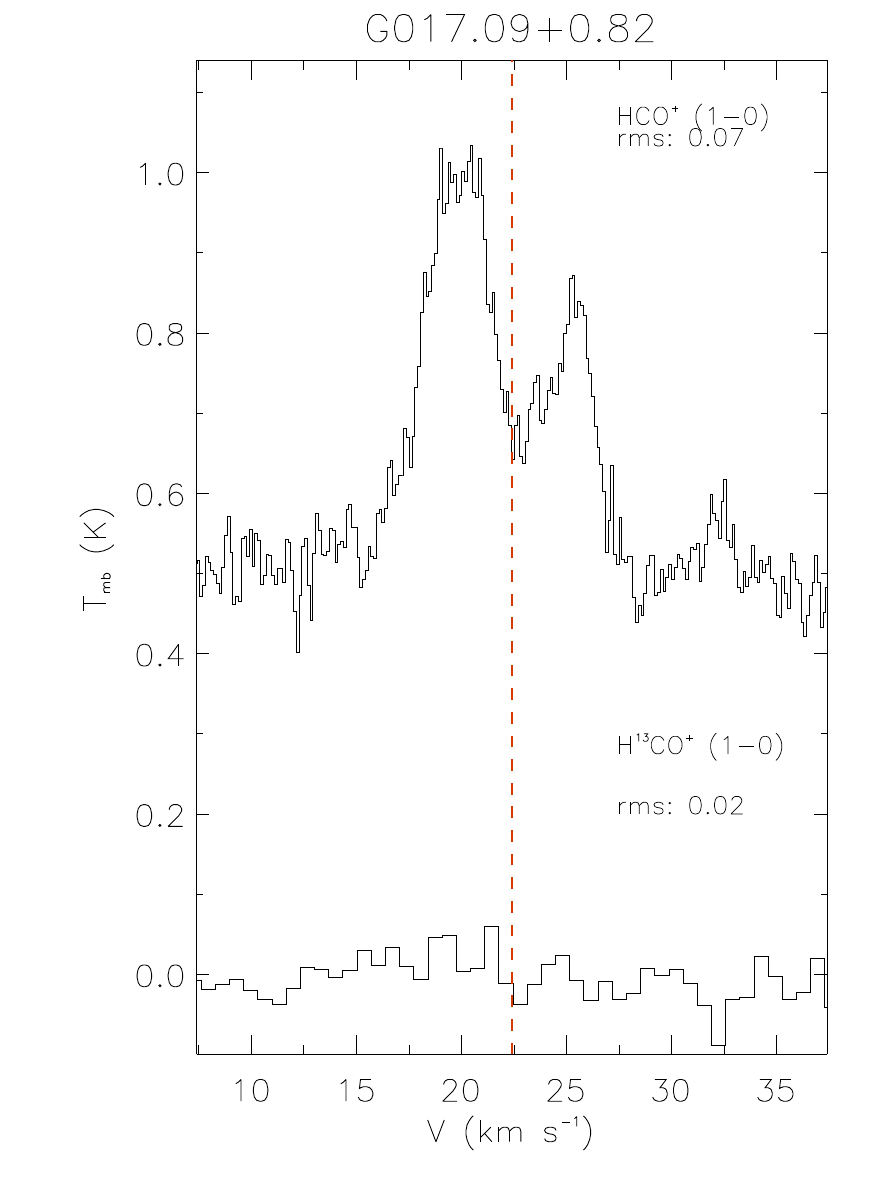}
  \end{minipage}%
  \begin{minipage}[t]{0.19\linewidth}
  \centering
   \includegraphics[width=39mm]{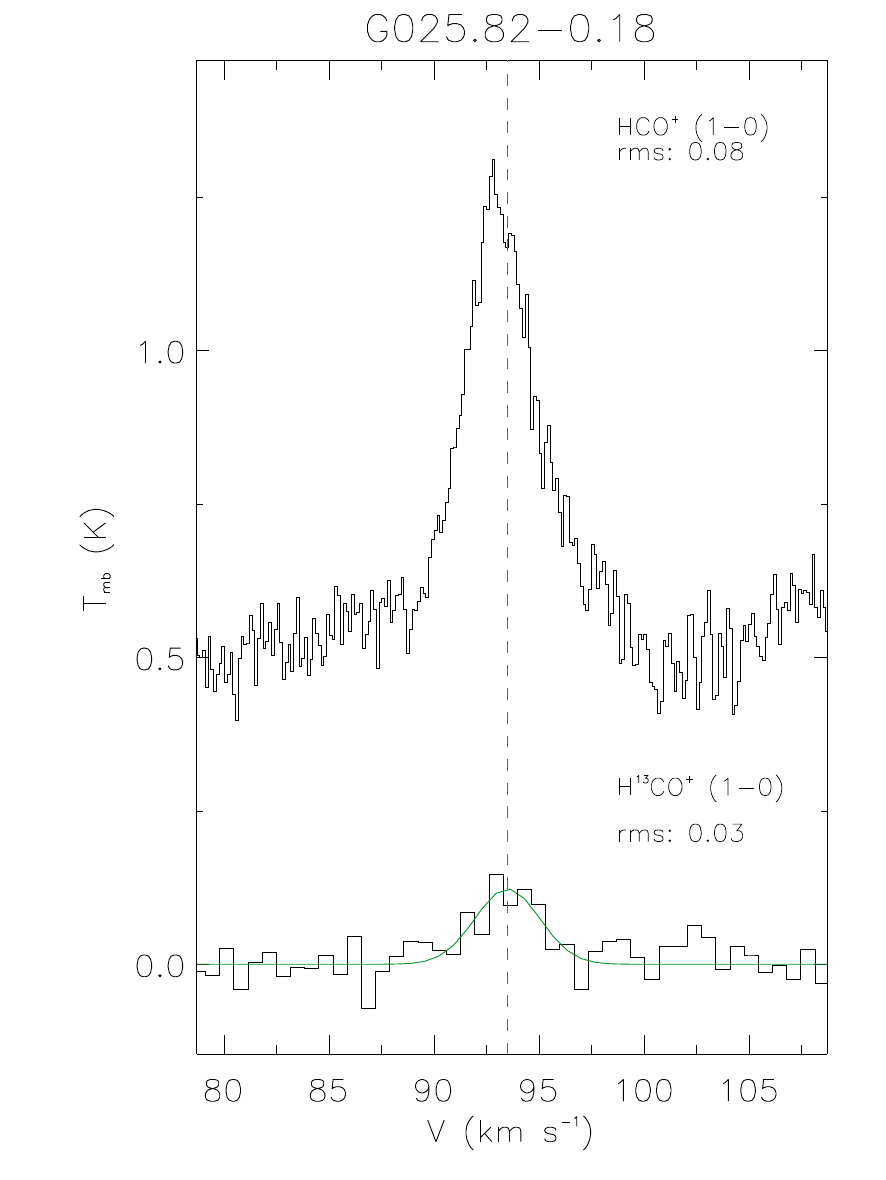}
  \end{minipage}%
  \begin{minipage}[t]{0.19\linewidth}
  \centering
   \includegraphics[width=39mm]{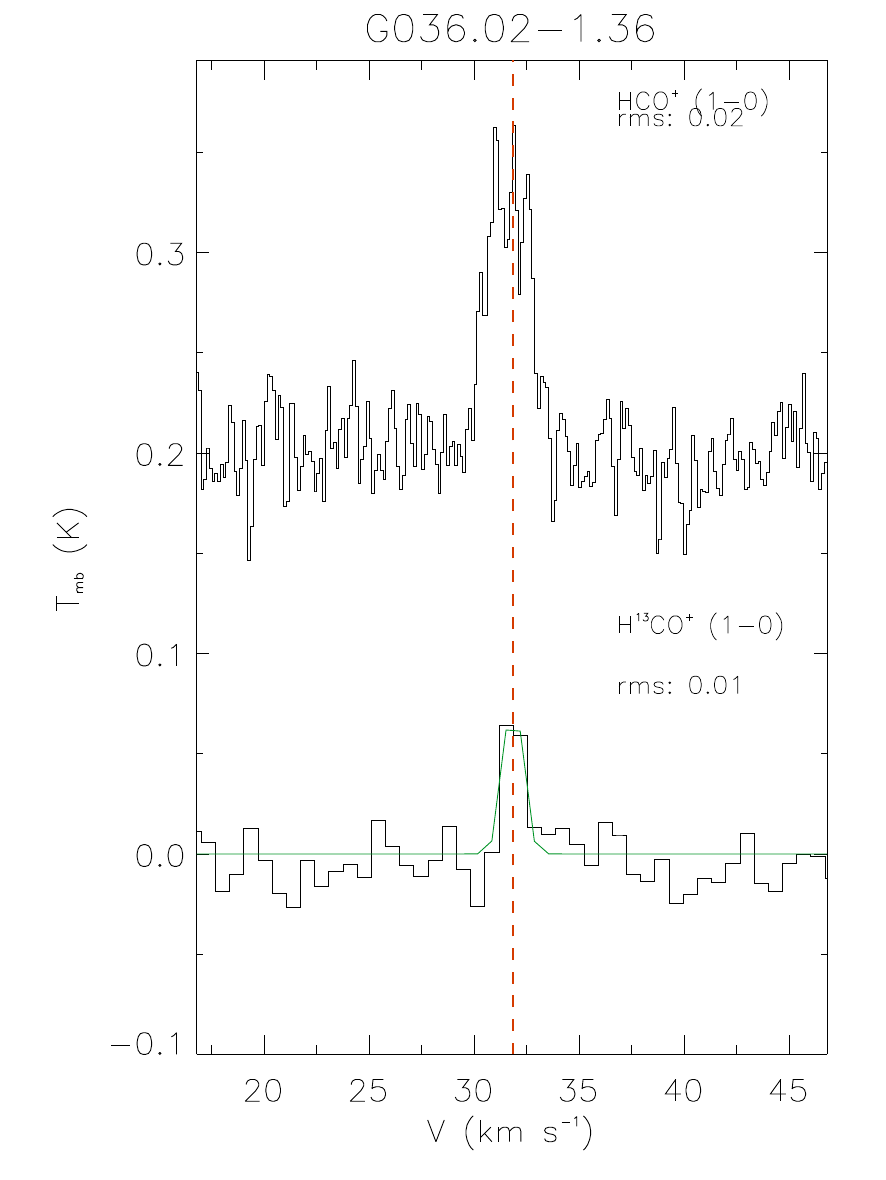}
  \end{minipage}%
  \begin{minipage}[t]{0.19\linewidth}
  \centering
   \includegraphics[width=39mm]{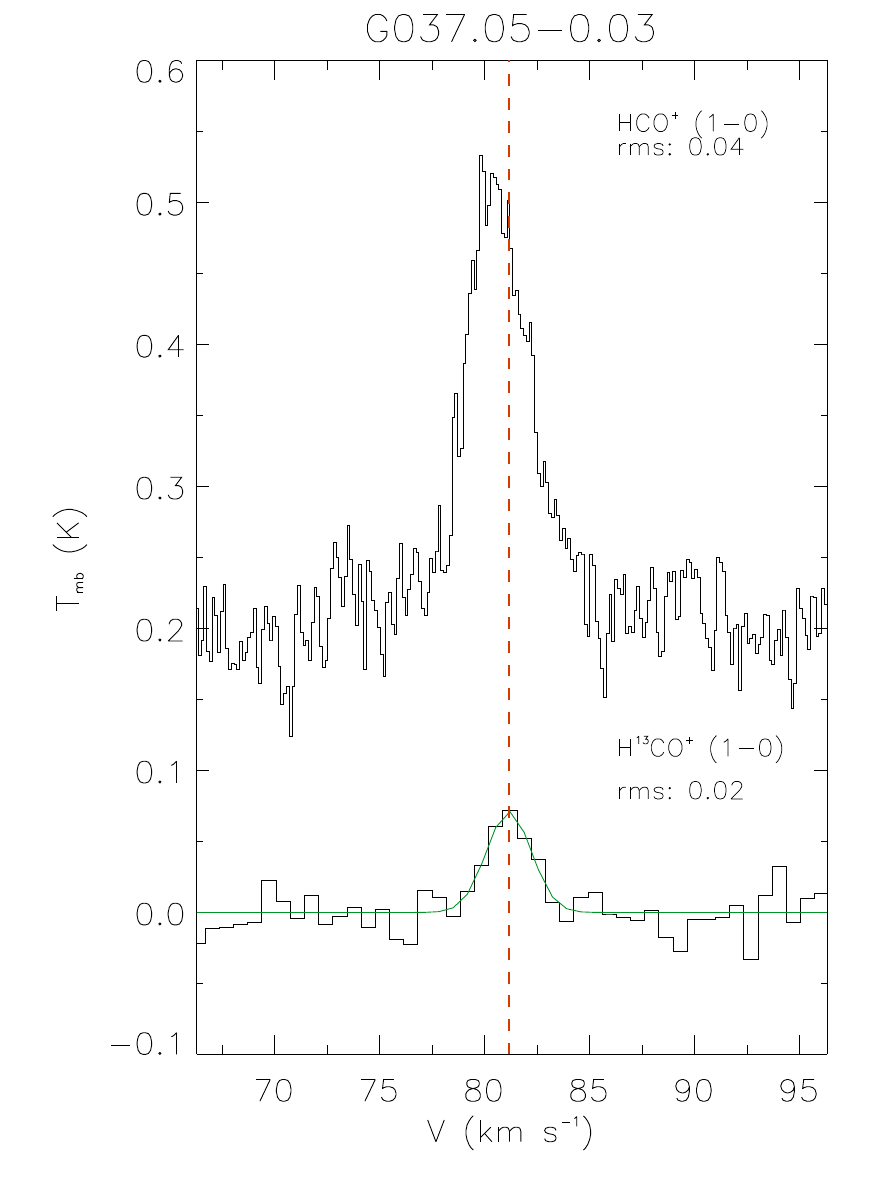}
  \end{minipage}%
  \begin{minipage}[t]{0.19\linewidth}
  \centering
   \includegraphics[width=39mm]{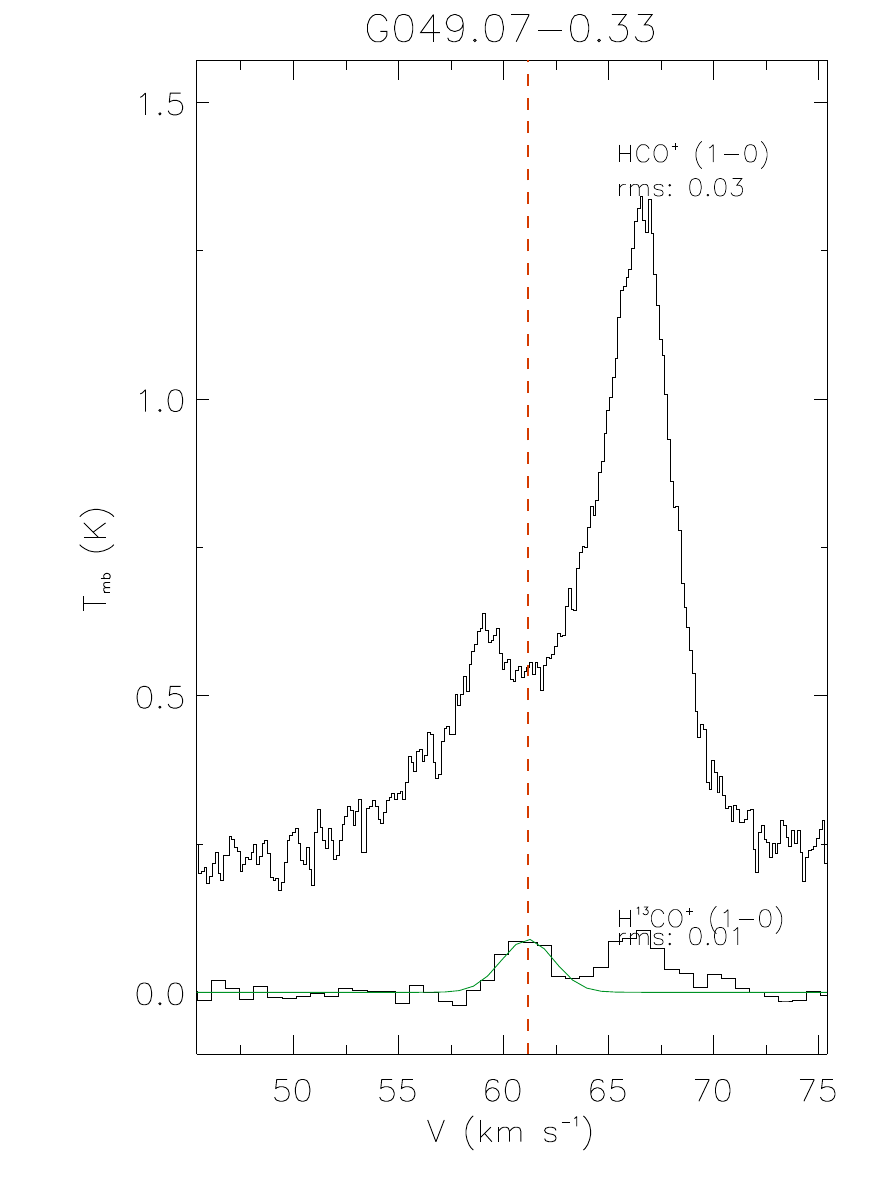}
  \end{minipage}%  
  
  \begin{minipage}[t]{0.19\linewidth}
  \centering
   \includegraphics[width=39mm]{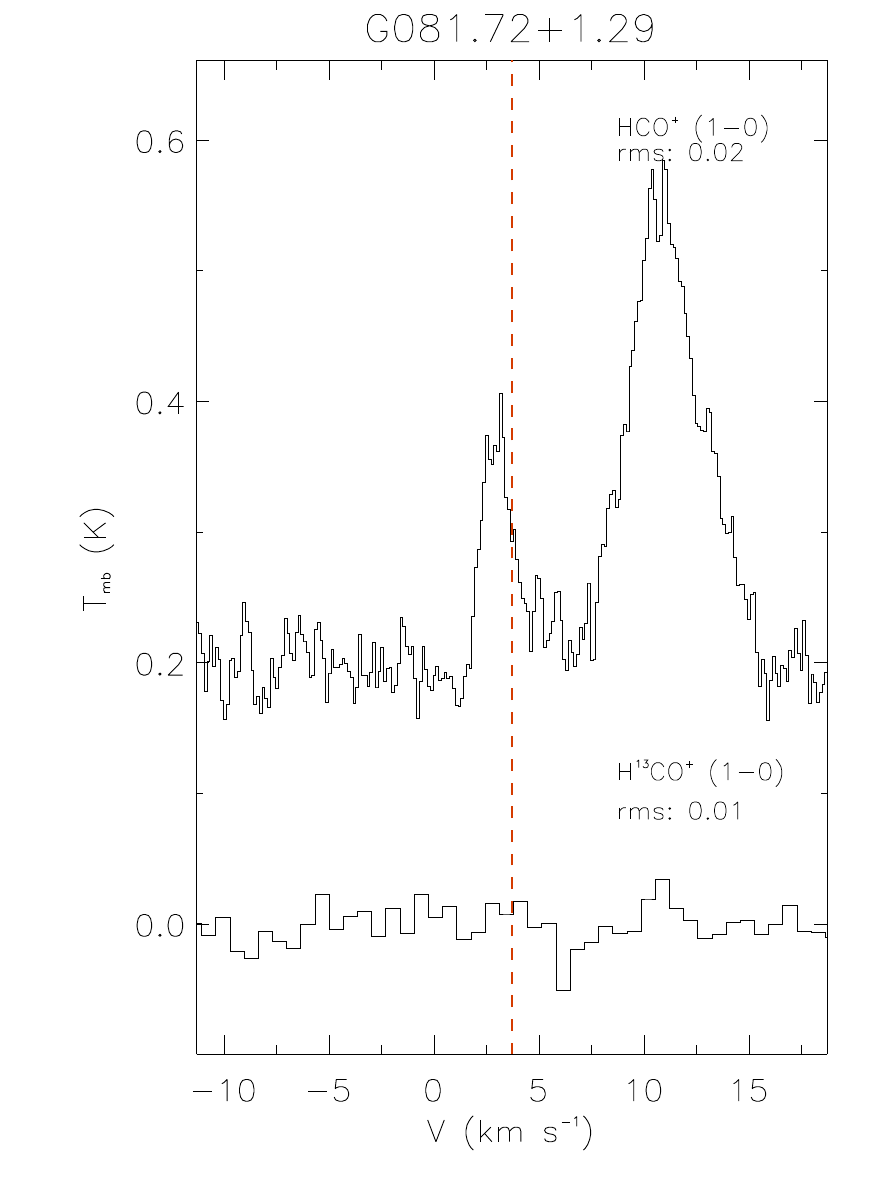}
  \end{minipage}%
  \begin{minipage}[t]{0.19\linewidth}
  \centering
   \includegraphics[width=39mm]{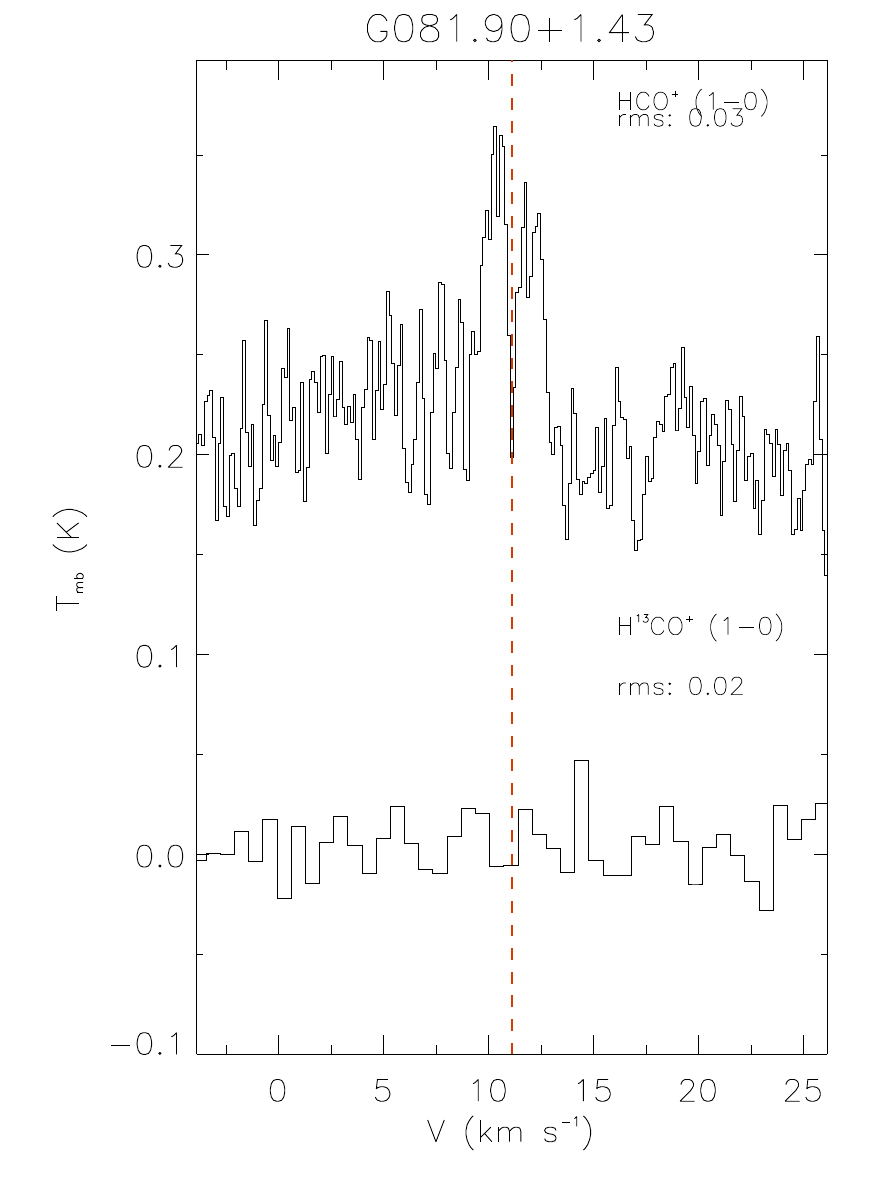}
  \end{minipage}%
    \begin{minipage}[t]{0.19\linewidth}
  \centering
   \includegraphics[width=39mm]{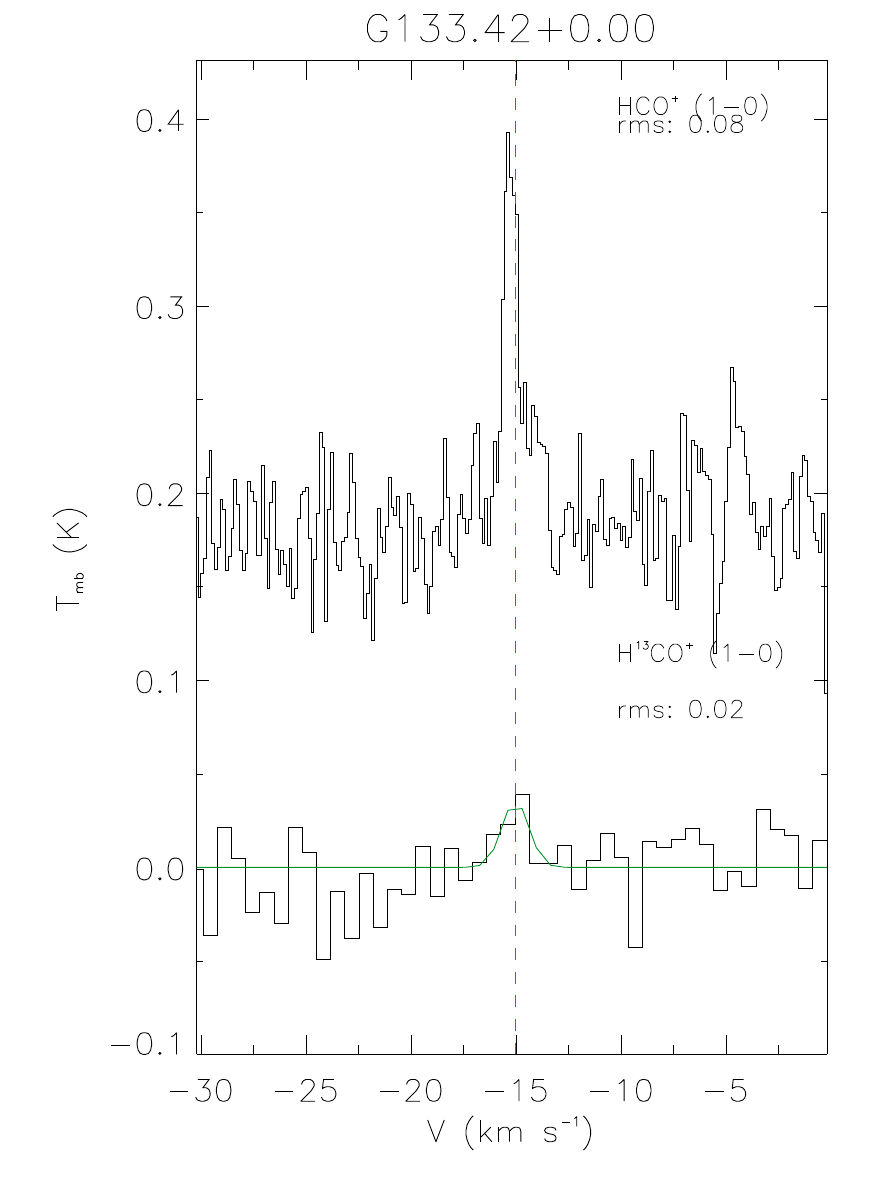}
  \end{minipage}%
\caption{The average spectra of HCO$^+$ (1-0) and H$^{13}$CO$^+$ (1-0) over the entire mapping area. The green lines show the results of Gaussian fitting of H$^{13}$CO$^+$ (1-0), and the dashed red lines indicate the central radial velocity of H$^{13}$CO$^+$ (1-0) (if H$^{13}$CO$^+$ emissions are not detected, we use C$^{18}$O data to trace the central radial velocity).
\label{fig:ave}}
\end{figure*}

Figure \ref{fig:ave} shows the average spectra of the entire mapping areas for the HCO$^+$ (1-0) and H$^{13}$CO$^+$ (1-0) lines. HCO$^+$ (1-0) emissions are detected in all targets, while H$^{13}$CO$^+$ (1-0) emissions are detected in ten of them (excluding G017.09+0.82, G081.72+1.29 and G081.90+1.43). The green lines in the figure represent the Gaussian fitting results of the H$^{13}$CO$^+$ (1-0) lines, while the dashed red lines indicate the central radial velocities. For the three targets where H$^{13}$CO$^+$ emissions were not detected, we adopt the central radial velocities of the C$^{18}$O (1-0) lines provided in Paper II. All the parameters of these two spectral lines are listed in Table \ref{Tab:result}. For the G049.07-0.33 and G081.72+1.29 targets, the optically thick lines show an additional velocity component near the velocity component we focused on, showing a stronger emission. And their optically thin lines also show emissions at corresponding velocity positions. This suggests the presence of an additional molecular cloud, which falls outside the scope of this study and will not be further analyzed.

\begin{figure*}[h]  
  \begin{minipage}[t]{0.5\linewidth}
  \centering
   \includegraphics[width=100mm]{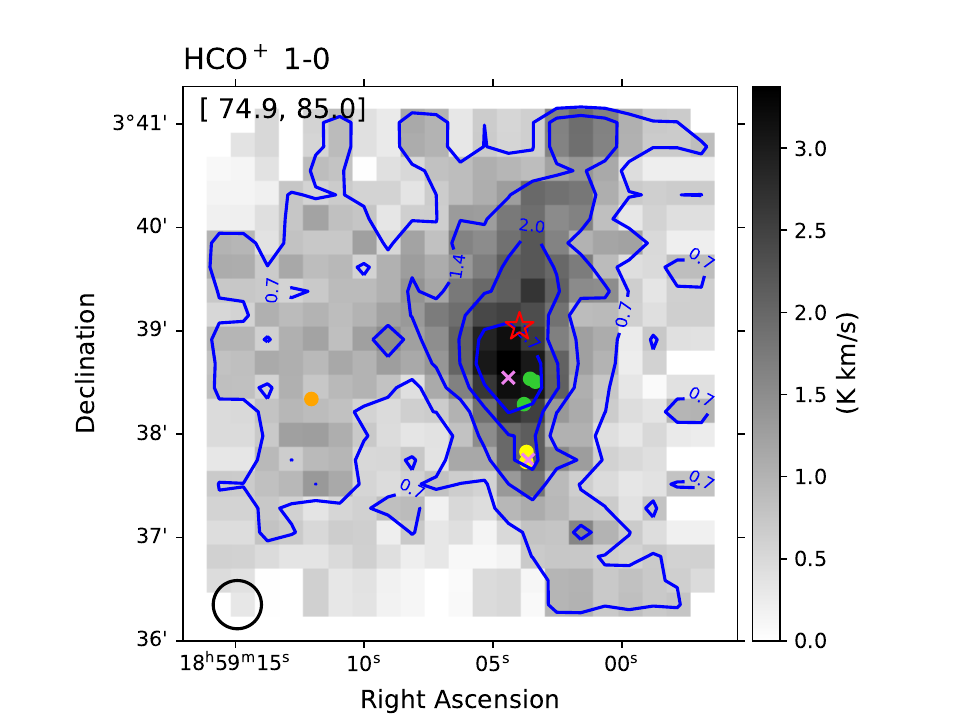}
  \end{minipage}%
  \begin{minipage}[t]{0.5\linewidth}
  \centering
   \includegraphics[width=100mm]{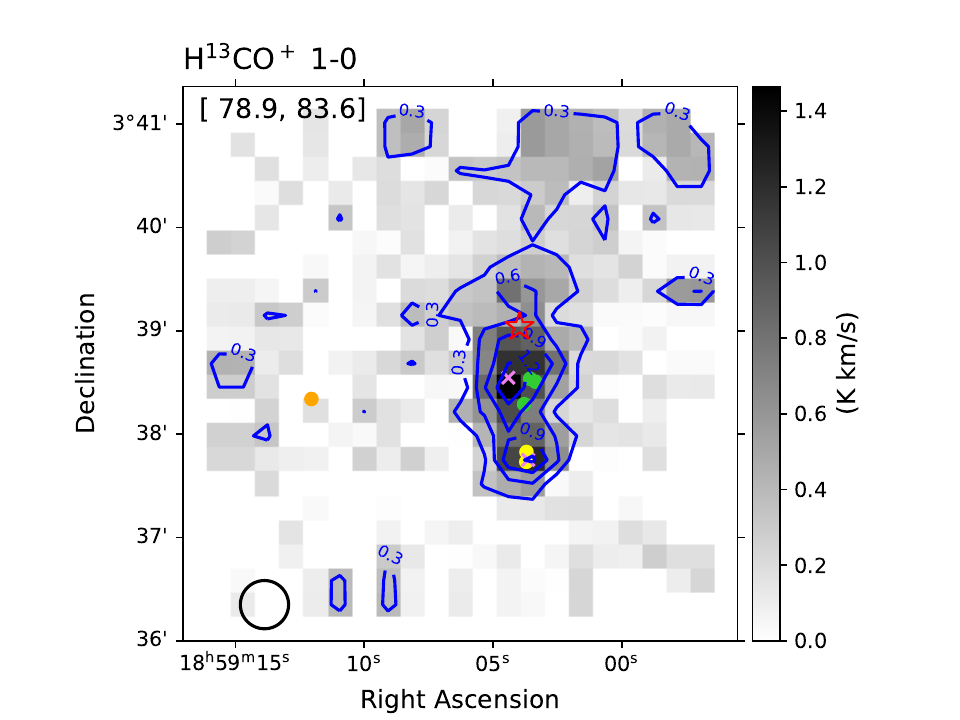}
  \end{minipage}% 
\caption{Example of target G037.05-0.03: HCO$^+$ (1-0) and H$^{13}$CO$^+$ (1-0) integrated intensity maps. The contours (blue) are the [0.2, 0.4, 0.6, 0.8] of the maximum value of its integrated intensity. The pentagram symbol denotes the position of the infall candidate identified in Paper II. The green, yellow, and orange points denote the Class I, Flat-spectrum, Class II YSOs, respectively. The magenta crosses denote the 6.7 GHz methanol masers. The black circles on the lower left indicate the beam size of the IRAM 30-m telescope.
\label{fig:map_eg}}
\end{figure*}

The HCO$^+$ (1-0) and H$^{13}$CO$^+$ (1-0) integrated intensity maps of all targets are shown in the Appendix \ref{sec:Appendix1}. Among them, ten targets show clumpy structures in the HCO$^+$ integrated intensity maps, while nine targets show clumpy structures in the H$^{13}$CO$^+$ integrated intensity maps (also listed in Table \ref{Tab:result}). Figure \ref{fig:map_eg} shows the maps of G037.05-0.03 as an example. In the figure, the pentagram symbol denotes the position of infall candidate identified in Paper II. Additionally, we display the positions of young stellar objects (YSOs) and masers on the observing field (the velocities of the masers are close to the V$\rm_{LSR}$ of the sources). The YSOs are identified in the Spitzer/IRAC candidate YSOs (SPICY) catalog \citep{Kuhn+etal+2021}, and we mark the Class I, Flat-spectrum, and Class II YSOs with green, yellow, and orange points, respectively. If the coordinates of our targets are outside the coverage range of the SPICY catalog, we use the AllWISE sources \citep{vizier:II328} and classify them based on the YSOs criteria suggested by \citet{Koenig+etal+2012}. The positions of the masers are sourced from \citet{Yang+etal+2017,Yang+etal+2019,Valdettaro+etal+2001,Anglada+etal+1996,Qiao+etal+2016,Qiao+etal+2018,Qiao+etal+2020}, etc., marked with crosses of different colors: blue for 95 GHz methanol maser, magenta for 6.7 GHz methanol maser, orange for H$_2$O maser, and red for OH maser).

In most cases, the spatial distributions of the clumps traced by the HCO$^+$ and its isotope H$^{13}$CO$^+$ lines are similar. For G012.87-0.22, G012.96-0.23, G014.00-0.17, and G017.09+0.82, it seems to have two distinct clumps in the observation areas. These clumps are labeled as A and B (as shown in the Figure \ref{fig:map}), and they are analyzed separately in the following sections. On the other hand, although the observation area was selected based on the MWISP CO mapping data, there are a few targets (i.e. G036.02-1.36, G081.72+1.29, and G081.90+1.43) that show weak emissions in the center of the observation area. This may be caused by HCO$^+$ and CO tracing different density regions, as well as the differences in angular resolution among the telescopes used.

%________________________________________ Table 3: Line parameters

\begin{table}
\begin{center}
  \caption{Average spectral line parameters}\label{Tab:result}  
%Please Capitalize the First Letter of Each Notional Word in table's caption
 \setlength{\tabcolsep}{1mm}{
\begin{tabular}{cccccccccc}
  \hline\noalign{\smallskip}
Source&  \multicolumn{5}{c}{H$^{13}$CO$^+$ (1-0)}  &  \multicolumn{4}{c}{HCO$^+$ (1-0)} \\
\cmidrule(lr){2-6}  \cmidrule(lr){7-10}
Name & V$_{peak}$ & FWHM & T$_{peak}$ & RMS & Clump & V$_{peak}$ & T$_{peak}$ & RMS & Clump \\
 & (km s$^{-1}$) & (km s$^{-1}$) & (K) & (K) & & (km s$^{-1}$) & (K) & (K) & \\
  \hline\noalign{\smallskip}
G012.79-0.20  &  35.56(0.02)   &  4.12(0.02)   &  0.57(0.03)   &  0.01  &  yes  &  35.1  &  2.59  &  0.02  &  yes  \\
G012.87-0.22  &  36.46(0.01)   &  3.24(0.03)   &  0.34(0.03)   &  0.02  &  yes  &  38.4  &  0.85  &  0.04  &  yes  \\
G012.96-0.23  &  35.16(0.01)   &  3.99(0.07)   &  0.23(0.01)   &  0.02  &  yes  &  32.6  &  0.56  &  0.02  &  yes  \\
G014.00-0.17  &  40.53(0.02)   &  2.96(0.03)   &  0.15(0.02)   &  0.03  &  yes  &  42.3  &  0.81  &  0.05  &  yes  \\
G014.25-0.17  &  38.17(0.02)   &  2.69(0.02)   &  0.13(0.02)   &  0.02  &  yes  &  35.6  &  0.46  &  0.05  &  yes  \\
G017.09+0.82  &  Nd  &  Nd  &  Nd  &  0.02  &  no  &  20.5  &  0.53  &  0.04  &  yes  \\
G025.82-0.18  &  93.49(0.01)   &  3.64(0.01)   &  0.12(0.01)   &  0.03  &  yes  &  92.8  &  0.81  &  0.05  &  yes  \\
G036.02-1.36  &  31.86(0.01)   &  1.06(0.02)   &  0.08(0.02)   &  0.01  &  no  &  31.9  &  0.16  &  0.02  &  no  \\
G037.05-0.03  &  81.17(0.01)   &  2.52(0.04)   &  0.07(0.01)   &  0.02  &  yes  &  79.8  &  0.33  &  0.03  &  yes  \\
G049.07-0.33  &  61.16(0.01)   &  2.93(0.01)   &  0.09(0.03)   &  0.01  &  yes  &  66.6  &  1.14  &  0.04  &  yes  \\
G081.72+1.29  &  Nd  &  Nd  &  Nd  &  0.01  &  no  &  10.9  &  0.38  &  0.02  &  no  \\
G081.90+1.43  &  Nd  &  Nd  &  Nd  &  0.02  &  no  &  10.3  &  0.16  &  0.02  &  no  \\
G133.42+0.00  &  -15.03(0.02)   &  1.50(0.01)   &  0.05(0.02)   &  0.02  &  yes  &  -15.4  &  0.19  &  0.03  &  yes  \\
  \hline\noalign{\smallskip}
\end{tabular}}
\end{center}
\tablecomments{
Nd denotes non-detection with H$^{13}$CO$^+$ (1-0) emission. H$^{13}$CO$^+$ line parameters are derived from Gaussian fitting. The values in parentheses are the uncertainty of the Gaussian fitting. HCO$^+$ line parameters are directly derived from the observed data.}
\end{table}

\section{Analysis and Discussion} \label{sec:analysis}

\subsection{Physical Parameters of the clumps} \label{subsec:clumps}

The obtained mapping data allows us to estimate the physical parameters of the observed clumps. Table \ref{Tab:clump_para} lists the properties of the clumps, including the radius, kinetic temperature, H$_2$ density, and mass. The radius values are obtained by performing 2-dimensional Gaussian fitting on the H$^{13}$CO$^+$ integrated intensity maps. Since the clumps are often non-spherical, we adopt the average radius as $R = \sqrt{a \times b}$, where $a$ and $b$ represent the semimajor and semiminor axes obtained from the fitting results, respectively. The aspect ratios are also listed in the Table \ref{Tab:clump_para}. All angular sizes have been converted to the linear sizes. The kinetic temperature and H$_2$ column densities presented in columns (4) and (5) are derived from the MWISP CO data (see Section 3.1 of Paper II for more details). The mass values of the clumps are roughly estimated by $M=\pi R^2 \mu m_H N(H_2)$, where $R$ is the average radius, $\mu$ is the mean molecular weight of the interstellar medium (adopted as $\mu=2.8$), $m_H$ is the mass of a hydrogen atom, and $N(H_2)$ is the H$_2$ column density. The H$_2$ column density is the mean column density value given in Paper II. The H$_2$ densities can estimated from the H$_2$ column densities by $\rho=\frac{3}{4R} N(H_2)$ under assumption that the clump is a uniform sphere. However, since the beam size of the 13.7-m telescope is approximately twice that of the IRAM 30-m telescope, beam dilution may make us underestimate the density and mass of the clump. Furthermore, the uncertainty in distance estimates can introduce a significant error in determining the clump radius. Therefore, the values given here are only approximate estimates, providing the order of magnitude for the clump masses.

We use the 1D non-LTE RADEX radiative transfer code \citep{vanderTak+etal+2007} to estimate the excitation temperatures and column densities of H$^{13}$CO$^+$. The RADEX code required input parameters such as the spectral range, the cosmic microwave background temperature (i.e., 2.73 K), the kinetic temperature, H$_2$ volume density of the clump, and the molecular line width. We iterate over the molecular column density in a loop, until the simulated line intensity matches the peak intensity of the observed H$^{13}$CO$^+$ lines. The results show that the excitation temperatures of H$^{13}$CO$^+$ are in the range of several to ten Kelvin, while the H$^{13}$CO$^+$ column densities range from $10^{12}$ to $10^{13}$ cm$^{-2}$. Combining these with the H$_2$ column densities, we obtain abundance ratios of H$^{13}$CO$^+$: [H$^{13}$CO$^+$]/[H$_2$] $\approx$ $10^{-12}$ -- $10^{-10}$. In order to estimate the uncertainties in column densities and excitation temperatures due to errors in the input parameters (mainly from the kinetic temperature and H$_2$ density), we conduct multiple calculations by slightly varying the input parameters (considering an uncertainty of approximately 20\% as mentioned in Paper II). The obtained results are listed in Table \ref{Tab:clump_para}. However, it is important to note that the uncertainty could be further amplified due to the uncertainty in radius which for the sources with kinematic distances could be high, as well as the local thermal equilibrium assumption and the uncertainty in the [C$^{18}$O]/[H$_2$] abundance ratio used for estimating the H$_2$ density.

On the other hand, we do not estimate the excitation temperatures and column densities of HCO$^+$ in this way, because the HCO$^+$ (1-0) line profiles consistently show self-absorption. This makes it difficult to determine the observed intensity accurately, resulting in significant uncertainty in the physical parameters. Therefore, we estimate the HCO$^+$ column densities using the C/$^{13}$C abundance ratio. In this paper, we adopt a ratio of 69 \citep{Wilson+1999}. However, some studies have indicated a broader range of C/$^{13}$C ratios in diffuse interstellar clouds \citep[e.g.,][]{Yan+etal+2019}, which may introduce additional uncertainty to the obtained results.

%________________________________________ Table 4: Physical parameters of the clumps

\begin{table}
\begin{center}
\caption{Physical parameters of the clumps.}\label{Tab:clump_para}
\setlength{\tabcolsep}{0.5mm}{
\begin{tabular}{cccccccccccccc}   
  \hline\noalign{\smallskip}
Source & Radius & Aspect Ratio & $T_{kin}$ $^1$ & log($\frac{N({\rm H}_2)}{{\rm cm}^{-2}}$)$^1$ & log($\frac{\rho}{{\rm cm}^{-3}}$) & log($\frac{M}{M_{\odot}}$) & $T_{ex}$(H$^{13}$CO$^+$) & log($\frac{N({\rm H^{13}CO}^+)}{{\rm cm}^{-2}}$)  & log($\frac{N({\rm HCO}^+)}{{\rm cm}^{-2}}$)  & log($\frac{[\rm H^{13}CO^+]}{[\rm H_2]}$) \\
Name  & (pc) & & (K) &  &  &  & (K) &  &  &  \\ 
  \hline\noalign{\smallskip} 
G012.79-0.20   & 0.42(0.07) & 1.28 & 33.0  & 23.1  & 4.9  & 3  & 9.3$_{-2.0}^{+3.3}$ & 13.0$_{-0.05}^{+0.09}$ & 14.8 & -10.1 \\
G012.87-0.22 A   & 0.36(0.15) & 3.38 & 22.3  & 23.2  & 5.1  & 3  & 9.2$_{-2.5}^{+4.1}$ & 12.2$_{-0.03}^{+0.06}$ & 14.0 & -11.0 \\
G012.87-0.22 B   & 0.32(0.23) & 1.26 & 25.5  & 23.0  & 4.9  & 3  & 7.1$_{-1.5}^{+2.2}$ & 12.1$_{-0.05}^{+0.08}$ & 13.9 & -10.9 \\
G012.96-0.23 A  & 0.13(0.03) & 1.64 & 17.1  & 23.1  & 5.3  & 2  & 12.3$_{-4.1}^{+7.7}$ & 11.7$_{-0.01}^{+0.04}$ & 13.5 & -11.4 \\
G012.96-0.23 B  & 0.25(0.17) & 2.23 & 17.1  & 23.1  & 5.1  & 3  & 7.3$_{-1.7}^{+2.4}$ & 12.1$_{-0.04}^{+0.07}$ & 13.9 & -11.0 \\
G014.00-0.17 A & 0.31(0.15) & 1.44  & 26.0  & 22.8  & 4.7  & 3  & 7.1$_{-1.5}^{+2.1}$ & 12.0$_{-0.05}^{+0.08}$ & 13.8 & -10.9 \\
G014.00-0.17 B & 0.43(0.16) & 1.78 & 22.3  & 22.9  & 4.7  & 3  & 6.2$_{-1.1}^{+1.5}$ & 12.0$_{-0.06}^{+0.09}$ & 13.8 & -11.0 \\
G014.25-0.17   & 0.54(0.28) & 1.99 & 20.9  & 22.9  & 4.5  & 3  & 4.8$_{-0.6}^{+0.8}$ & 12.0$_{-0.07}^{+0.11}$ & 13.8 & -10.9 \\
G017.09+0.82 A & 0.32(0.23) & 2.44 & 25.0  & 22.6  & 4.5  & 2  & Nd & Nd & ... & Nd \\
G017.09+0.82 B  & 0.16(0.06) & 1.05  & 25.0  & 22.6  & 4.8  & 2  & Nd & Nd & ... & Nd \\
G025.82-0.18   & 0.53(0.12) & 1.02  & 17.4  & 22.8  & 4.4  & 3  & 4.3$_{-0.4}^{+0.5}$ & 12.6$_{-0.09}^{+0.14}$ & 14.4 & -10.1 \\
G036.02-1.36   & ... & & 10.0  & 21.9  & ... & ... & ... & ... & ... & ... \\
G037.05-0.03   & 0.36(0.14) & 2.26  & 13.2  & 22.5  & 4.3  & 2  & 3.8$_{-0.5}^{+0.9}$ & 12.3$_{-0.11}^{+0.16}$ & 14.1 & -10.2 \\
G049.07-0.33   & 0.51(0.22) & 1.28 & 19.5  & 22.5  & 4.1  & 3  & 3.7$_{-0.4}^{+1.2}$ & 12.7$_{-0.12}^{+0.16}$ & 14.5 & -9.8 \\
G081.72+1.29   & ... & & 13.4  & 22.1  & ... & ... & Nd & Nd & ... & Nd \\
G081.90+1.43   & ... & & 13.7  & 21.7  & ... & ... & Nd & Nd & ... & Nd \\
G133.42+0.00   & 0.07(0.04) & 1.38 & 10.6  & 21.7  & 4.2  & $\textless$ 1 & 3.6$_{-0.5}^{+1.3}$ & 11.9$_{-0.13}^{+0.19}$ & 13.7 & -9.8 \\
  \hline\noalign{\smallskip}

\end{tabular}}
\end{center}
\tablecomments{
Nd denotes non-detection with H$^{13}$CO$^+$ (1-0) emissions. $^1$ The values of $T_{kin}$ and N(H$_2$) for these sources are obtained from Paper II. $T_{kin}$ is derived from the excitation temperature of CO, with an uncertainty of less than 20\%. The N(H$_2$) value is obtained from the column density of C$^{18}$O through the abundance ratio of [C$^{18}$O]/[H$_2$], which can have an error several times larger than the N(H$_2$) value.
}
\end{table}

\subsection{Line Profiles and Asymmetries} \label{subsec:profiles}

In order to obtain the spatial distributions of the optically thick line profiles for the 13 sources, we use the mapping data to draw HCO$^+$ (1-0) map grids. Figure \ref{fig:mapgrid_eg} shows an example, while the map grids for all targets are provided in Appendix \ref{sec:Appendix2} Figure \ref{fig:mapgrid}. The grid size corresponds to half of the beam size of the IRAM 30-m telescope, which is 14$\arcsec$. 

A dimensionless parameter, $\delta V$, is commonly used to quantify the asymmetry of spectral line profiles \citep{Mardones+etal+1997}: $\delta V = (V_{thick} - V_{thin})/\Delta V_{thin}$, where $V_{thick}$ and $V_{thin}$ represent the peak velocities of the optically thick and thin lines, and $\Delta V_{thin}$ is the FWHM of the optically thin line. Among them, the $V_{thick}$ value is derived from the brightest emission peak position of HCO$^+$ (1-0) line, while $V_{thin}$ and $\Delta V_{thin}$ are obtained from the Gaussian fitting results of H$^{13}$CO$^+$ (1-0) line. For the three targets G017.09+0.82, G081.72+1.29, and G081.90+1.43, where H$^{13}$CO$^+$ emissions are not detected, we use the C$^{18}$O (1-0) data from Paper II instead. Based on the values of the dimensionless parameter $\delta V$, which can be negative, positive, or close to zero, the line profiles of HCO$^+$ can be categorized into blue profiles, red profiles, and others (e.g., symmetry profiles and multi-peaked profiles). In Figure \ref{fig:mapgrid_eg}, the grids that show a blue profile in the HCO$^+$ line and H$^{13}$CO+ shows significant emission (signal-to-noise ratio greater than 3) are marked with a ``B" in the upper-left corner, while those with other types of profiles remain unmarked.

Except for G081.72+1.29 and G081.90+1.43, the HCO$^+$ line profiles of the other targets show obvious evidence of gas infall motions, as denoted by the red boxes in the map grids. Among these infall sources, the HCO$^+$ line profiles of G012.79-0.20 show complexity. The emission lines in the central area of the clump show a three-peaked profile, with velocity ranges approximately [28.4, 32.9], [32.9, 38.9], and [38.9, 60.0] km s$^{-1}$. Its H$^{13}$CO$^+$ line profile is mostly unimodal, with a center velocity of 35.56 km s$^{-1}$. But its line width is relatively wide, suggesting the possibility of multiple velocity components overlapping. This source may have complex internal structures or gas motions, such as infall and outflow. However, due to the insufficient spatial resolution in this study, it is difficult to provide a definitive conclusion. For G012.87-0.22 region 1, the HCO$^+$ line profiles show both red and blue profiles. Blue profiles are predominantly observed in the central region of the clump, while red profiles are found in the southeastern region of the clump. In the case of G014.00-0.17, the HCO$^+$ blue profiles are predominantly distributed in the region surrounding clumps A and B, while the HCO$^+$ lines in the central region of these two clumps show red profiles, which may indicate the presence of outflow \citep[e.g.,][]{Li+etal+2019}. For G049.07-0.33, since the presence of another closely located velocity component, the blue profiles can only be observed in the northern part of the red box region. The blue profiles of the remaining infall sources are mainly concentrated on the clumps, and six clumps of them show significant global collapse (i.e. infall signatures are observed all across the clump, marked in Table \ref{Tab:infall}). In the following section, we will use two models to further analyze the infall properties of these clumps.

\begin{figure*}[h]  
  \centering
   \includegraphics[width=200mm]{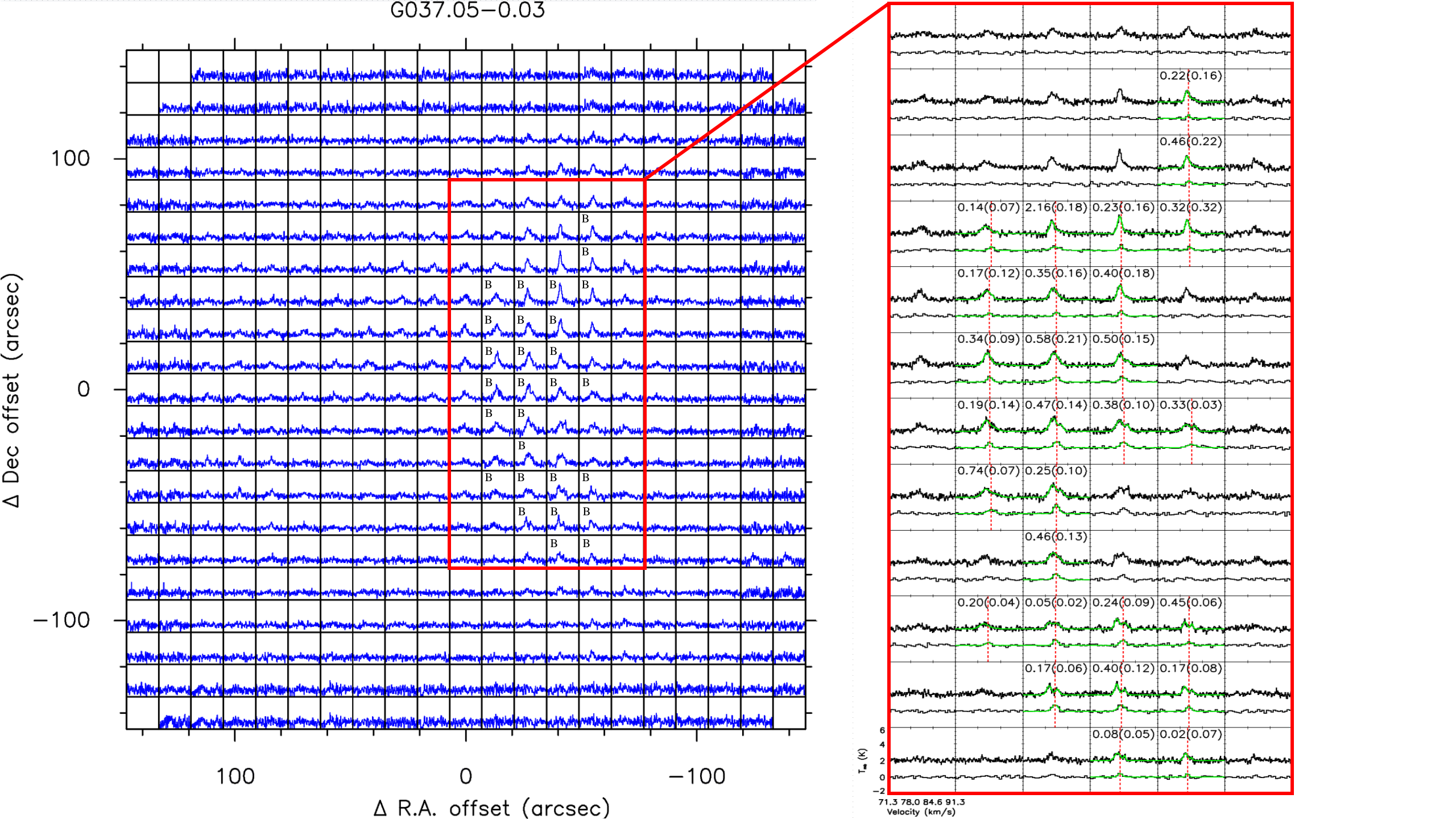}
\caption{The left panel: An example of target G037.05-0.03 HCO$^+$ (1-0) map grid (gridded to half of beam size). The axes plot the offsets $\Delta$ R.A. and $\Delta$ Dec relative to the coordinates from Table \ref{Tab:src-catalog}. The right panel: A HCO$^+$ and H$^{13}$CO$^+$ map grids that zooms in the red box area of the top left panel, and the gas infall velocity and its uncertainty (in parentheses) are marked in each grid. %The lower panel: The radius vs the infall velocity of G037.05-0.03.
\label{fig:mapgrid_eg}}
\end{figure*}

\subsection{Infall Velocity and Distribution} \label{subsec:Vin}

\begin{figure*}[h]  

  \begin{minipage}[t]{0.5\linewidth}
  \centering
   \includegraphics[width=90mm]{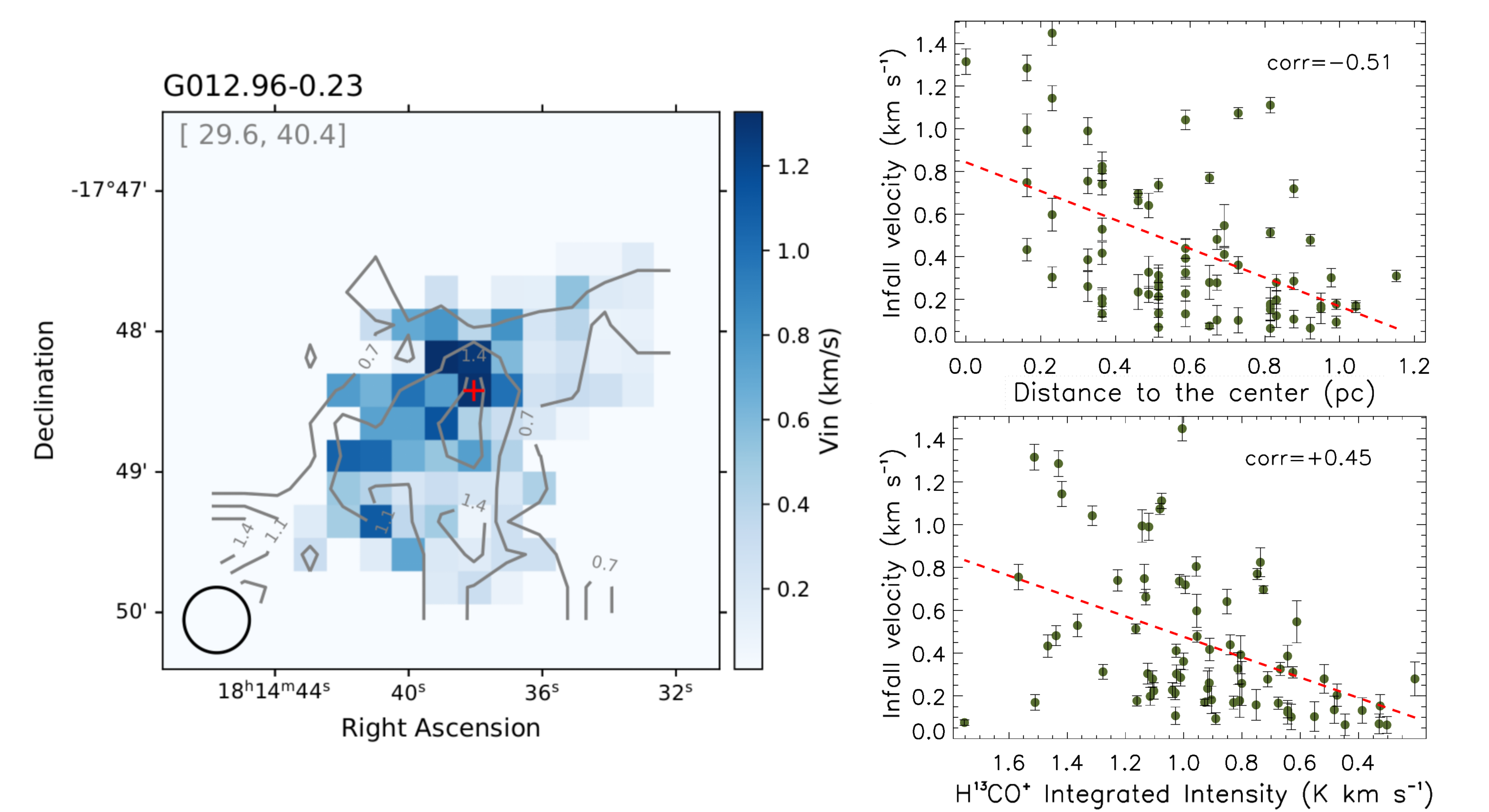}
  \end{minipage}% 
  \begin{minipage}[t]{0.5\linewidth}
  \centering
   \includegraphics[width=90mm]{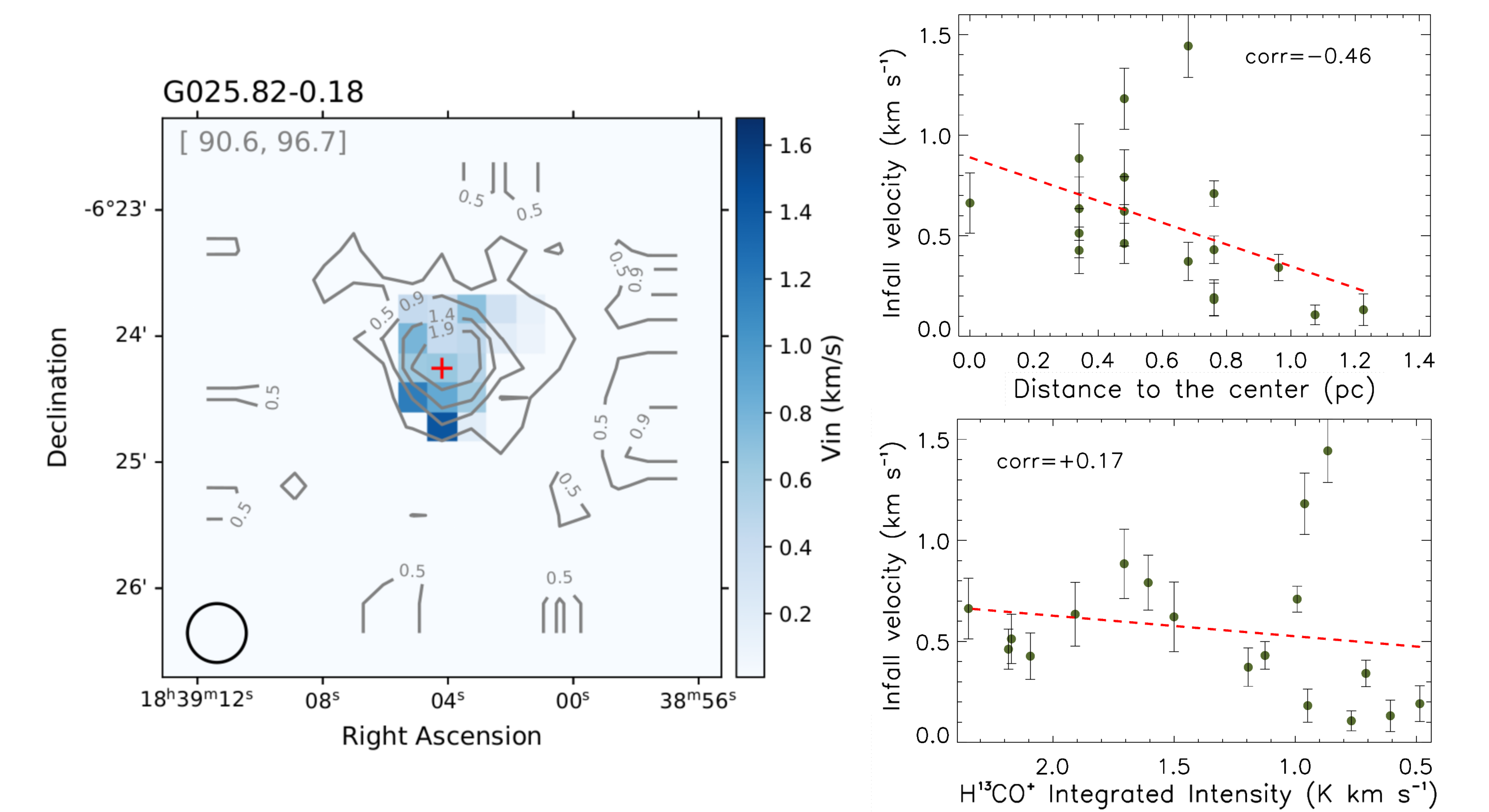}
  \end{minipage}%
  
  \begin{minipage}[t]{0.5\linewidth}
  \centering
   \includegraphics[width=90mm]{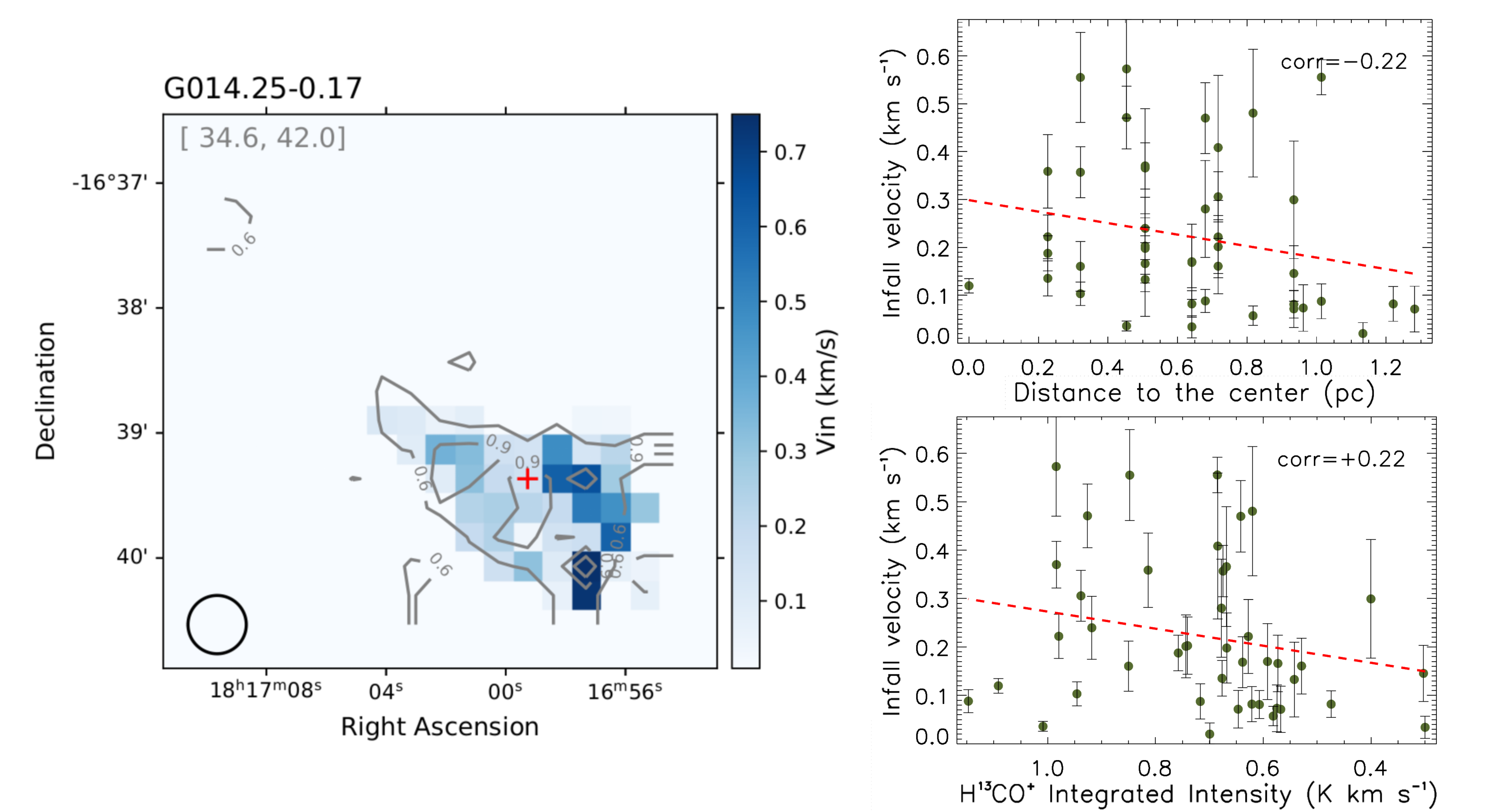}
  \end{minipage}%
  \begin{minipage}[t]{0.5\linewidth}
  \centering
   \includegraphics[width=90mm]{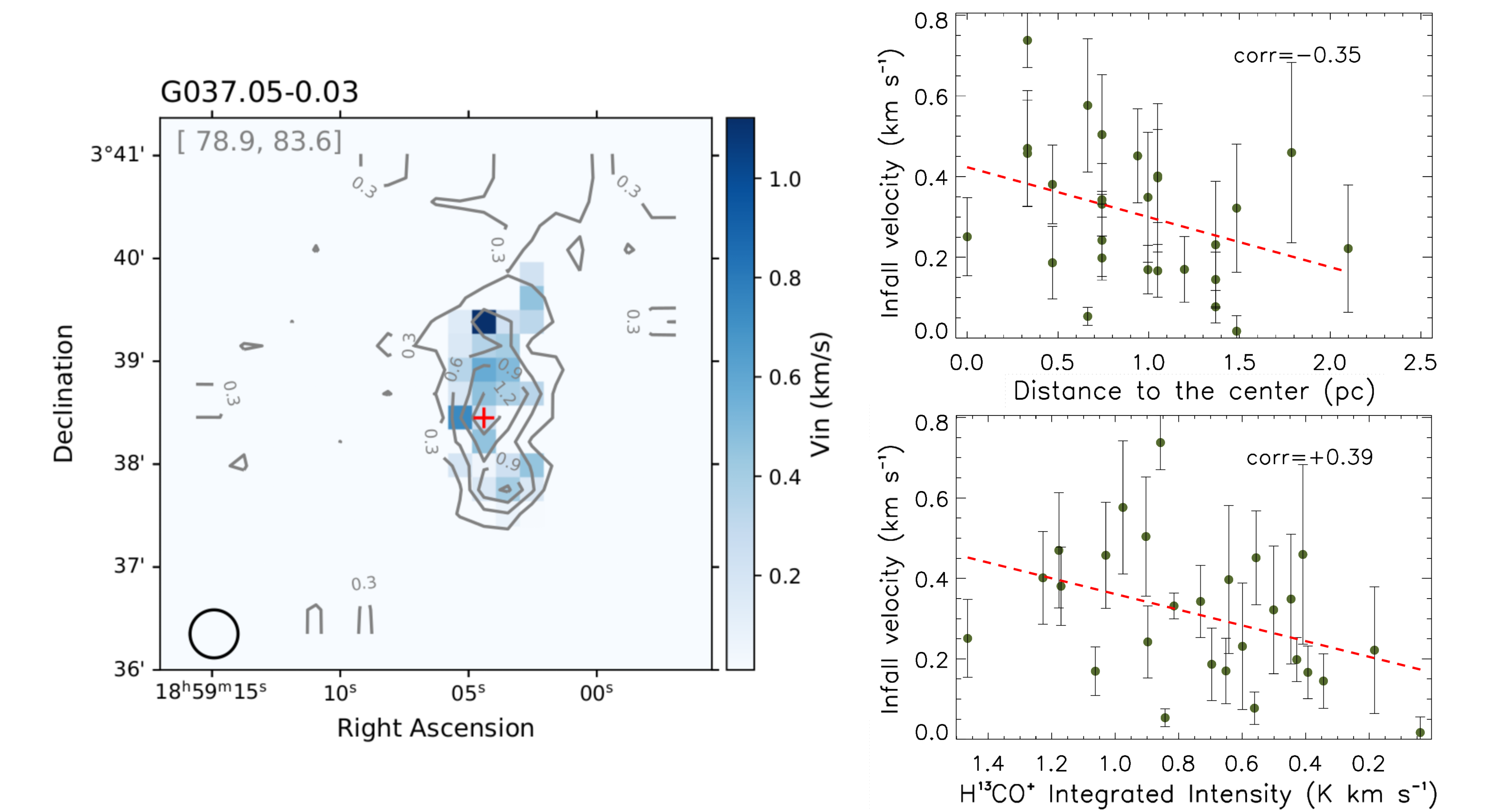}
  \end{minipage}% 

  \begin{minipage}[t]{0.6\linewidth}
  \centering
   \includegraphics[width=118mm]{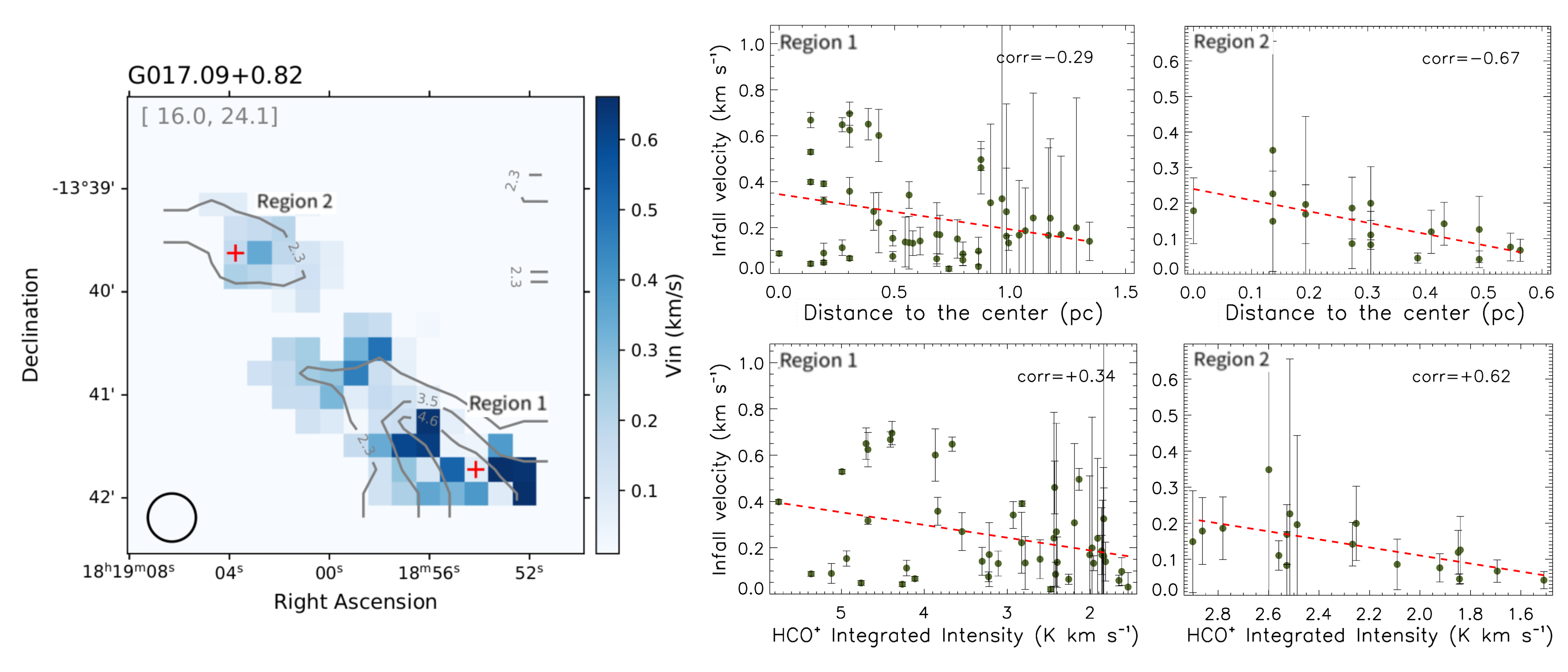}
  \end{minipage}% 

\caption{The spatial distribution maps of the infall velocity (the left panel of each subfigure) and the variation images of infall velocity along the distance to the clump center and the H$^{13}$CO$^+$ integrated intensity (the right panel of each subfigure, from top to bottom). The contours (gray) are the H$^{13}$CO$^+$ integrated intensity values of each sources. For the sources without H$^{13}$CO$^+$ emissions detected, the integrated intensity of HCO$^+$ is used. The red plus sign marked the position of clump center in each velocity distribution map. The red dotted line represents the regression line of the two variables in the right panel of each subfigure.
\label{fig:vin_r}}
\end{figure*}

\begin{figure*}[h]       
  \begin{minipage}[t]{0.29\linewidth}
  \centering
   \includegraphics[width=55mm]{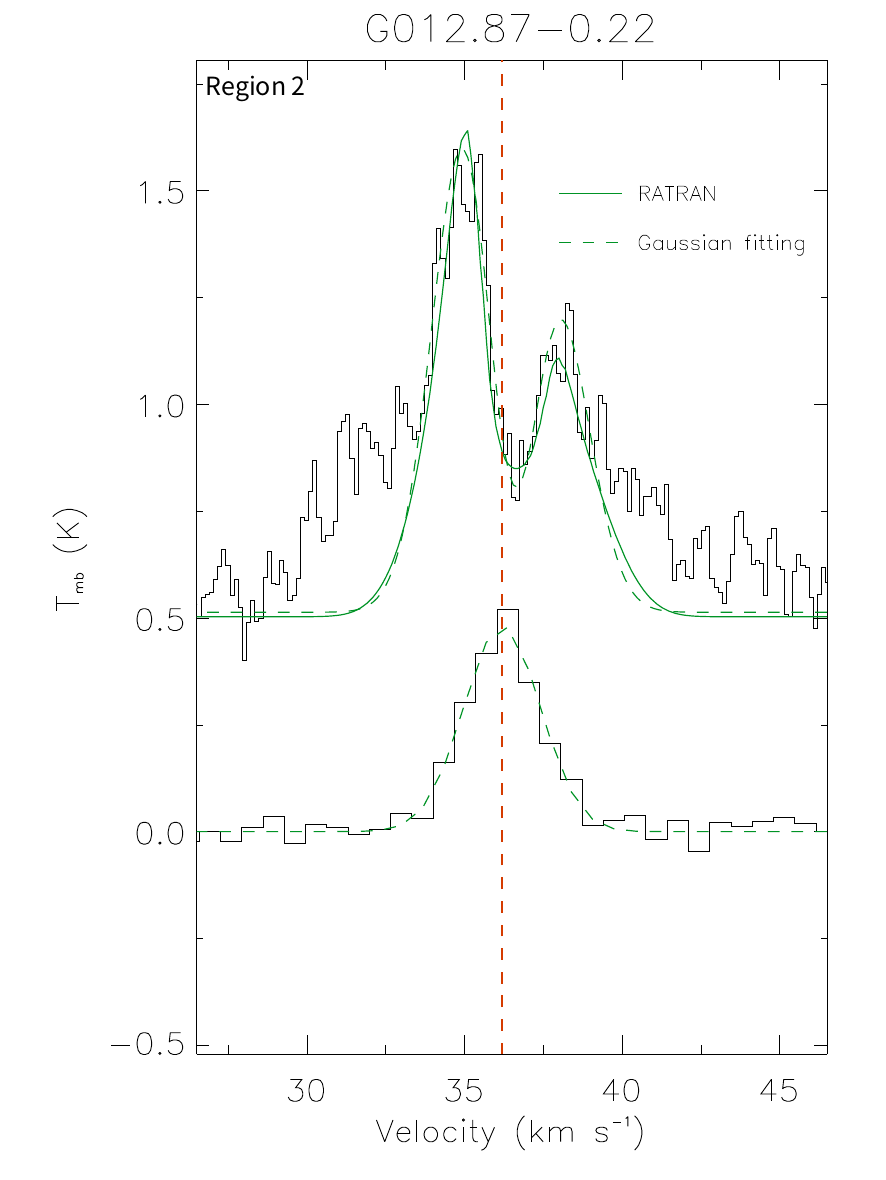}
  \end{minipage}%
  \begin{minipage}[t]{0.29\linewidth}
  \centering
   \includegraphics[width=55mm]{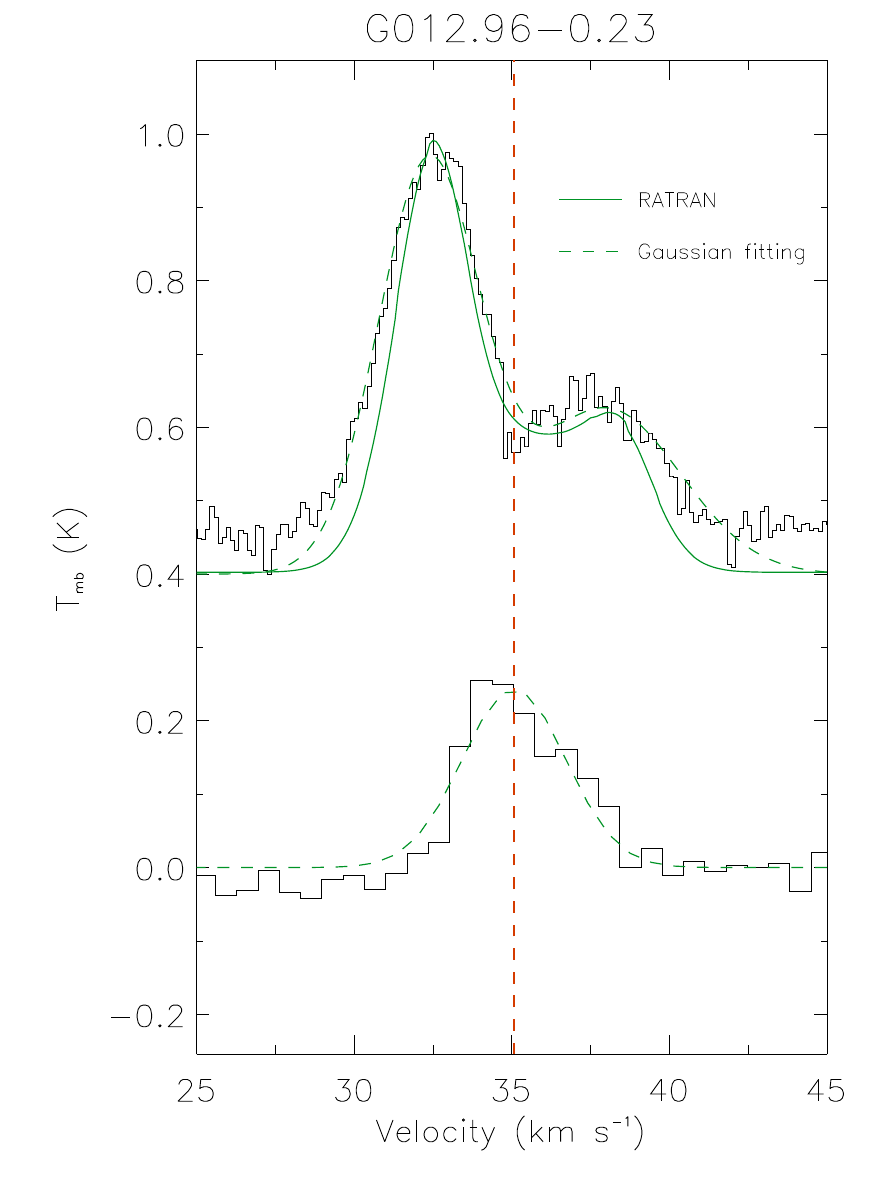}
  \end{minipage}%
  \begin{minipage}[t]{0.29\linewidth}
  \centering
   \includegraphics[width=55mm]{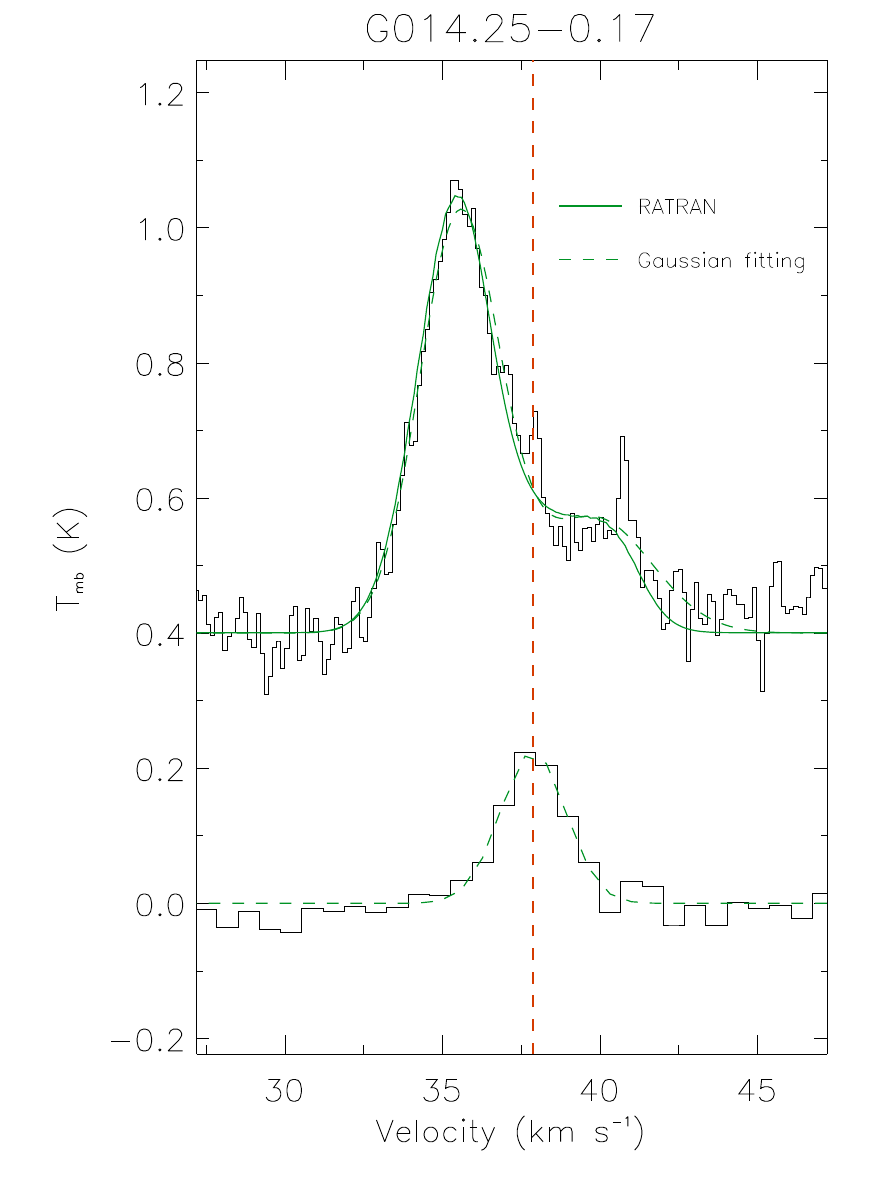}%*
  \end{minipage}%
  
  \begin{minipage}[t]{0.29\linewidth}
  \centering
   \includegraphics[width=55mm]{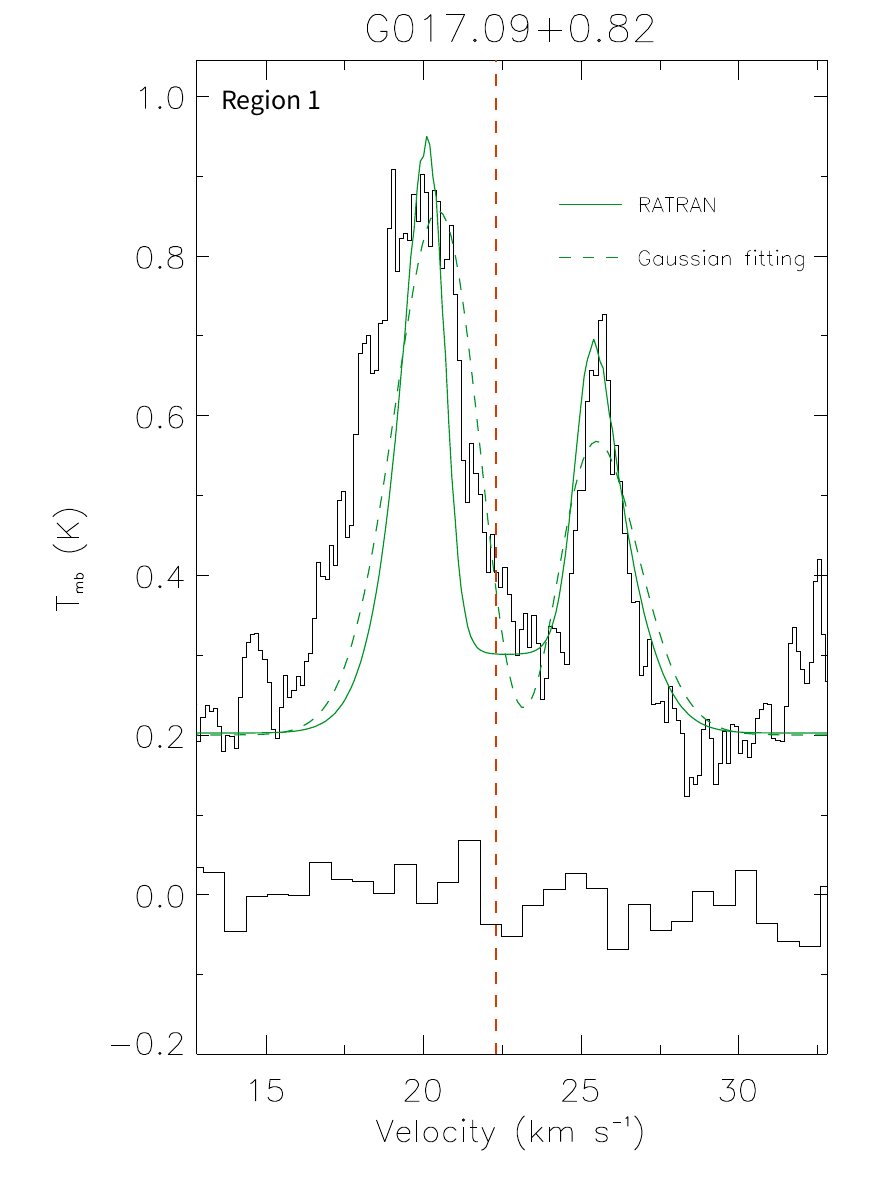}
  \end{minipage}%
  \begin{minipage}[t]{0.29\linewidth}
  \centering
   \includegraphics[width=55mm]{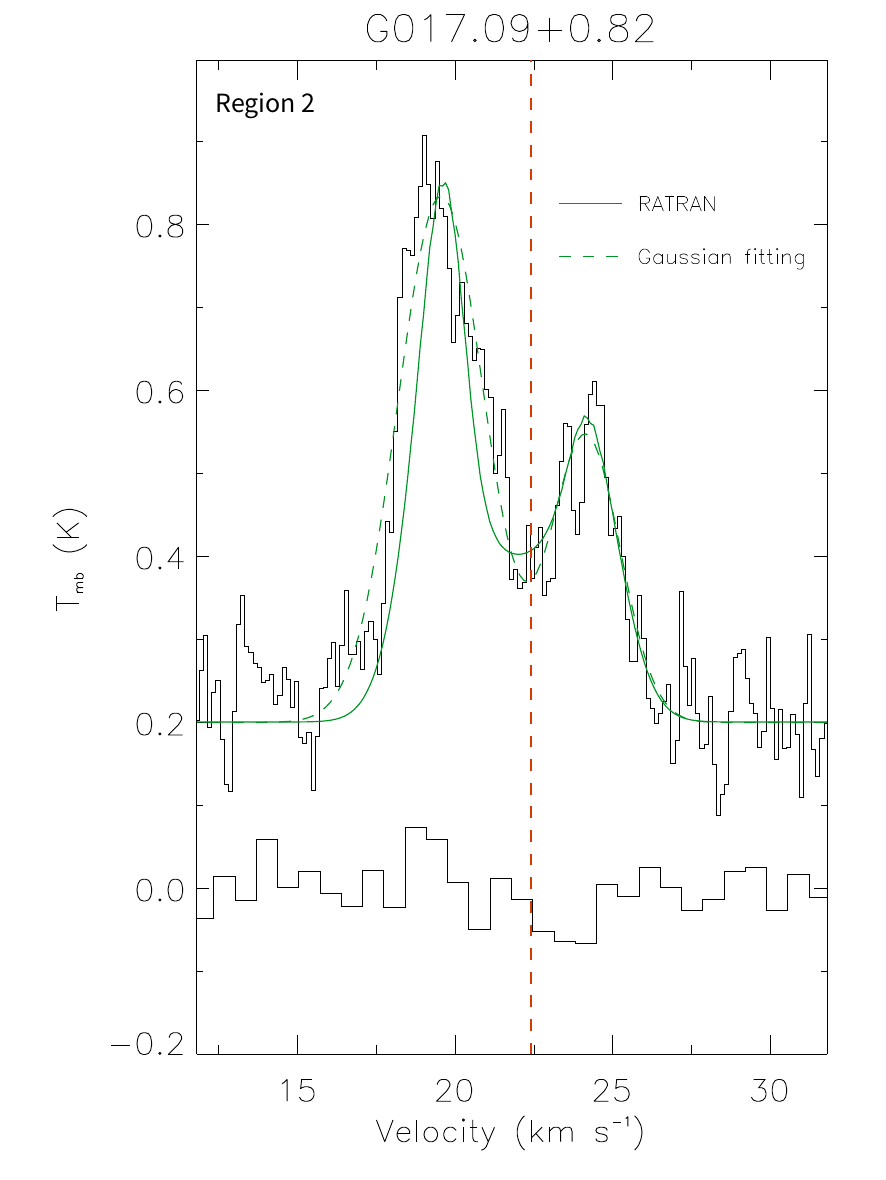}
  \end{minipage}%
  \begin{minipage}[t]{0.29\linewidth}
  \centering
   \includegraphics[width=55mm]{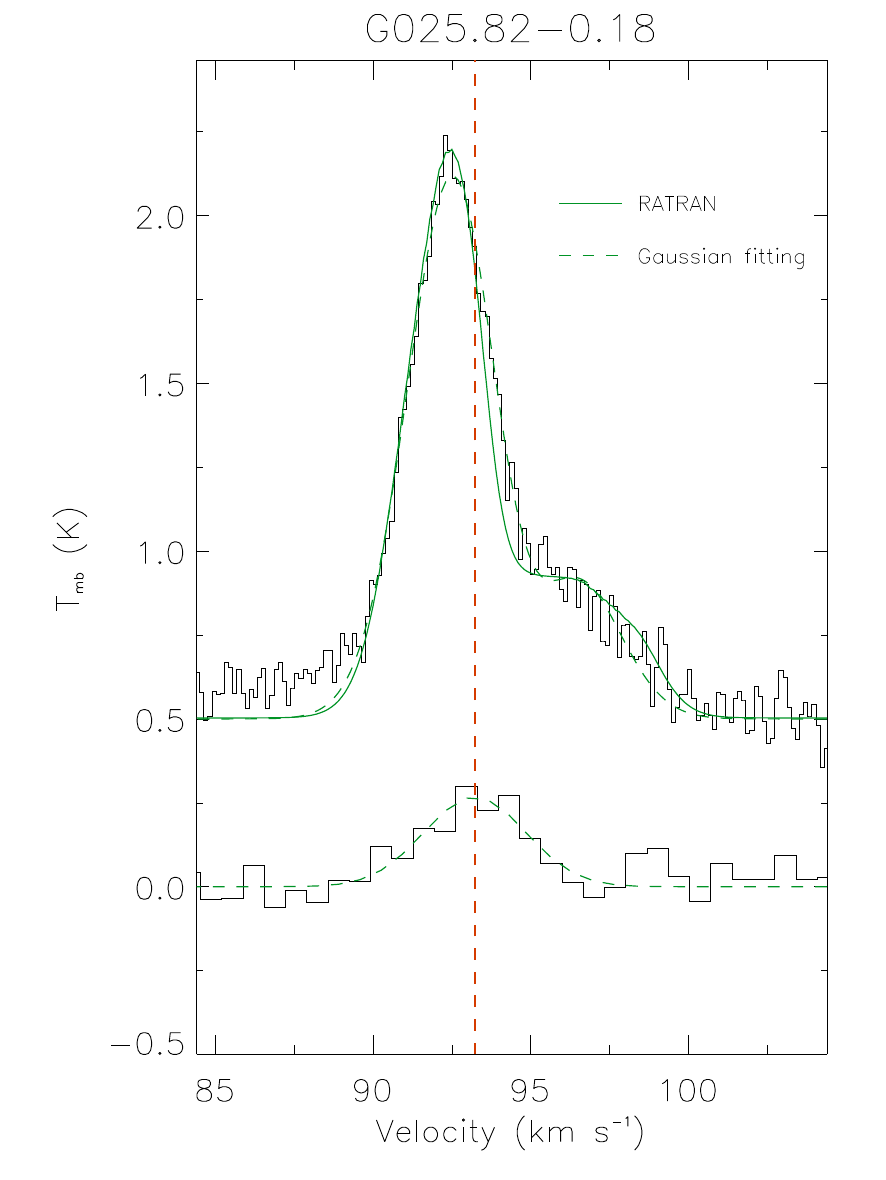}%*
  \end{minipage}%

  \begin{minipage}[t]{0.29\linewidth}
  \centering
   \includegraphics[width=55mm]{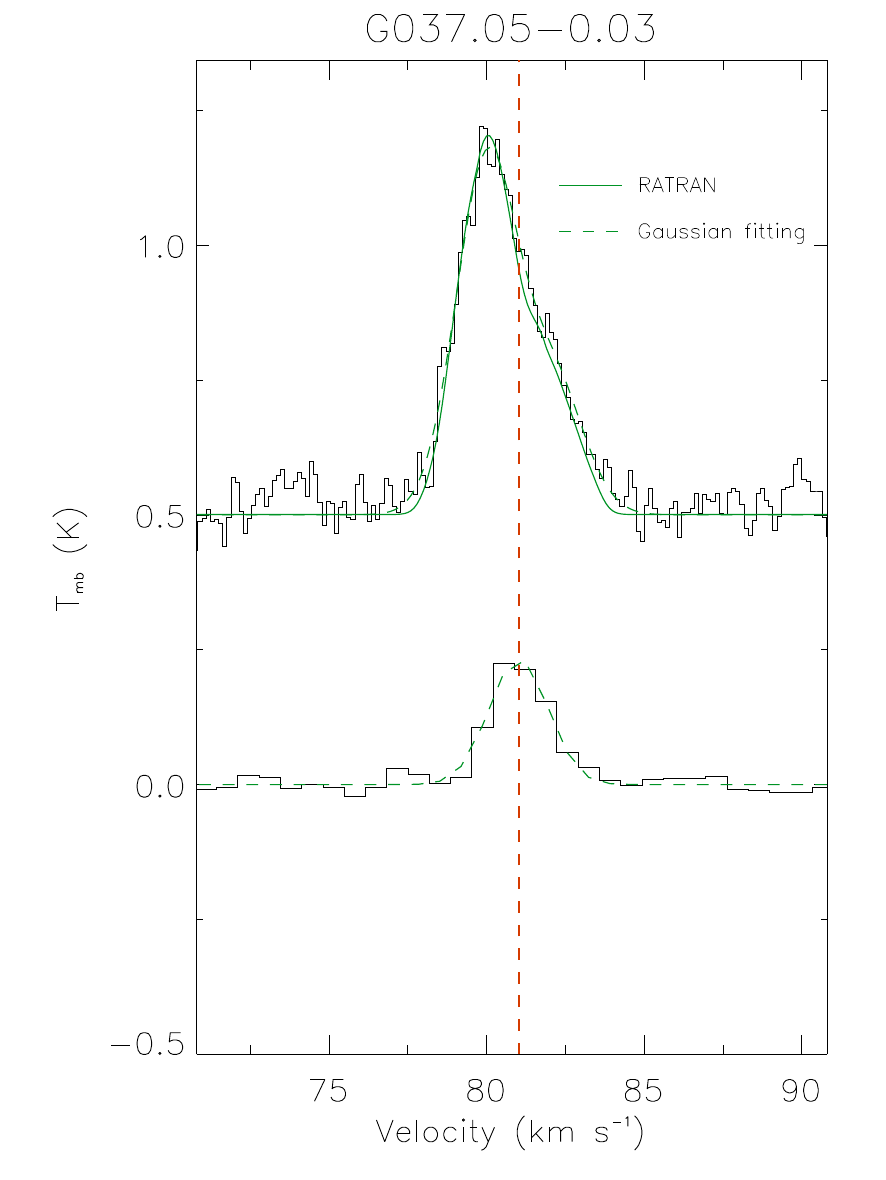}
  \end{minipage}%
  \begin{minipage}[t]{0.29\linewidth}
  \centering
   \includegraphics[width=55mm]{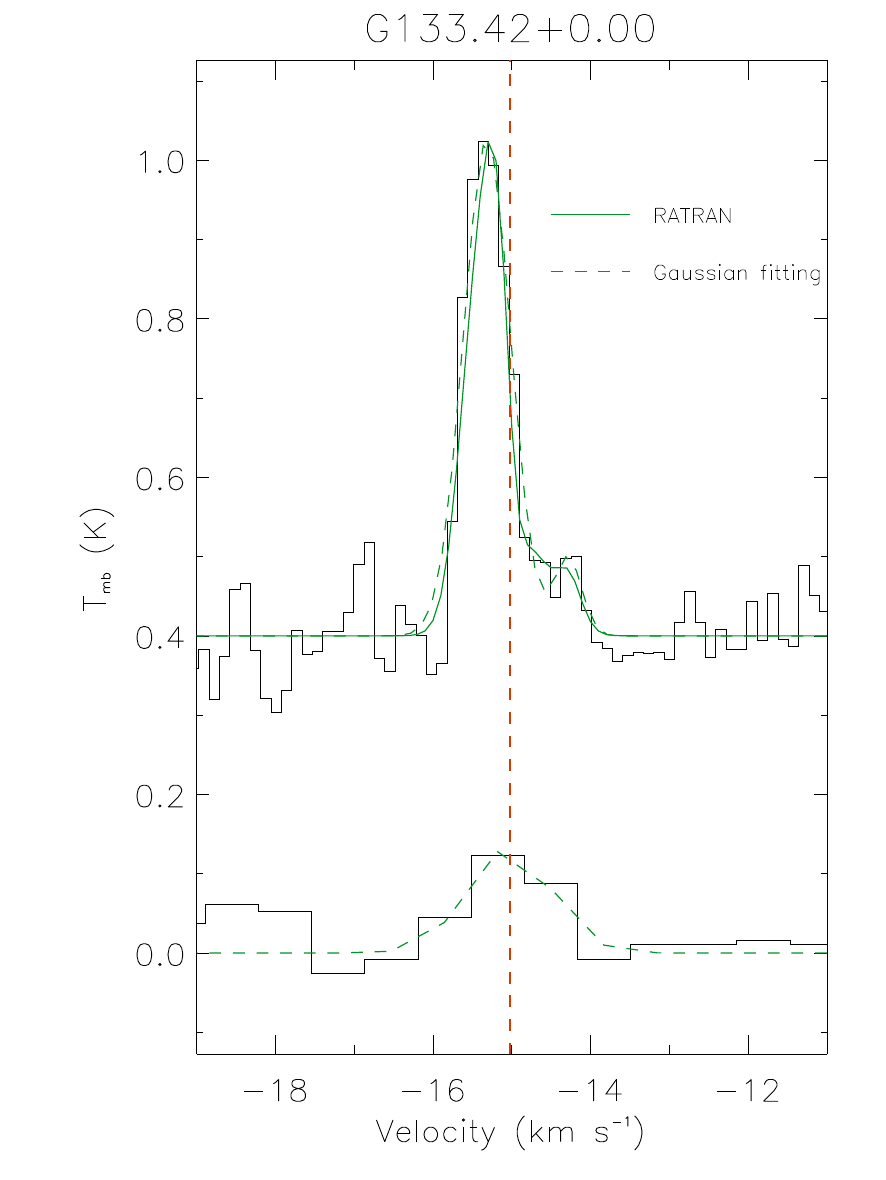}%*
  \end{minipage}%
\caption{HCO$^+$ (1-0) and H$^{13}$CO$^+$ (1-0) average spectra obtained from RATRAN modelling of a collapsing cloud (see text). The observed HCO$^+$ (1-0) spectrum is plotted in black, and the average area is the red box area of each target. The green line is the spectrum obtained by RATRAN model, while the dashed green line is the result of Gaussian fitting of HCO$^+$ (1-0) and H$^{13}$CO$^+$ (1-0). The dashed red line indicates the central radial velocity of H$^{13}$CO$^+$. \label{fig:infall}}
\end{figure*}

Based on the observed HCO$^+$ (1-0) line profiles, we have identified 11 targets that show obvious evidence of gas infall motions. In this analysis, we will estimate the gas infall velocity and analyze the spatial distribution of the infall velocity for these sources. However, we will exclude G036.02-1.36 from further analysis, the blue line profiles of this target are only shown at the upper edge of the observation area. This exclusion is necessary as the observation area did not fully cover this clump.

The two-layer model is used to estimate the infall velocity for each grid and obtain the spatial distribution of the gas infall velocity. This model assumes two uniform regions along the line of sight, with a constant excitation temperature at the front \citep{Myers+etal+1996}. The infall velocity can be estimated through the optically thick and thin lines data:
\begin{equation}
  V_{in}=\frac{\sigma^{2}}{V_{red}-V_{blue}}\ln\left(\frac{1+e(T_{bd}/T_{dip})}{1+e(T_{rd}/T_{dip})}\right)
\label{eq:equ2}
\end{equation}
where $\sigma$ is the velocity dispersion of gas derived from $FWHM$ of the optically thin line ($\sigma=FWHM/\sqrt{8ln(2)}$). $V_{red}$ and $V_{blue}$ are the velocities of red and blue peaks of the optically thick line, $T_{dip}$ is the intensity of the self-absorption dip. $T_{bd}$ and $T_{rd}$ are the temperatures of the blue and red peaks above the dip, respectively. The parameters of optically thin lines are obtained through Gaussian fitting of the H$^{13}$CO$^+$ (1-0) data, while the parameters of optically thick line are obtained through double Gaussian fitting (a combination of an emission line and an absorption line) for HCO$^+$ (1-0) data. The positions of the blue peak, red peak, and dip are determined based on the positions of the local maximum and minimum points of the fitted spectral line. In cases where the spectral line profile shows a peak-shoulder profile or a single-peaked profile with the peak skewed to the blue, it is difficult to identify the red peak and the dip. %\textbf{In such situations, we use the inflection points of the first derivative of spectral lines (i.e. the local maximum and minimum points of the second derivative lines) to determine the positions of the red peak and dip.} 
In such situations, two local maximum points and the saddle point between them are used to determine the blue peak, red peak and dip \citep{Jiang+etal+2023}. 
The uncertainty of these parameters is mainly caused by the noise of the spectral line and fitting error. In the right panel of Figure \ref{fig:mapgrid_eg}, we have zoomed in the spectral line profiles toward the region associated with gas infall of the G037.05-0.03 HCO$^+$ map grids, and marked the gas infall velocity on each grid. The two spectral lines in this subfigures, from top to bottom, correspond to the HCO$^+$ (1-0) and H$^{13}$CO$^+$ (1-0) lines. The green lines show the Gaussian fitting results of H$^{13}$CO$^+$ (1-0) lines with a signal-to-noise ratio greater than 3 and HCO$^+$ (1-0) lines with blue profiles. The dashed red lines indicate the central radial velocity of H$^{13}$CO$^+$ lines in each grid.

In order to analyze the spatial distribution characteristics of the infall velocity, we draw an infall velocity distribution map and plot the variation of infall velocity along the distance to the clump center and H$^{13}$CO$^+$ integrated intensity for the sources displaying significant global infall evidence (as shown in the Figure \ref{fig:vin_r}). As can be seen from the velocity distribution map, for each target, while a few grids show relatively high infall velocities, the variation in gas infall velocity among most grids is not significant. Specifically, for G014.25-0.17, the gas infall velocity distribution is higher in the western region of the source and lower in the eastern region, and for G025.82-0.18, the infall velocity is relatively high in the southeastern region of the source. The remaining sources generally show higher infall velocities in the center and lower velocities in the periphery. In the distance to the clump center vs V$\rm{_{in}}$ image, we use the position of the maximum value obtained from the 2-dimensional Gaussian fitting on the H$^{13}$CO$^+$ integrated intensity map as the zero point of the distance. This position is marked with a red plus sign on each velocity distribution map. The distance is defined as the linear scale between the grids showing an infall line profile and the zero point. We calculate the correlation coefficients and regression lines (represented by red dotted line) for the relationship between the V$\rm{_{in}}$ and distance. The results indicate correlation coefficients ranging from -0.67 to -0.22, suggesting a negative correlation between the infall velocity and distance for most sources. Among them, G012.96-0.23 and G017.09+0.82 region 2 show a clear trend of decreasing infall velocity with increasing distance. In addition, there is a modest and similar trend observed between the infall velocity and H$^{13}$CO$^+$ integrated intensity. In the central regions with higher integral intensity, the infall velocity tends to be higher. The correlation coefficients for these two variables range from 0.17 to 0.62. However, the position of the maximum infall velocity does not strictly coincide with the position of the maximum integral intensity. These sources with global infall evidence may contain more complex internal gas motions that are difficult to determine at the current spatial resolution.

In addition, we use the radiative transfer and molecular excitation (RATRAN) 1D Monte Carlo model \citep{Hogerheijde+etal+2000} for fitting the line profile of the HCO$^+$ average spectrum of each clump (extracted from the red box area) to estimate their average infall velocities. The RATRAN model allows us to input a multi-layer shell to simulate a collapsing cloud, which is more suitable for analyzing the global properties rather than grid by grid. The input parameters are the shell number, the radius, the distribution of H$_2$ density and the molecular abundance, the kinetic temperature distribution, turbulent velocity dispersion, and infall velocity distribution. To model the envelope, we use 10 concentric spherical shells with a power-law density and velocity distribution, where the power index were set to -1.5 and -0.5, respectively. The radius, density, and kinetic temperature of the clump we used are quoted in Table \ref{Tab:clump_para}. The HCO$^+$ abundance used in this paper is derived from the [H$^{13}$CO$^+$]/[H$_2$] ratio. We run the RATRAN code varying the velocity dispersion and the infall velocity, to simulate a series of models to match the observed HCO$^+$ lines. We use the Kolmogorov-Smirnov (K-S) test to verify the similarity between simulation and observation results. The one with the highest probability in a series of models was adopted as the best fit for the HCO$^+$ line. Since the variation of H$^{13}$CO$^+$ infall velocity and velocity dispersion has no significant impact on the fitting results, we do not fit H$^{13}$CO$^+$ lines in this analysis (refer to Paper III Section 4.2). The fitting results are presented in Figure \ref{fig:infall}. On the other hand, we also use the two-layer model to estimate the average infall velocity of the clumps for comparison. All the results are listed in Table \ref{Tab:infall}.

The average gas infall velocities obtained by fitting the average spectra with RATRAN and two-layer model are 0.24 -- 1.85 and 0.28 -- 1.45 km s$^{-1}$, respectively. Comparing the results estimated by these two models, we found that the infall velocity values of some clumps (i.e., G012.96-0.23 and G037.05-0.03) estimated by RATRAN are higher than those estimated by the two-layer model. The infall velocities of the remaining clumps are roughly consistent within the margin of error. In cases where the spectral line profile of a source does not show a significant red peak, the infall velocity values estimated by these two models often have significant differences. This discrepancy may be attributed to unsatisfactory fitting results. It is difficult to accurately determine the positions of the red peak and dip when the spectral line profile shows a peak-shoulder profile or a single-peaked profile with the peak skewed to the blue. Consequently, fitting the blue profile and line wing of the spectral line becomes difficult. Another possible reason is that since most clumps have large radius, there may be some complex structures and gas motions in the clumps, which may cause the observation results do not match well with the simple gas infalling model.

%________________________________________ Table 5: infall parameters of the clumps

\begin{table}
\begin{center}
\caption{Infall Velocities and Mass Infall Rates of Infall Candidates with Blue Profiles}\label{Tab:infall}
\setlength{\tabcolsep}{1mm}{
\begin{tabular}{lcccccc}   
  \hline\noalign{\smallskip}
 & \multicolumn{3}{c}{Two-layer Model} & \multicolumn{3}{c}{RATRAN Model} \\
 Source Name & V$\rm{_{in}}$ (grid)$^1$ & V$\rm{_{in}}$ (ave)$^2$ & $\rm{\dot M_{in}}$ & V$\rm{_{in}}$ (ave)$^3$ & $\rm{\sigma}$ & $\rm{\dot M_{in}}$ \\ 
  & ($\mathrm{km\,s}^{-1}$) & ($\mathrm{km\,s}^{-1}$) & ($\times10^{-4} M_{\odot}\,\mathrm{yr}^{-1}$) & ($\mathrm{km\,s}^{-1}$) & ($\mathrm{km\,s}^{-1}$) & ($\times10^{-4} M_{\odot}\,\mathrm{yr}^{-1}$) \\
  \hline\noalign{\smallskip} 
G012.79-0.20 Region 1	&	0.44--1.45(0.95)	&	...	&	...	&	...	&	...	&	...	\\
G012.79-0.20 Region 2	&	0.05--2.29(0.24)	&	...	&	...	&	...	&	...	&	...	\\
G012.87-0.22 Region 1	&	0.06--1.37(0.57)	&	...	&	...	&	...	&	...	&	...	\\
G012.87-0.22 Region 2	&	0.08--0.53(0.26)	&	0.28(0.04)	&	14	&	0.24(0.01)	&	1.6(0.2)	&	12	\\
G012.96-0.23$^4$	&	0.01--0.83(0.26)	&	0.59(0.08)	&	27	&	1.46(0.11)	&	3.8(0.2)	&	66	\\
G014.00-0.17 Region 1	&	0.02--0.16(0.04)	&	...	&	...	&	...	&	...	&	...	\\
G014.00-0.17 Region 2	&	0.03--0.23(0.06)	&	...	&	...	&	...	&	...	&	...	\\
G014.25-0.17$^4$	&	0.02--1.99(0.18)	&	0.54(0.21)	&	32	&	0.94(0.40)	&	2.2(0.3)	&	55	\\
G017.09+0.82 Region 1$^4$	&	0.03--0.56(0.12)	&	0.49(0.06)	&	9.4	&	0.45(0.11)	&	2.5(0.1)	&	8.7	\\
G017.09+0.82 Region 2$^4$	&	0.04--0.35(0.14)	&	0.69(0.09)	&	6.6	&	0.56(0.08)	&	2.5(0.1)	&	5.4	\\
G025.82-0.18$^4$ 	&	0.11--1.44(0.51)	&	1.45(0.16)	&	67	&	1.85(0.53)	&	2.6(0.3)	&	85	\\
%G036.02-1.36	&	0.01--0.30(0.09)	&	...	&	...	&	...	&	...	&	...	\\
G037.05-0.03$^4$	&	0.02--2.16(0.33)	&	0.65(0.23)	&	12	&	1.58(0.40)	&	1.0(0.2)	&	28	\\
%G049.07-0.33$^4$	&	0.04--0.36(0.12)	&	0.22(0.11)	&	5.2	&	0.62(0.21)	&	2.8(0.2)	&	15	\\
G133.42+0.00	&	0.04--0.55(0.14)	&	0.58(0.08)	&	0.30 	&	0.47(0.19)	&	0.4(0.1)	&	0.24	\\
  \hline\noalign{\smallskip}

\end{tabular}}
\end{center}
\tablecomments{
$^1$ The infall velocity values of each grid in the red box of the HCO$^+$ (1-0) map grids, and the value in parentheses is the median value of the infall velocity. $^2$ The average infall velocities estimated from the average spectra, and the values in parentheses represent the uncertainty. $^3$ V$\rm{_{in}}$ (RAT) is the mean infall velocity of the clump obtained by the RATRAN model. Since the shell width in RATRAN increases with radius, we used weighted mean value (weighted by the shell’s width). $^4$ The sources with global infall evidence.
}
\end{table}

\subsection{Mass Infall Rate} \label{subsec:Min}

In the previous section, we have estimated the average infall velocities of the clumps using RATRAN and two-layer models. Assuming the clump to be a uniform sphere, it is possible to obtain a rough estimation of its mass infall rate \citep{Lopez-Sepulcre+etal+2010}:
\begin{equation}
   \dot M_{in}=4 \pi R^2 V_{in} \mu m_H \rho
\label{eq:equ3}
\end{equation}
where R is the clump radius estimated from the H$^{13}$CO$^+$ integrated intensity map, and $\rho$ is the H$_2$ density of the clump (see Table \ref{Tab:clump_para}). $V_{in}$ is the average infall velocity obtained from the RATRAN and/or two-layer model (see Section \ref{subsec:Vin}). The estimated mass infall rates for the clumps are listed in Table \ref{Tab:infall}. The results indicate that the mass infall rates of these sources range from approximately $10^{-5}$ to $10^{-2}$ M$_{\odot}$ yr$^{-1}$. With the exception of G133.42+0.00, the mass infall rates of other clumps are greater than $10^{-4}$ M$_{\odot}$ yr$^{-1}$, which is consistent with previous studies on high-mass star formation \citep[e.g.][]{Palla+Stahler+1993, Whitney+etal+1997, Kirk+etal+2005}. The mass infall rate of G133.42+0.00 is about $10^{-5}$ M$_{\odot}$ yr$^{-1}$, which suggests that intermediate-mass star may be forming in this clump. For most clumps, there is not much difference in the magnitude of the mass infall rate estimated by the RATRAN and two-layer models. However, since most of these sources do not have parallax distances, while the uncertainty of kinematic distances can reach approximately 70\%. Also, the kinematic distances depend on model assumptions, which can introduce errors that exceed expectations in the mass infall rate. Moreover, assuming isotropic collapse of the clump may result in overestimation or underestimation of the local mass infall rate. Obtaining higher-resolution data in the future would greatly contribute to a more detailed investigation of the local gas infalling properties of these clumps.

%% Putting eqnarrays or equations inside the mathletters environment groups
%% the enclosed equations by letter. For instance, the eqnarray below, instead
%% of being numbered, say, (4) and (5), would be numbered (4a) and (4b).
%% LaTeX the paper and look at the output to see the results.

\section{Summary} \label{sec:summary}

In previous studies, we identified an infall sample (Paper II) and conducted mapping observations and analysis for 24 targets of them (Paper III). In this paper, we report the mapping results of an additional 13 targets and analyze the gas infall motions observed in these sources. The summarized results are as follows:

(i) All 13 targets show HCO$^+$ emissions, and ten of them show H$^{13}$CO$^+$ emissions, with detected rates of 100\% and 77\%, respectively. The detection rate of H$^{13}$CO$^+$ is consistent with that reported in Paper III. Among these targets, the HCO$^+$ mapping observations of ten sources show clear clumpy structures, while the H$^{13}$CO$^+$ mapping of nine sources show clumpy structures. 

(ii) Using RADEX radiative transfer code, we have made rough estimations of the excitation temperatures and column densities of H$^{13}$CO$^+$ in the clumps. The results show that the excitation temperatures range from several to ten Kelvin, and the H$^{13}$CO$^+$ column densities range from $10^{12}$ to $10^{13}$ cm$^{-2}$. Using these and the $N(H_2)$ column densities, we have calculated the abundance ratio [H$^{13}$CO$^+$]/[H$_2$], which is about 10$^{-12}$ -- 10$^{-10}$. The column densities of HCO$^+$ have been estimated using the ratio C/$^{13}$C = 69 \citep{Wilson+1999}, and they are approximately $10^{14}$ to $10^{15}$ cm$^{-2}$.

(iii) Based on the classification of optically thick line profiles, 11 targets show blue profiles in HCO$^+$ lines, which may be the confirmed infall sources. At the spatial resolution used in this study, six clumps show evidence of global collapse. The HCO$^+$ lines of G012.87-0.22 and G014.00-0.17 show both red and blue profiles. And the HCO$^+$ line profile of G012.79-0.20 is quite complex, suggesting that this source may have complex clumpy structure or gas motions.

(iv) We use the two-layer model to obtain the distribution of infall velocity. Within each region, the variations in gas infall velocity among different grids are not significant. In the case of G014.25-0.17, the gas infall velocity distribution shows relatively high values in the western region of the source. For G025.82-0.18, the infall velocity is relatively high in the southeastern part of the source. For the remaining sources, the infall velocity tends to be higher in the central region and lower in the periphery. Additionally, we observe a slight decrease in the infall velocity with increasing distance to the clump center, while it shows an increasing trend with the integral intensity of H$^{13}$CO$^+$.

(v) For the analysis of infall properties, we use both the RATRAN and two-layer model to fit the average spectrum of each clump. The average infall velocities obtained from these two models range from 0.24 to 1.85 and 0.28 to 1.45 km s$^{-1}$, respectively. The estimated mass infall rates range from $10^{-5}$ to $10^{-2}$ M$_{\odot}$ yr$^{-1}$, which are consistent with the results of intermediate- or high-mass star formation in previous studies. With the exception of three clumps, the infall velocities estimated by these two models for the remaining sources are roughly consistent within the margin of error. Moreover, for most sources, there is no significant difference in the order of magnitude of the mass infall rate values calculated by the RATRAN and two-layer models.

\acknowledgments

We are grateful to the staff of the Institut de Radioastronomie Millim{\'e}trique (IRAM) for their assistance and support during the observations. This work has been supported by the National Key R\&D Program of China (No. 2022YFA603102), and the National Natural Science Foundation of China (NSFC) Grant Nos. U2031202, 11903083, 11873093.

\vspace{5mm}

\facility{IRAM:30m}

\software{GILDAS \citep{Pety+2005, Gildas+2013}, RADEX \citep{vanderTak+etal+2007}, RATRAN \citep{Hogerheijde+etal+2000}}


%\vspace{5mm}
%\facilities{HST(STIS), Swift(XRT and UVOT), AAVSO, CTIO:1.3m, CTIO:1.5m,CXO}


%\software{astropy \citep{2013A&A...558A..33A},  
%          Cloudy \citep{2013RMxAA..49..137F}, 
%          SExtractor \citep{1996A&AS..117..393B}
%          }



\clearpage

\restartappendixnumbering

\appendix

\section{HCO$^+$ (1-0) and H$^{13}$CO$^+$ (1-0) Integrated Intensity Maps} \label{sec:Appendix1}

\begin{figure}
%\figurenum{1}
\plotone{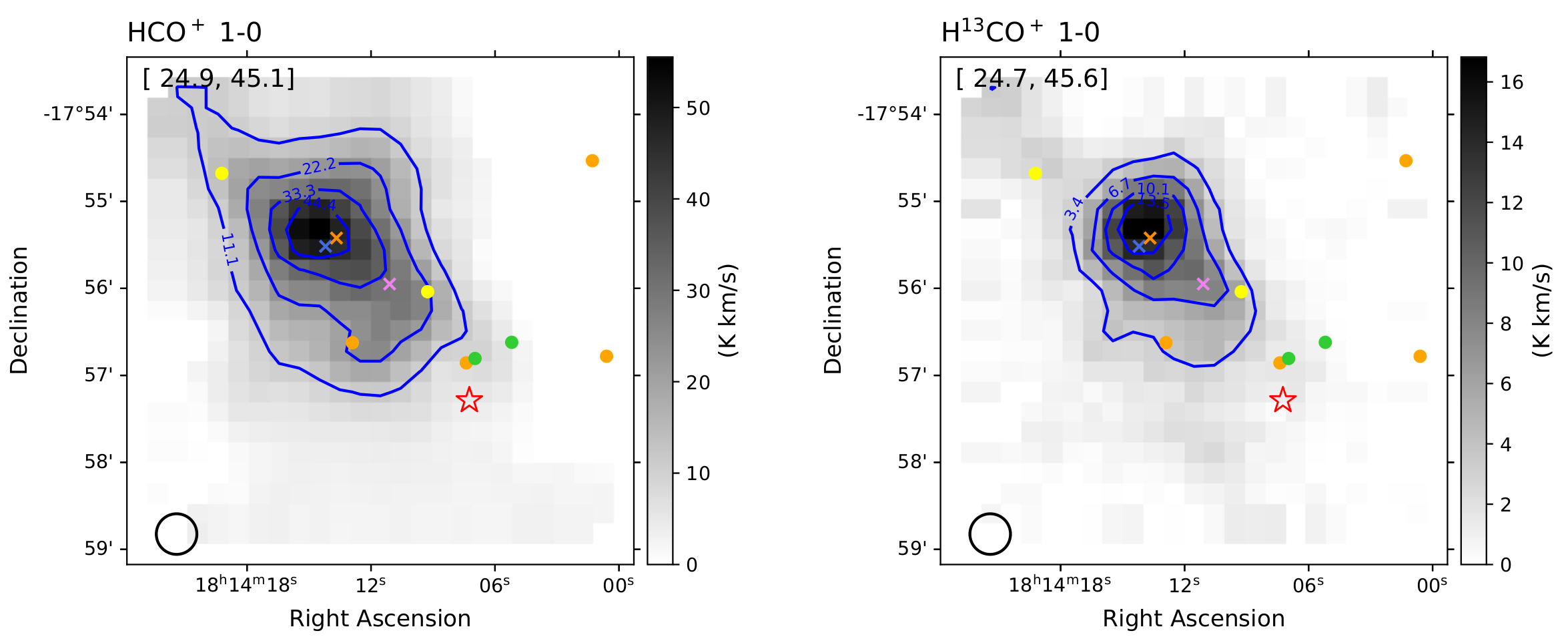}
\caption{HCO$^+$ (1-0) and H$^{13}$CO$^+$ (1-0) integrated intensity maps. The pentagram symbol denotes the position of the infall candidate identified in Paper II. The green, yellow, and orange points denote the Class I, Flat-spectrum, Class II YSOs \citep{Kuhn+etal+2021}, respectively. The blue, magenta, orange, and red crosses denote the 95 GHz methanol masers \citep{Yang+etal+2017}, 6.7 GHz methanol masers \citep{Yang+etal+2019}, H$_2$O masers \citep{Anglada+etal+1996,Valdettaro+etal+2001}, and OH masers \citep{Qiao+etal+2016,Qiao+etal+2018,Qiao+etal+2020}. For targets that show two distinct clumps within the observed areas, we have labeled these clumps as A and B. The black circles on the lower left indicate the beam sizes of the IRAM 30-m telescope.}
\label{fig:map}
\end{figure}

Figure \ref{fig:map} presents the HCO$^+$ (1-0) and H$^{13}$CO$^+$ (1-0) integrated intensity maps of 13 targets. The complete figure set is available in the online journal. More details can be found in Section \ref{sec:results}.

\figsetstart
%\figsetnum{1}
\figsettitle{HCO$^+$ (1-0) and H$^{13}$CO$^+$ (1-0) integrated intensity maps}

\figsetgrpstart
\figsetgrpnum{A1.1}
\figsetgrptitle{G012.79-0.20 }
\figsetplot{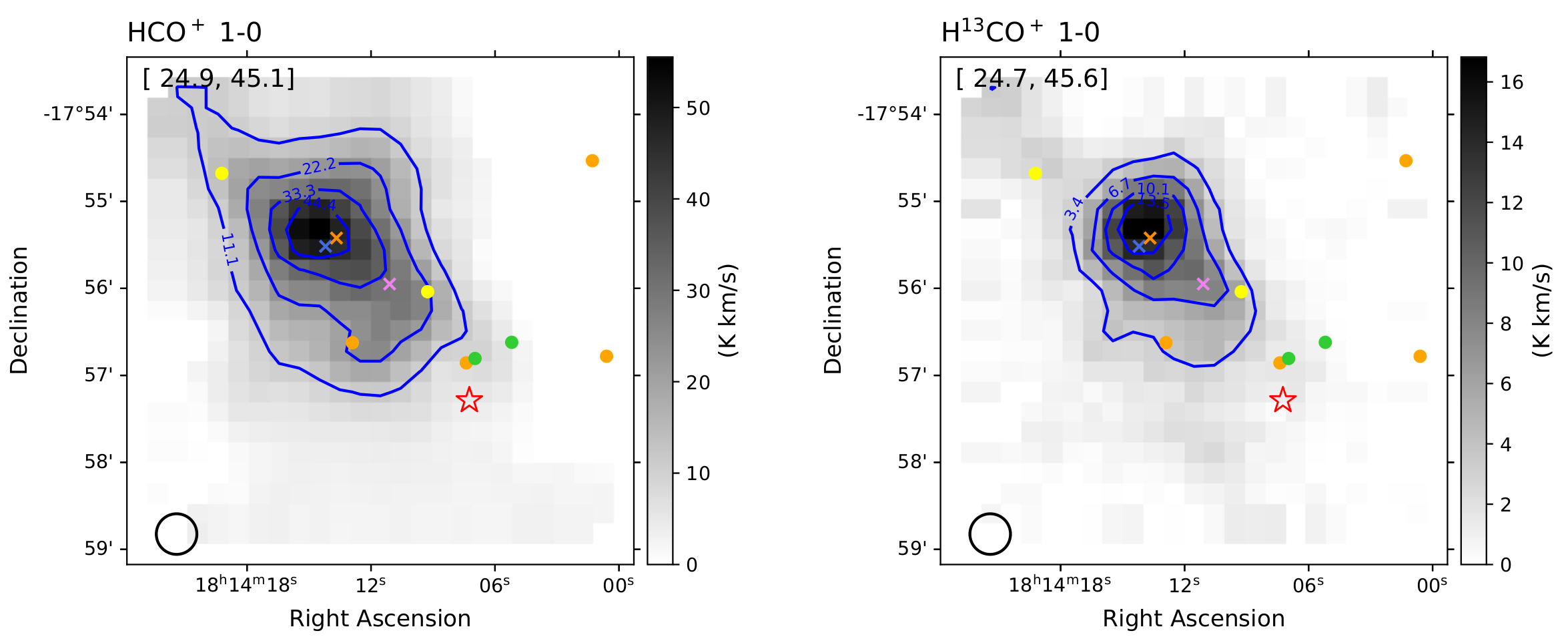}
\figsetgrpnote{HCO$^+$ (1-0) and H$^{13}$CO$^+$ (1-0) integrated intensity maps. The pentagram symbol denotes the position of the infall candidate identified in Paper II. The green, yellow, and orange points denote the Class I, Flat-spectrum, Class II YSOs \citep{Kuhn+etal+2021}, respectively. The blue, magenta, orange, and red crosses denote the 95 GHz methanol masers \citep{Yang+etal+2017}, 6.7 GHz methanol masers \citep{Yang+etal+2019}, H$_2$O masers \citep{Anglada+etal+1996,Valdettaro+etal+2001}, and OH masers \citep{Qiao+etal+2016,Qiao+etal+2018,Qiao+etal+2020}. For targets that show two distinct clumps within the observed areas, we have labeled these clumps as A and B. The black circles on the lower left indicate the beam sizes of the IRAM 30-m telescope.}
\figsetgrpend

\figsetgrpstart
\figsetgrpnum{A1.2}
\figsetgrptitle{G012.87-0.22}
\figsetplot{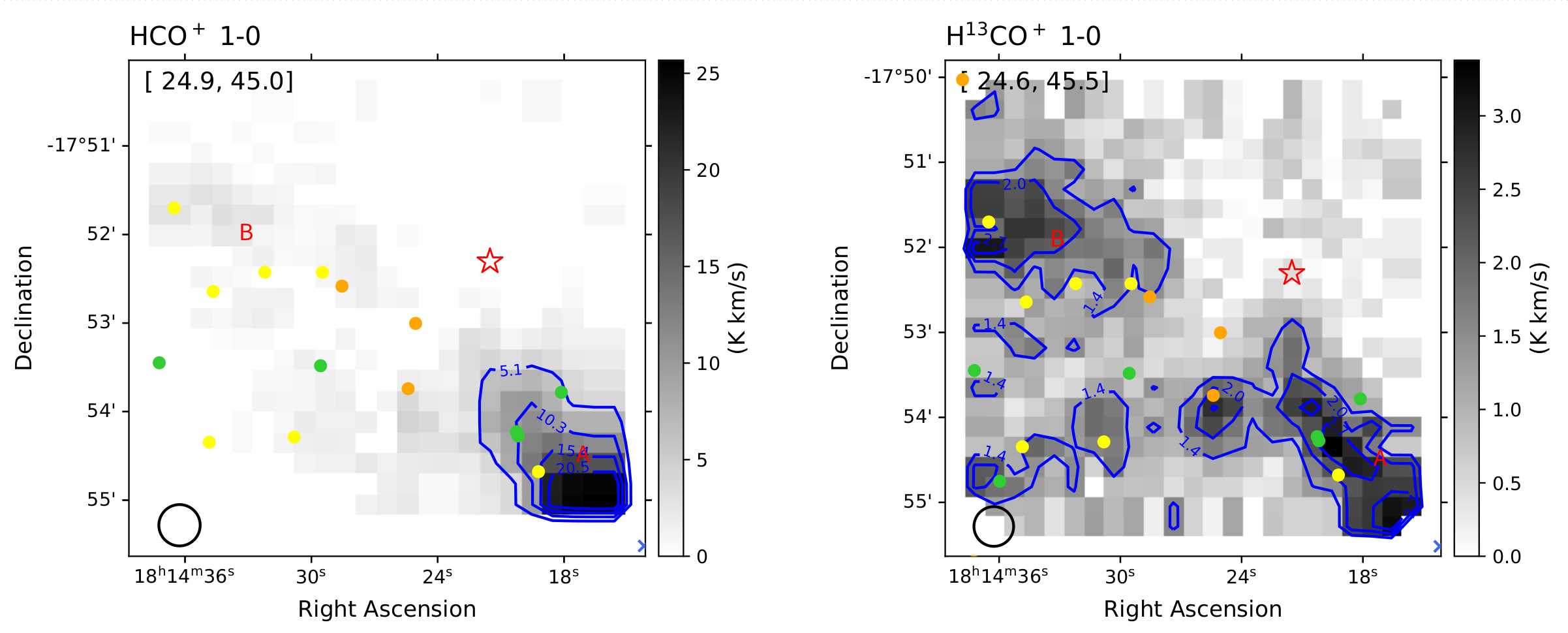}
\figsetgrpnote{HCO$^+$ (1-0) and H$^{13}$CO$^+$ (1-0) integrated intensity maps. The pentagram symbol denotes the position of the infall candidate identified in Paper II. The green, yellow, and orange points denote the Class I, Flat-spectrum, Class II YSOs \citep{Kuhn+etal+2021}, respectively. The blue, magenta, orange, and red crosses denote the 95 GHz methanol masers \citep{Yang+etal+2017}, 6.7 GHz methanol masers \citep{Yang+etal+2019}, H$_2$O masers \citep{Anglada+etal+1996,Valdettaro+etal+2001}, and OH masers \citep{Qiao+etal+2016,Qiao+etal+2018,Qiao+etal+2020}. For targets that show two distinct clumps within the observed areas, we have labeled these clumps as A and B. The black circles on the lower left indicate the beam sizes of the IRAM 30-m telescope.}
\figsetgrpend

\figsetgrpstart
\figsetgrpnum{A1.3}
\figsetgrptitle{G012.96-0.23}
\figsetplot{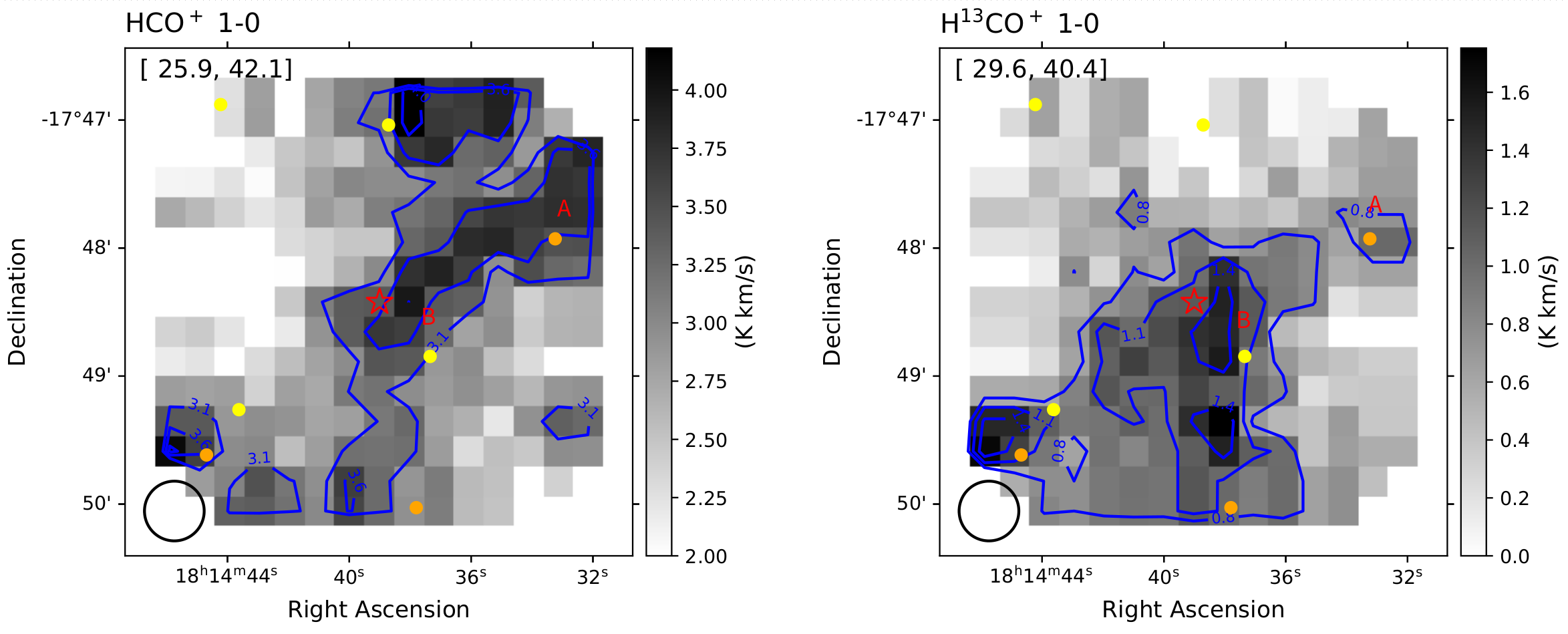}
\figsetgrpnote{HCO$^+$ (1-0) and H$^{13}$CO$^+$ (1-0) integrated intensity maps. The pentagram symbol denotes the position of the infall candidate identified in Paper II. The green, yellow, and orange points denote the Class I, Flat-spectrum, Class II YSOs \citep{Kuhn+etal+2021}, respectively. The blue, magenta, orange, and red crosses denote the 95 GHz methanol masers \citep{Yang+etal+2017}, 6.7 GHz methanol masers \citep{Yang+etal+2019}, H$_2$O masers \citep{Anglada+etal+1996,Valdettaro+etal+2001}, and OH masers \citep{Qiao+etal+2016,Qiao+etal+2018,Qiao+etal+2020}. For targets that show two distinct clumps within the observed areas, we have labeled these clumps as A and B. The black circles on the lower left indicate the beam sizes of the IRAM 30-m telescope.}
\figsetgrpend

\figsetgrpstart
\figsetgrpnum{A1.4}
\figsetgrptitle{G014.00-0.17}
\figsetplot{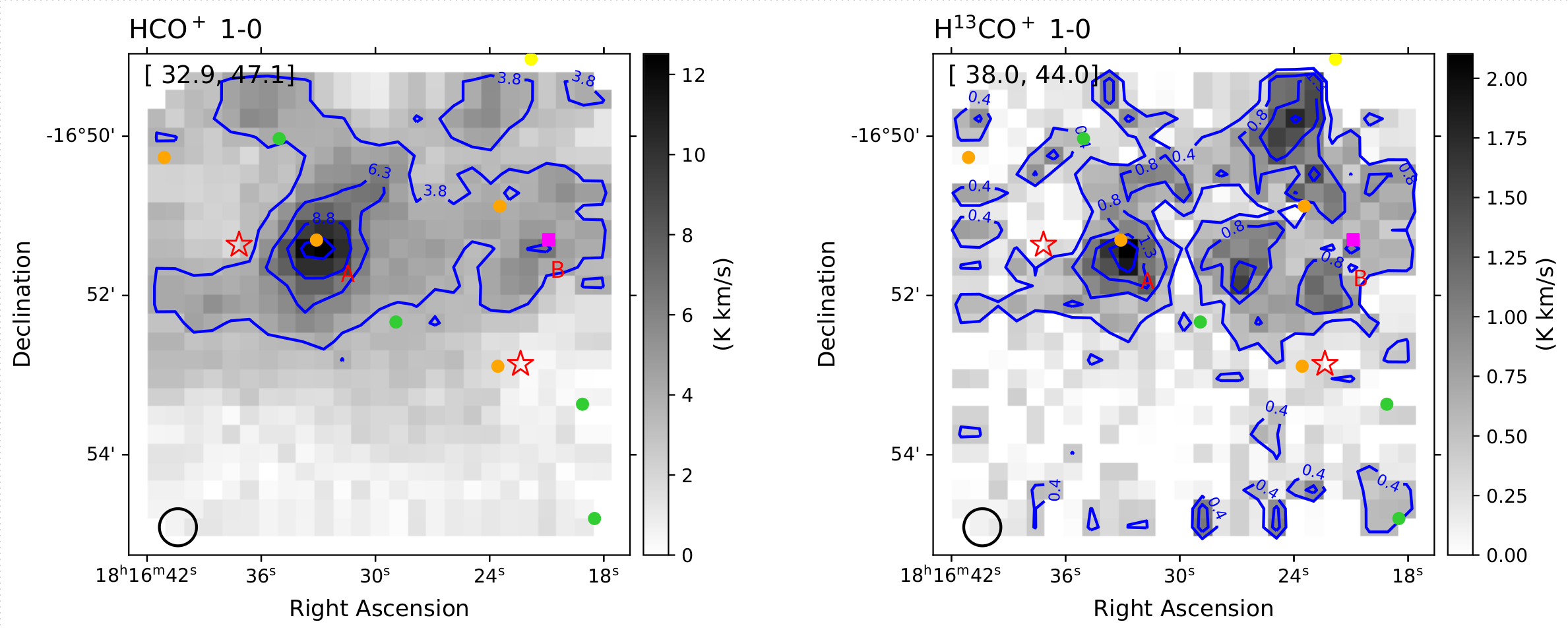}
\figsetgrpnote{HCO$^+$ (1-0) and H$^{13}$CO$^+$ (1-0) integrated intensity maps. The pentagram symbol denotes the position of the infall candidate identified in Paper II. The green, yellow, and orange points denote the Class I, Flat-spectrum, Class II YSOs \citep{Kuhn+etal+2021}, respectively. The blue, magenta, orange, and red crosses denote the 95 GHz methanol masers \citep{Yang+etal+2017}, 6.7 GHz methanol masers \citep{Yang+etal+2019}, H$_2$O masers \citep{Anglada+etal+1996,Valdettaro+etal+2001}, and OH masers \citep{Qiao+etal+2016,Qiao+etal+2018,Qiao+etal+2020}. For targets that show two distinct clumps within the observed areas, we have labeled these clumps as A and B. The black circles on the lower left indicate the beam sizes of the IRAM 30-m telescope.}
\figsetgrpend

\figsetgrpstart
\figsetgrpnum{A1.5}
\figsetgrptitle{G014.25-0.17}
\figsetplot{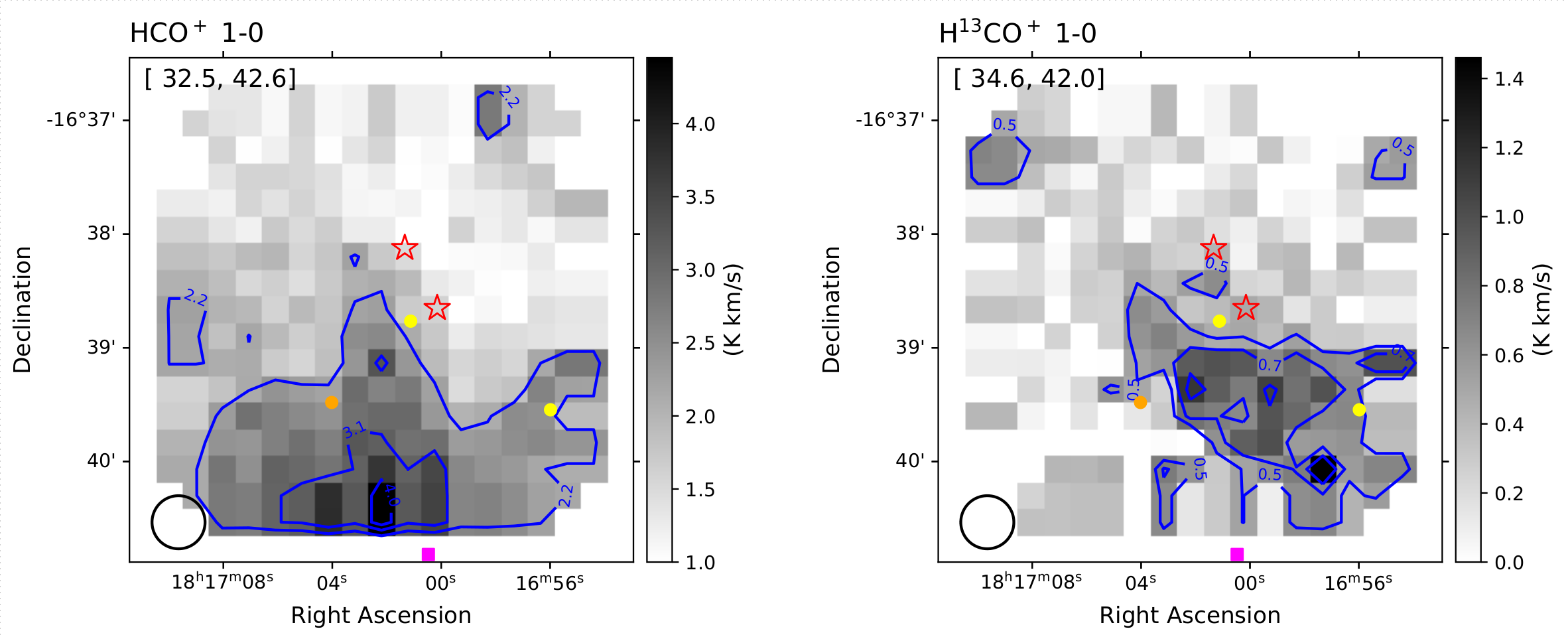}
\figsetgrpnote{HCO$^+$ (1-0) and H$^{13}$CO$^+$ (1-0) integrated intensity maps. The pentagram symbol denotes the position of the infall candidate identified in Paper II. The green, yellow, and orange points denote the Class I, Flat-spectrum, Class II YSOs \citep{Kuhn+etal+2021}, respectively. The blue, magenta, orange, and red crosses denote the 95 GHz methanol masers \citep{Yang+etal+2017}, 6.7 GHz methanol masers \citep{Yang+etal+2019}, H$_2$O masers \citep{Anglada+etal+1996,Valdettaro+etal+2001}, and OH masers \citep{Qiao+etal+2016,Qiao+etal+2018,Qiao+etal+2020}. For targets that show two distinct clumps within the observed areas, we have labeled these clumps as A and B. The black circles on the lower left indicate the beam sizes of the IRAM 30-m telescope.}
\figsetgrpend

\figsetgrpstart
\figsetgrpnum{A1.6}
\figsetgrptitle{G017.09+0.82}
\figsetplot{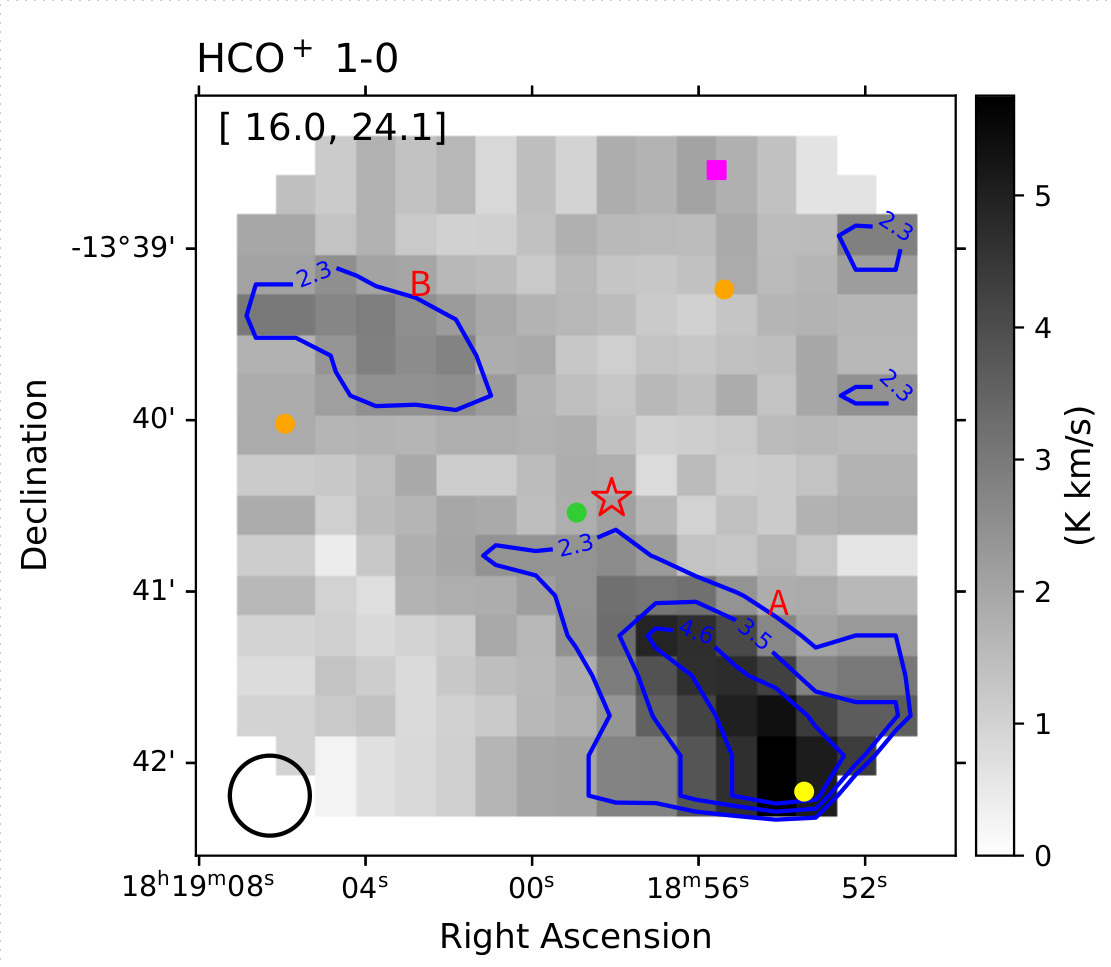}
\figsetgrpnote{HCO$^+$ (1-0) and H$^{13}$CO$^+$ (1-0) integrated intensity maps. The pentagram symbol denotes the position of the infall candidate identified in Paper II. The green, yellow, and orange points denote the Class I, Flat-spectrum, Class II YSOs \citep{Kuhn+etal+2021}, respectively. The blue, magenta, orange, and red crosses denote the 95 GHz methanol masers \citep{Yang+etal+2017}, 6.7 GHz methanol masers \citep{Yang+etal+2019}, H$_2$O masers \citep{Anglada+etal+1996,Valdettaro+etal+2001}, and OH masers \citep{Qiao+etal+2016,Qiao+etal+2018,Qiao+etal+2020}. For targets that show two distinct clumps within the observed areas, we have labeled these clumps as A and B. The black circles on the lower left indicate the beam sizes of the IRAM 30-m telescope.}
\figsetgrpend

\figsetgrpstart
\figsetgrpnum{A1.7}
\figsetgrptitle{G025.82-0.18}
\figsetplot{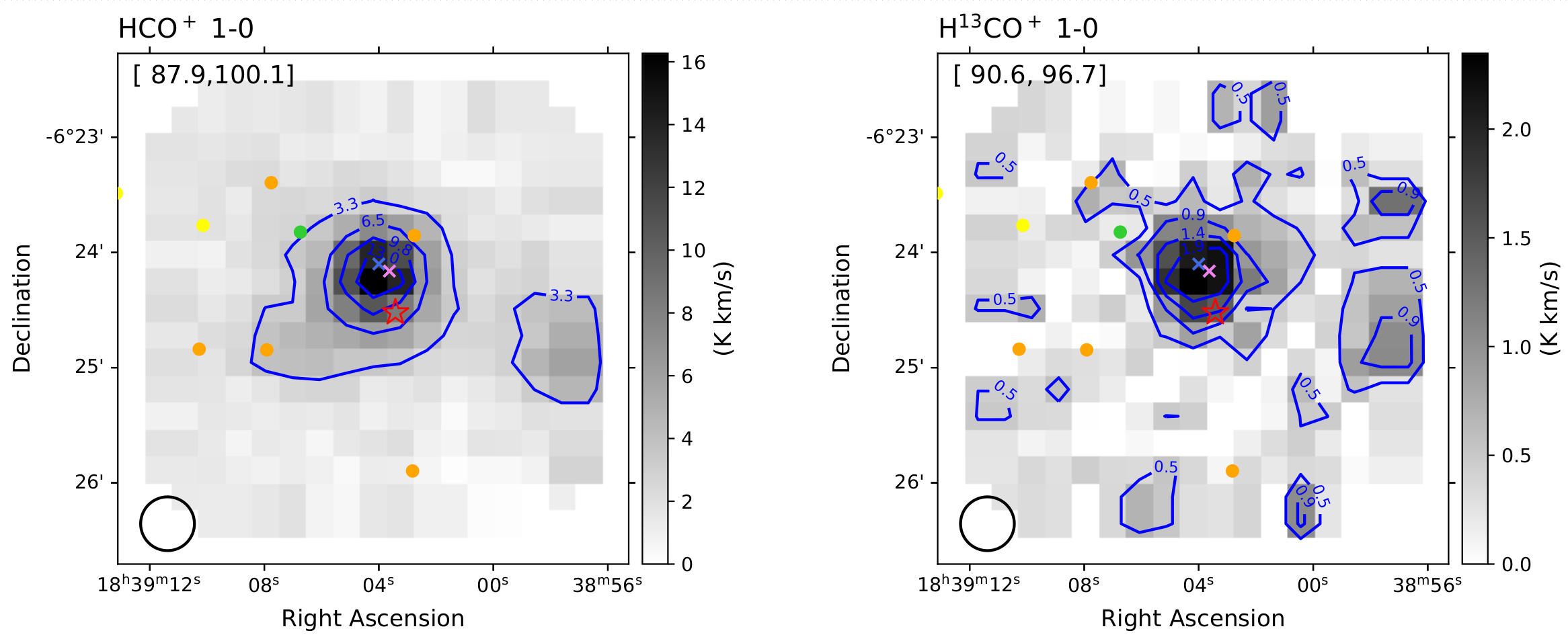}
\figsetgrpnote{HCO$^+$ (1-0) and H$^{13}$CO$^+$ (1-0) integrated intensity maps. The pentagram symbol denotes the position of the infall candidate identified in Paper II. The green, yellow, and orange points denote the Class I, Flat-spectrum, Class II YSOs \citep{Kuhn+etal+2021}, respectively. The blue, magenta, orange, and red crosses denote the 95 GHz methanol masers \citep{Yang+etal+2017}, 6.7 GHz methanol masers \citep{Yang+etal+2019}, H$_2$O masers \citep{Anglada+etal+1996,Valdettaro+etal+2001}, and OH masers \citep{Qiao+etal+2016,Qiao+etal+2018,Qiao+etal+2020}. For targets that show two distinct clumps within the observed areas, we have labeled these clumps as A and B. The black circles on the lower left indicate the beam sizes of the IRAM 30-m telescope.}
\figsetgrpend

\figsetgrpstart
\figsetgrpnum{A1.8}
\figsetgrptitle{G036.02-1.36}
\figsetplot{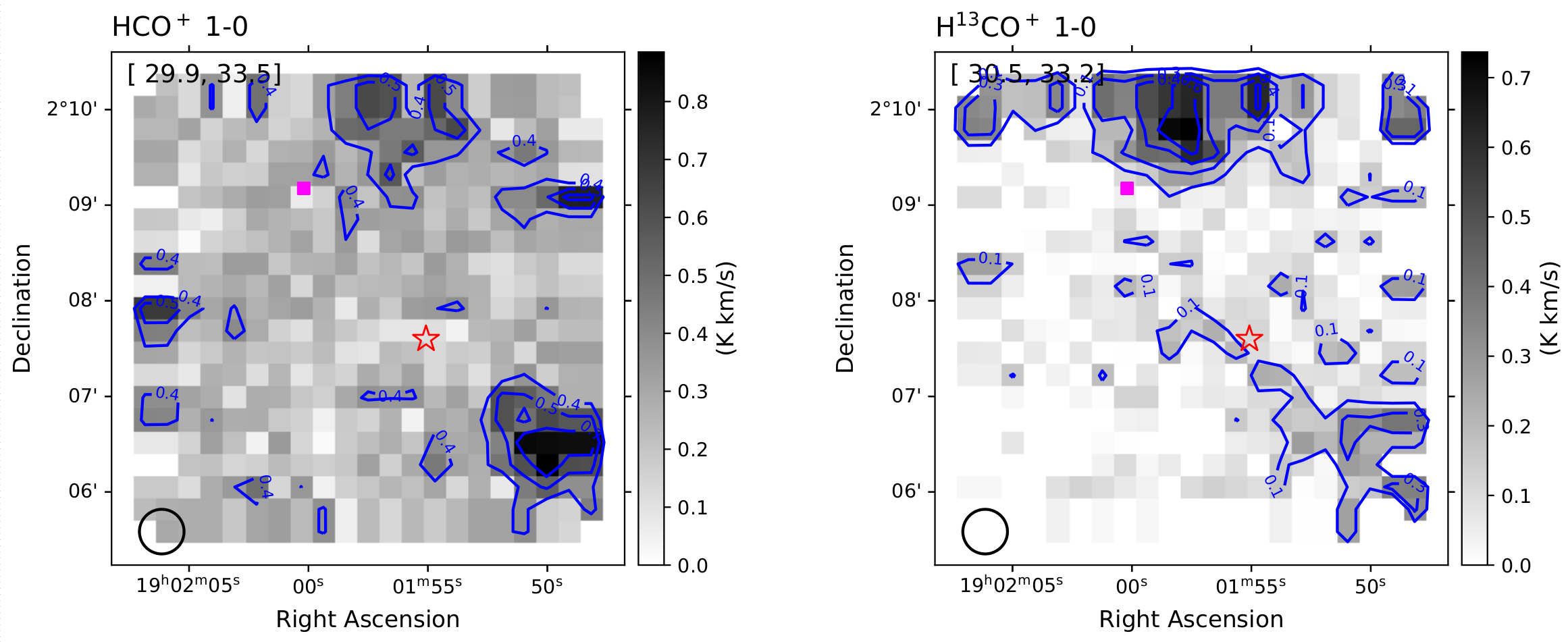}
\figsetgrpnote{HCO$^+$ (1-0) and H$^{13}$CO$^+$ (1-0) integrated intensity maps. The pentagram symbol denotes the position of the infall candidate identified in Paper II. The green, yellow, and orange points denote the Class I, Flat-spectrum, Class II YSOs \citep{Kuhn+etal+2021}, respectively. The blue, magenta, orange, and red crosses denote the 95 GHz methanol masers \citep{Yang+etal+2017}, 6.7 GHz methanol masers \citep{Yang+etal+2019}, H$_2$O masers \citep{Anglada+etal+1996,Valdettaro+etal+2001}, and OH masers \citep{Qiao+etal+2016,Qiao+etal+2018,Qiao+etal+2020}. For targets that show two distinct clumps within the observed areas, we have labeled these clumps as A and B. The black circles on the lower left indicate the beam sizes of the IRAM 30-m telescope.}
\figsetgrpend

\figsetgrpstart
\figsetgrpnum{A1.9}
\figsetgrptitle{G037.05-0.03}
\figsetplot{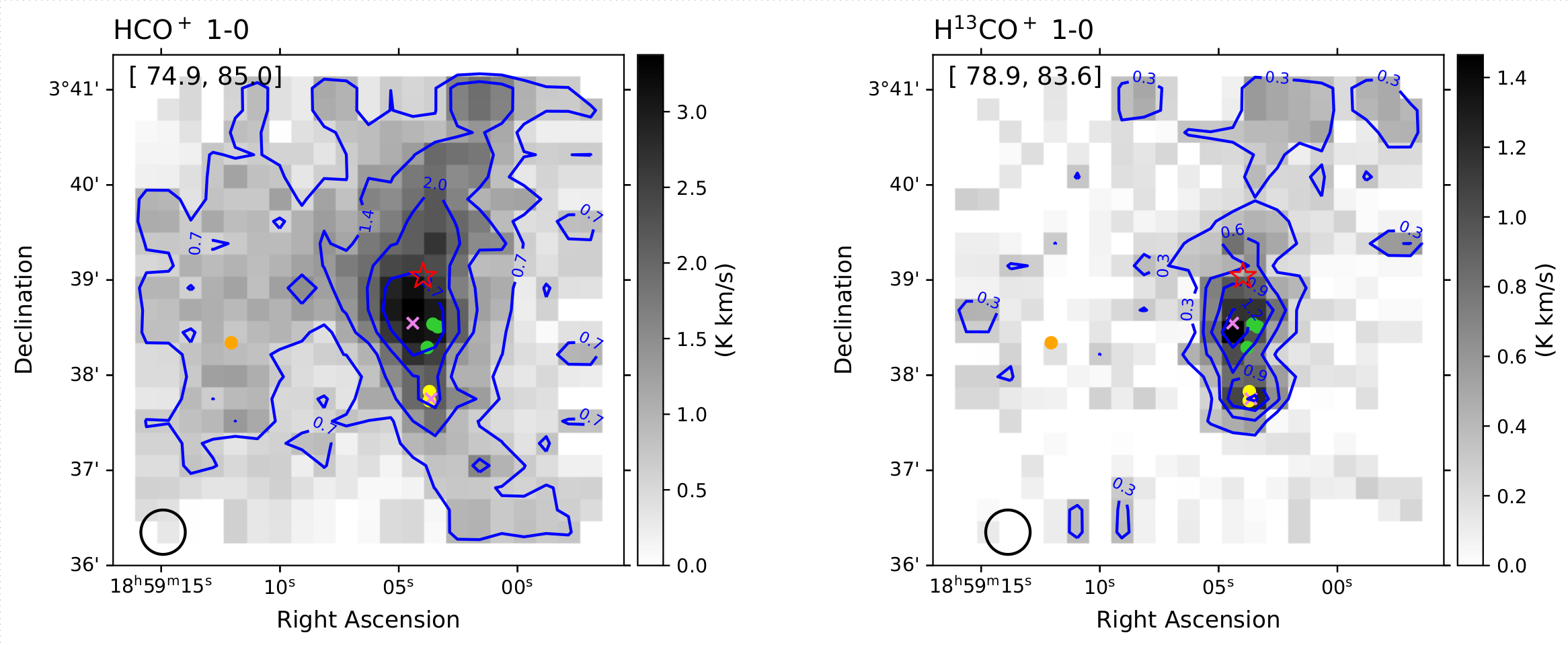}
\figsetgrpnote{HCO$^+$ (1-0) and H$^{13}$CO$^+$ (1-0) integrated intensity maps. The pentagram symbol denotes the position of the infall candidate identified in Paper II. The green, yellow, and orange points denote the Class I, Flat-spectrum, Class II YSOs \citep{Kuhn+etal+2021}, respectively. The blue, magenta, orange, and red crosses denote the 95 GHz methanol masers \citep{Yang+etal+2017}, 6.7 GHz methanol masers \citep{Yang+etal+2019}, H$_2$O masers \citep{Anglada+etal+1996,Valdettaro+etal+2001}, and OH masers \citep{Qiao+etal+2016,Qiao+etal+2018,Qiao+etal+2020}. For targets that show two distinct clumps within the observed areas, we have labeled these clumps as A and B. The black circles on the lower left indicate the beam sizes of the IRAM 30-m telescope.}
\figsetgrpend

\figsetgrpstart
\figsetgrpnum{A1.10}
\figsetgrptitle{G049.07-0.33}
\figsetplot{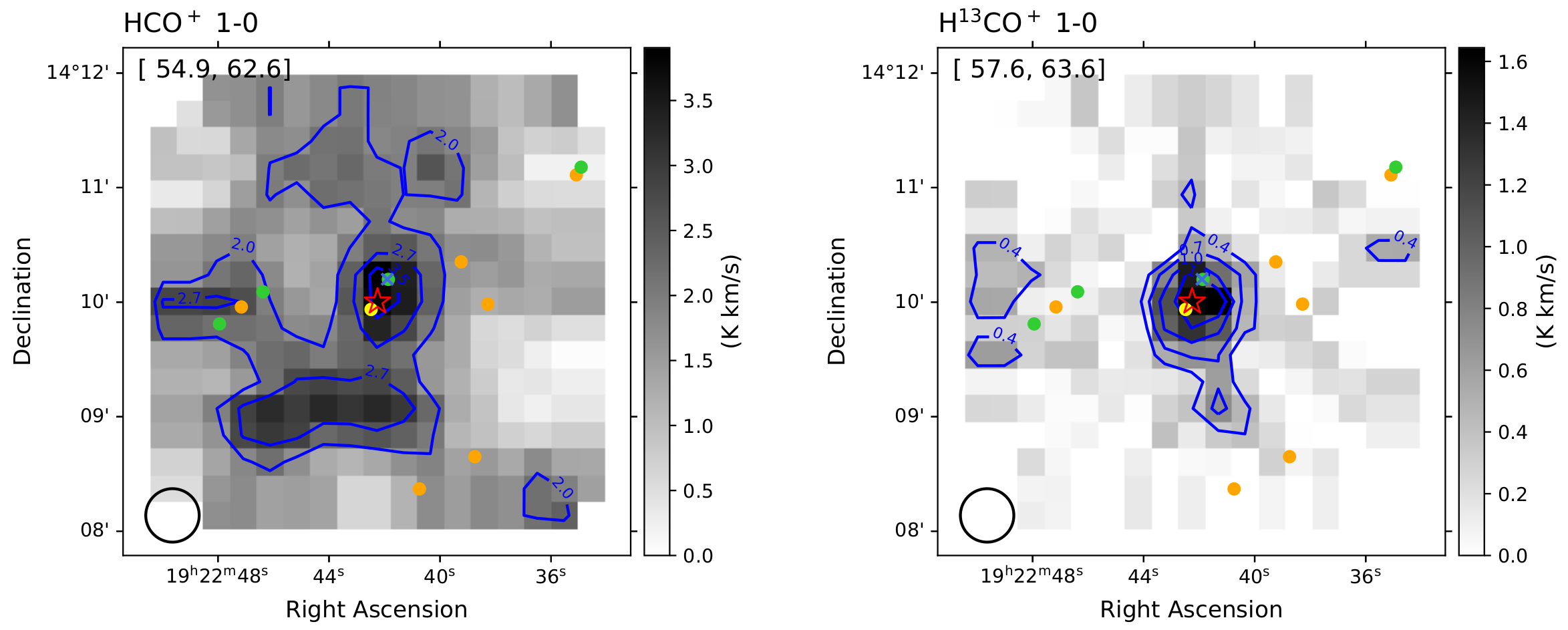}
\figsetgrpnote{HCO$^+$ (1-0) and H$^{13}$CO$^+$ (1-0) integrated intensity maps. The pentagram symbol denotes the position of the infall candidate identified in Paper II. The green, yellow, and orange points denote the Class I, Flat-spectrum, Class II YSOs \citep{Kuhn+etal+2021}, respectively. The blue, magenta, orange, and red crosses denote the 95 GHz methanol masers \citep{Yang+etal+2017}, 6.7 GHz methanol masers \citep{Yang+etal+2019}, H$_2$O masers \citep{Anglada+etal+1996,Valdettaro+etal+2001}, and OH masers \citep{Qiao+etal+2016,Qiao+etal+2018,Qiao+etal+2020}. For targets that show two distinct clumps within the observed areas, we have labeled these clumps as A and B. The black circles on the lower left indicate the beam sizes of the IRAM 30-m telescope.}
\figsetgrpend

\figsetgrpstart
\figsetgrpnum{A1.11}
\figsetgrptitle{G081.72+1.29}
\figsetplot{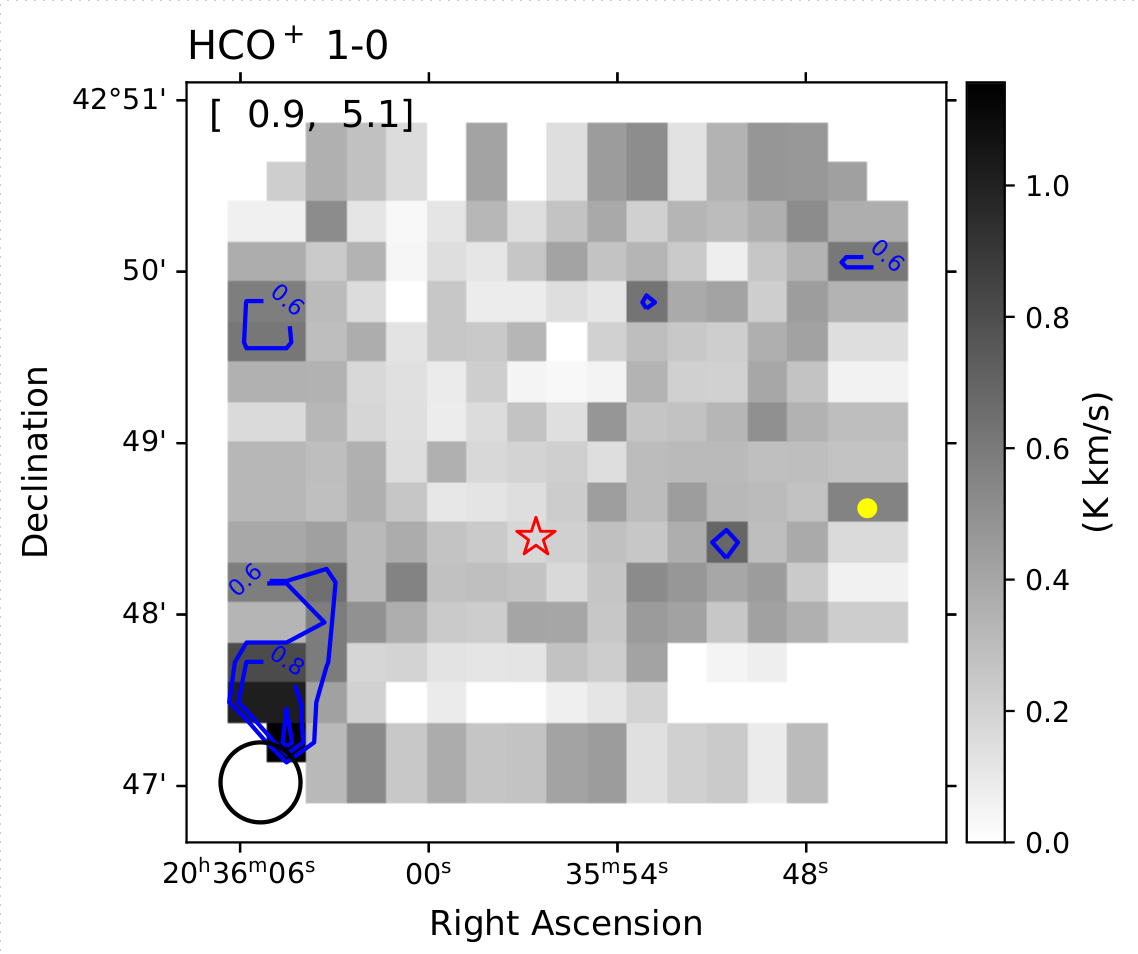}
\figsetgrpnote{HCO$^+$ (1-0) and H$^{13}$CO$^+$ (1-0) integrated intensity maps. The pentagram symbol denotes the position of the infall candidate identified in Paper II. The green, yellow, and orange points denote the Class I, Flat-spectrum, Class II YSOs \citep{Kuhn+etal+2021}, respectively. The blue, magenta, orange, and red crosses denote the 95 GHz methanol masers \citep{Yang+etal+2017}, 6.7 GHz methanol masers \citep{Yang+etal+2019}, H$_2$O masers \citep{Anglada+etal+1996,Valdettaro+etal+2001}, and OH masers \citep{Qiao+etal+2016,Qiao+etal+2018,Qiao+etal+2020}. For targets that show two distinct clumps within the observed areas, we have labeled these clumps as A and B. The black circles on the lower left indicate the beam sizes of the IRAM 30-m telescope.}
\figsetgrpend

\figsetgrpstart
\figsetgrpnum{A1.12}
\figsetgrptitle{G081.90+1.43}
\figsetplot{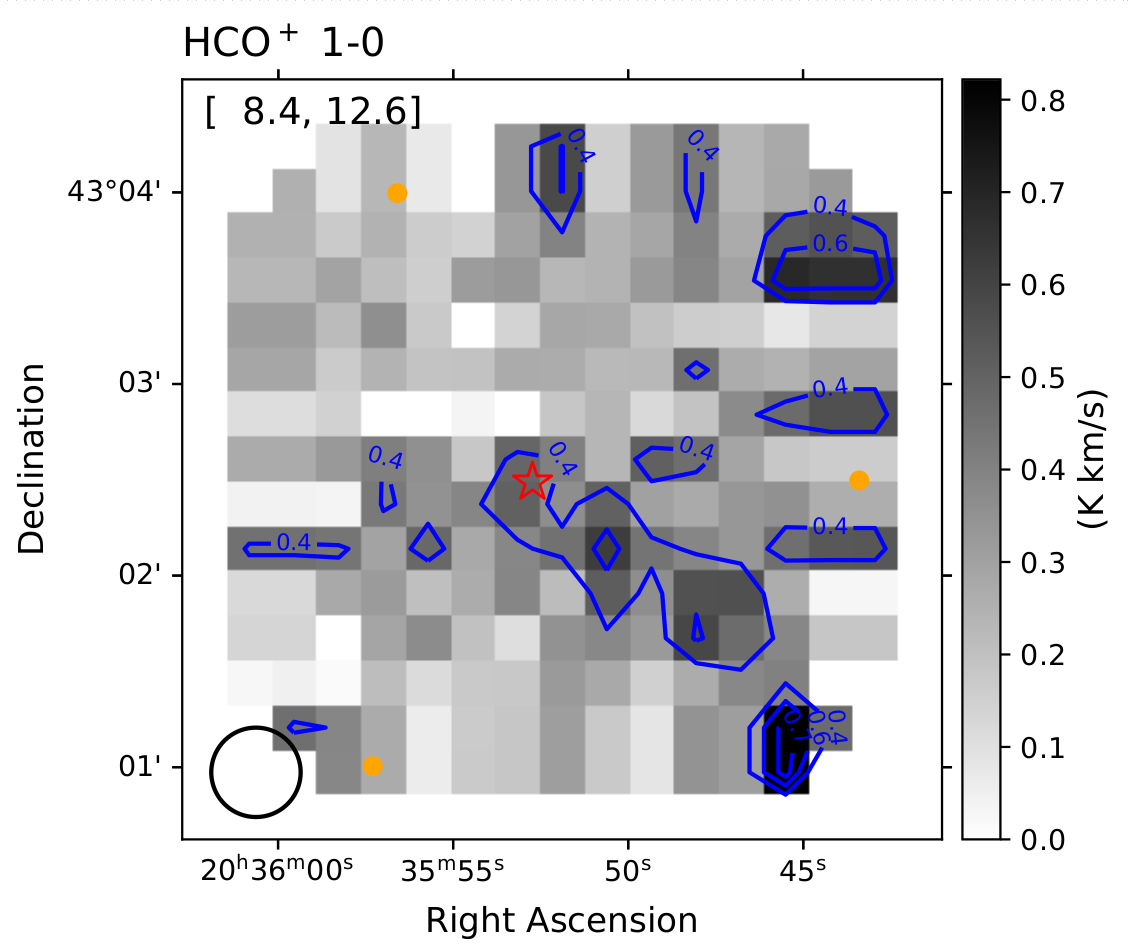}
\figsetgrpnote{HCO$^+$ (1-0) and H$^{13}$CO$^+$ (1-0) integrated intensity maps. The pentagram symbol denotes the position of the infall candidate identified in Paper II. The green, yellow, and orange points denote the Class I, Flat-spectrum, Class II YSOs \citep{Kuhn+etal+2021}, respectively. The blue, magenta, orange, and red crosses denote the 95 GHz methanol masers \citep{Yang+etal+2017}, 6.7 GHz methanol masers \citep{Yang+etal+2019}, H$_2$O masers \citep{Anglada+etal+1996,Valdettaro+etal+2001}, and OH masers \citep{Qiao+etal+2016,Qiao+etal+2018,Qiao+etal+2020}. For targets that show two distinct clumps within the observed areas, we have labeled these clumps as A and B. The black circles on the lower left indicate the beam sizes of the IRAM 30-m telescope.}
\figsetgrpend

\figsetgrpstart
\figsetgrpnum{A1.13}
\figsetgrptitle{G133.42+0.00}
\figsetplot{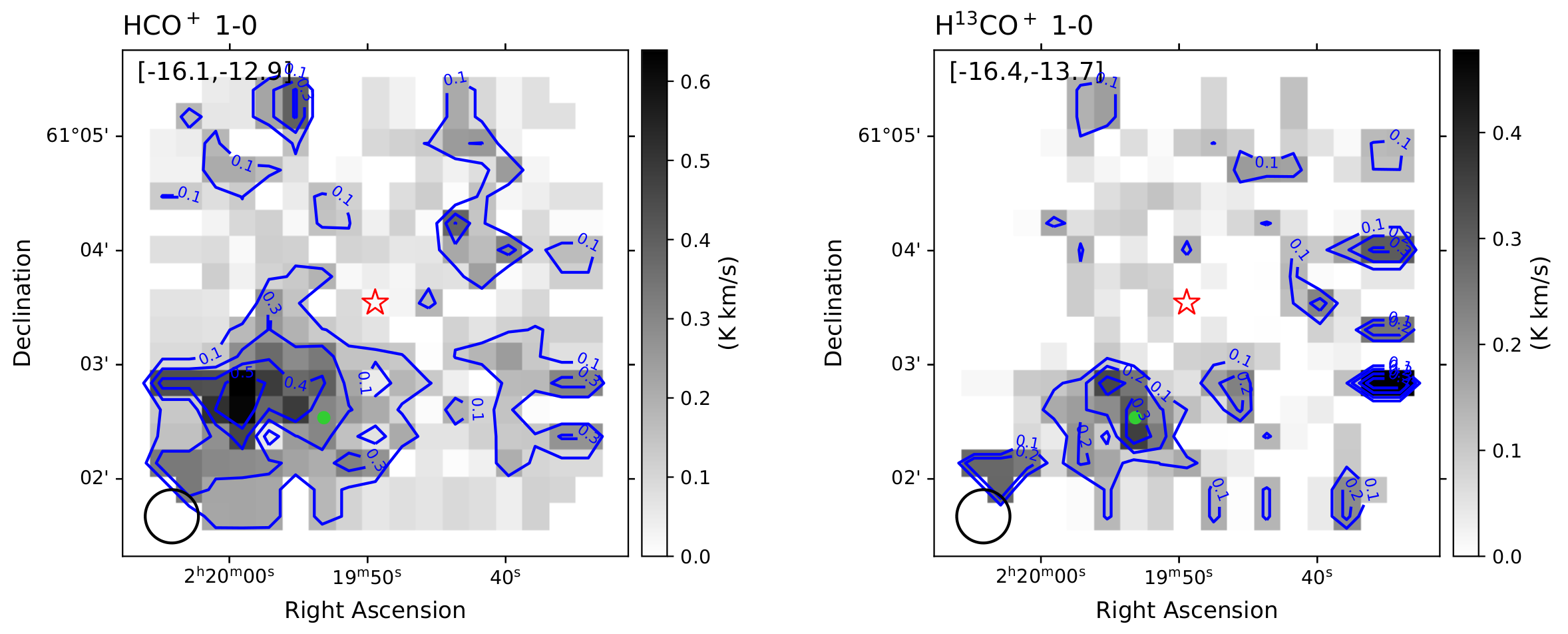}
\figsetgrpnote{HCO$^+$ (1-0) and H$^{13}$CO$^+$ (1-0) integrated intensity maps. The pentagram symbol denotes the position of the infall candidate identified in Paper II. The green, yellow, and orange points denote the Class I, Flat-spectrum, Class II YSOs \citep{Kuhn+etal+2021}, respectively. The blue, magenta, orange, and red crosses denote the 95 GHz methanol masers \citep{Yang+etal+2017}, 6.7 GHz methanol masers \citep{Yang+etal+2019}, H$_2$O masers \citep{Anglada+etal+1996,Valdettaro+etal+2001}, and OH masers \citep{Qiao+etal+2016,Qiao+etal+2018,Qiao+etal+2020}. For targets that show two distinct clumps within the observed areas, we have labeled these clumps as A and B. The black circles on the lower left indicate the beam sizes of the IRAM 30-m telescope.}
\figsetgrpend

\figsetend

\section{HCO$^+$ (1-0) and H$^{13}$CO$^+$ (1-0) Map Grids of Thirteen Targets} \label{sec:Appendix2}

\begin{figure}
%\figurenum{2}
\plotone{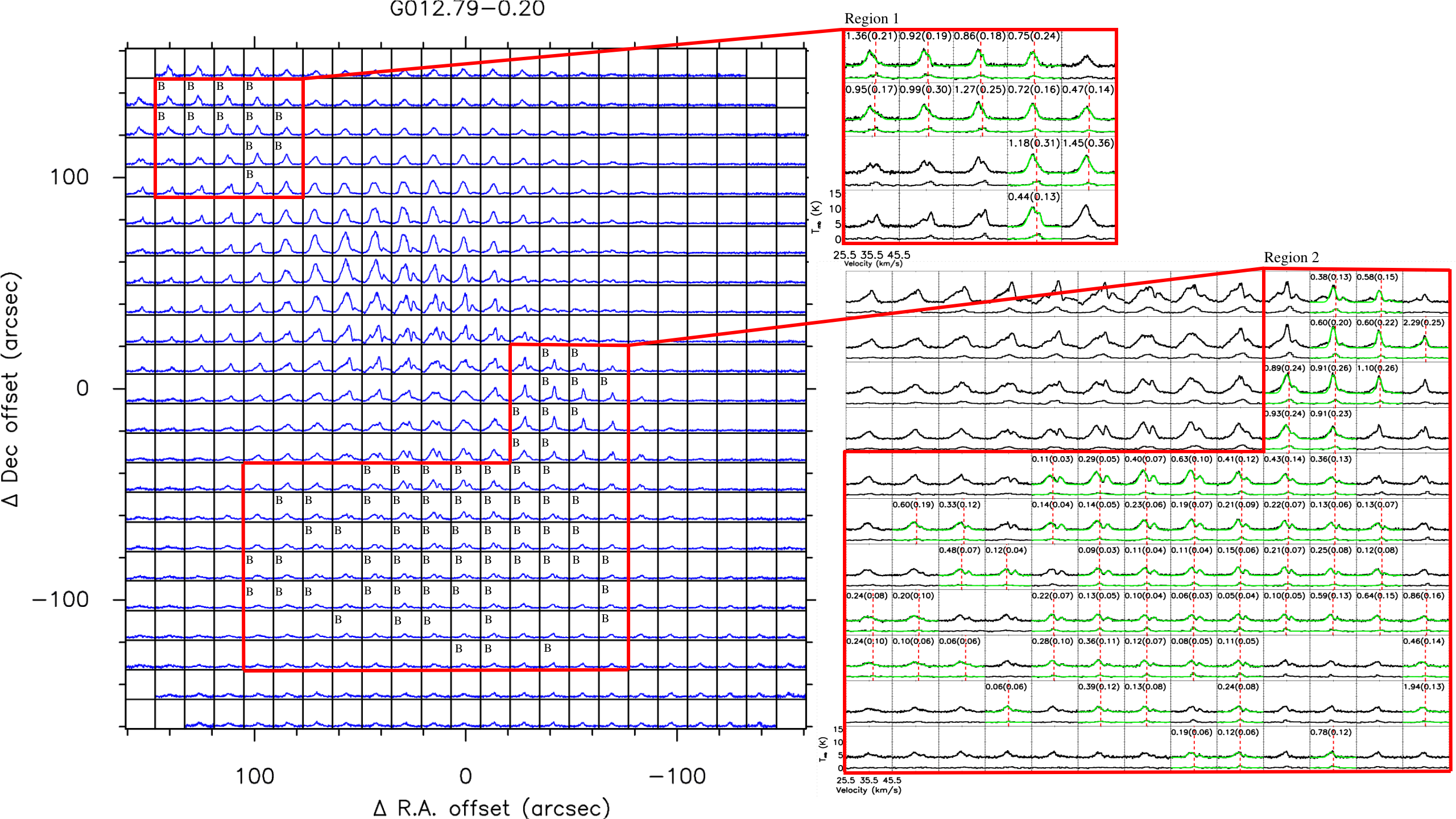}
\caption{HCO$^+$ (1-0) map grid (gridded to half of beam size). The axes plot the offsets $\Delta$ R.A. and $\Delta$ Dec relative to the coordinates from Table \ref{Tab:src-catalog}.}
\label{fig:mapgrid}
\end{figure}

Figure \ref{fig:mapgrid} presents the HCO$^+$ (1-0) map grids of 13 targets, which are gridded to half of beam size. The left panel of each subfigure shows the whole map grids, while the right panel zooms in the red box area of the left panel, and marks the gas infall velocity of each grid. The complete figure set (13 images) is available in the online journal. More details can be found in Section \ref{subsec:profiles} and \ref{subsec:Vin}.

\figsetstart
%\figsetnum{2}
\figsettitle{HCO$^+$ (1-0) map grid}

\figsetgrpstart
\figsetgrpnum{B2.1}
\figsetgrptitle{}
\figsetplot{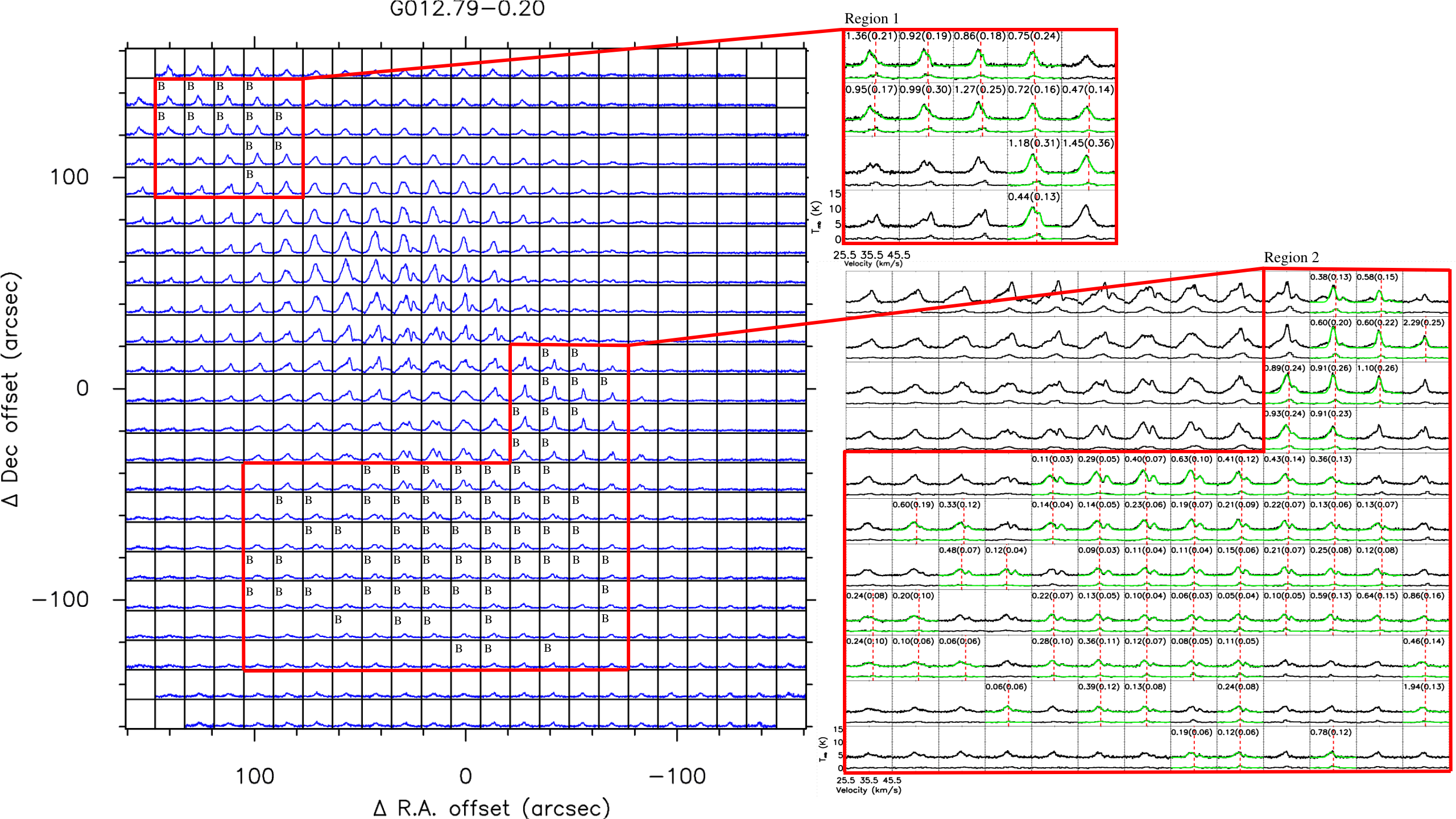}
\figsetgrpnote{HCO$^+$ (1-0) map grid (gridded to half of beam size). The axes plot the offsets $\Delta$ R.A. and $\Delta$ Dec relative to the coordinates from Table \ref{Tab:src-catalog}.}
\figsetgrpend

\figsetgrpstart
\figsetgrpnum{B2.2}
\figsetgrptitle{}
\figsetplot{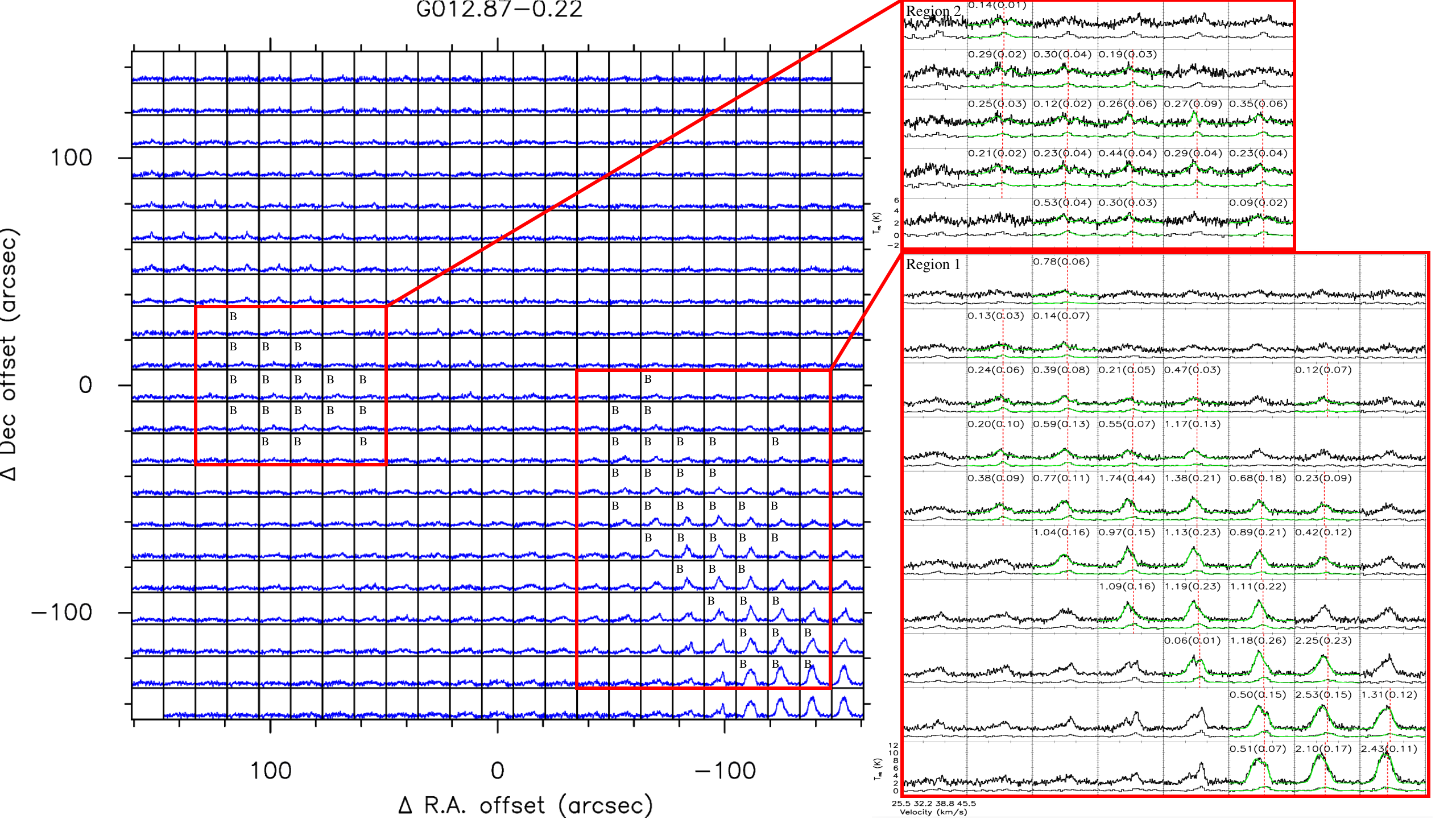}
\figsetgrpnote{HCO$^+$ (1-0) map grid (gridded to half of beam size). The axes plot the offsets $\Delta$ R.A. and $\Delta$ Dec relative to the coordinates from Table \ref{Tab:src-catalog}.}
\figsetgrpend

\figsetgrpstart
\figsetgrpnum{B2.3}
\figsetgrptitle{}
\figsetplot{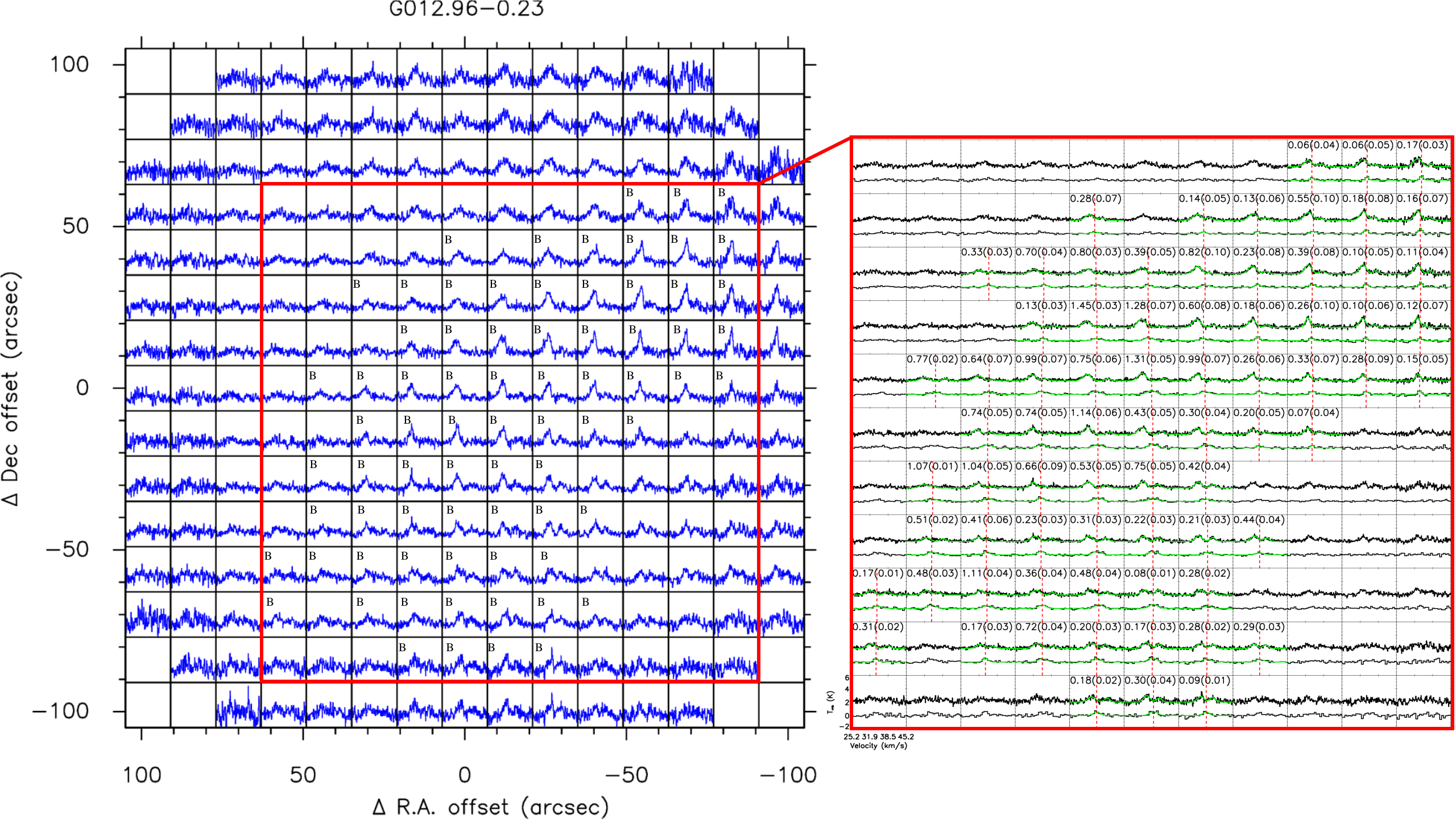}
\figsetgrpnote{HCO$^+$ (1-0) map grid (gridded to half of beam size). The axes plot the offsets $\Delta$ R.A. and $\Delta$ Dec relative to the coordinates from Table \ref{Tab:src-catalog}.}
\figsetgrpend

\figsetgrpstart
\figsetgrpnum{B2.4}
\figsetgrptitle{}
\figsetplot{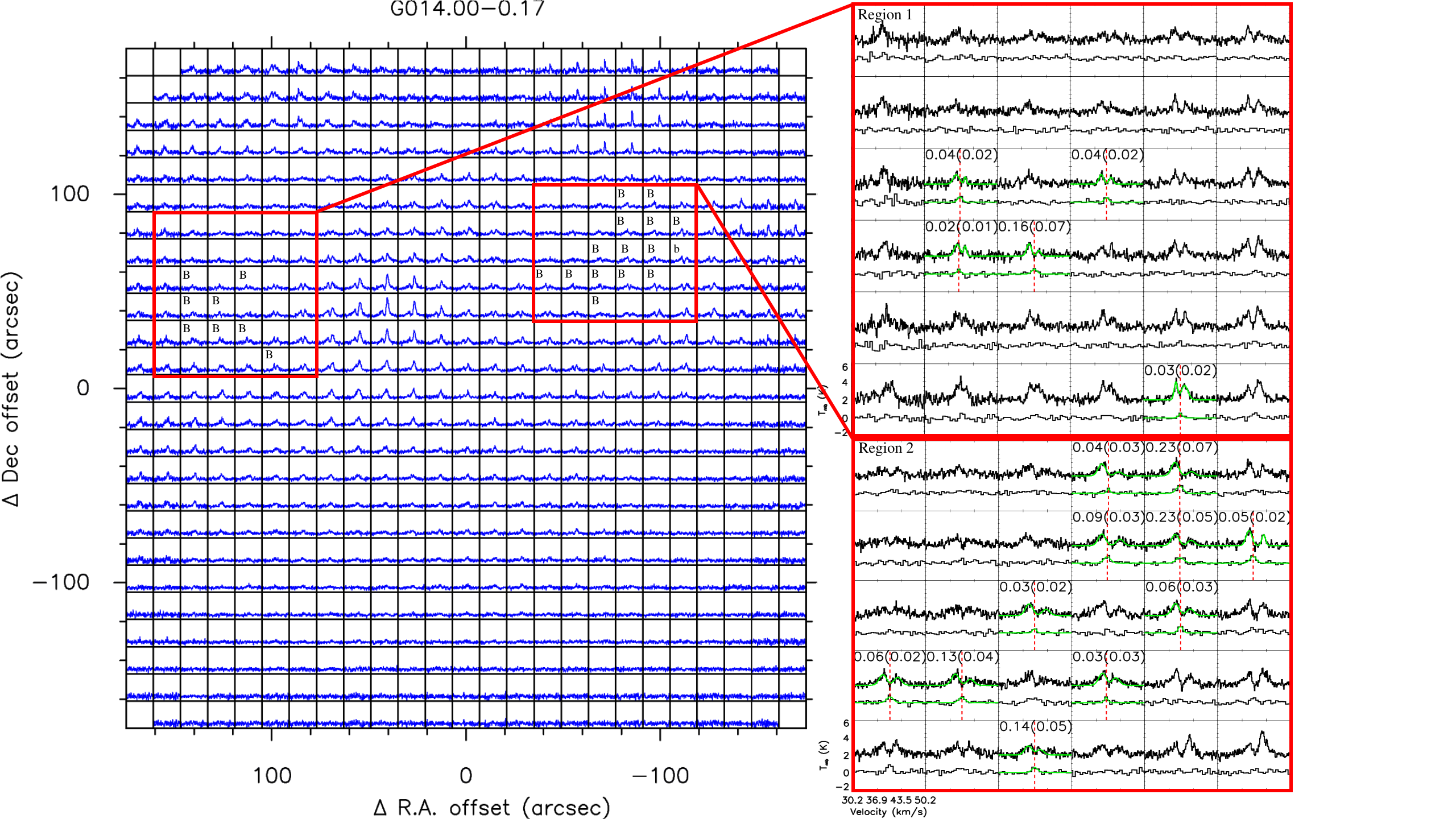}
\figsetgrpnote{HCO$^+$ (1-0) map grid (gridded to half of beam size). The axes plot the offsets $\Delta$ R.A. and $\Delta$ Dec relative to the coordinates from Table \ref{Tab:src-catalog}.}
\figsetgrpend

\figsetgrpstart
\figsetgrpnum{B2.5}
\figsetgrptitle{}
\figsetplot{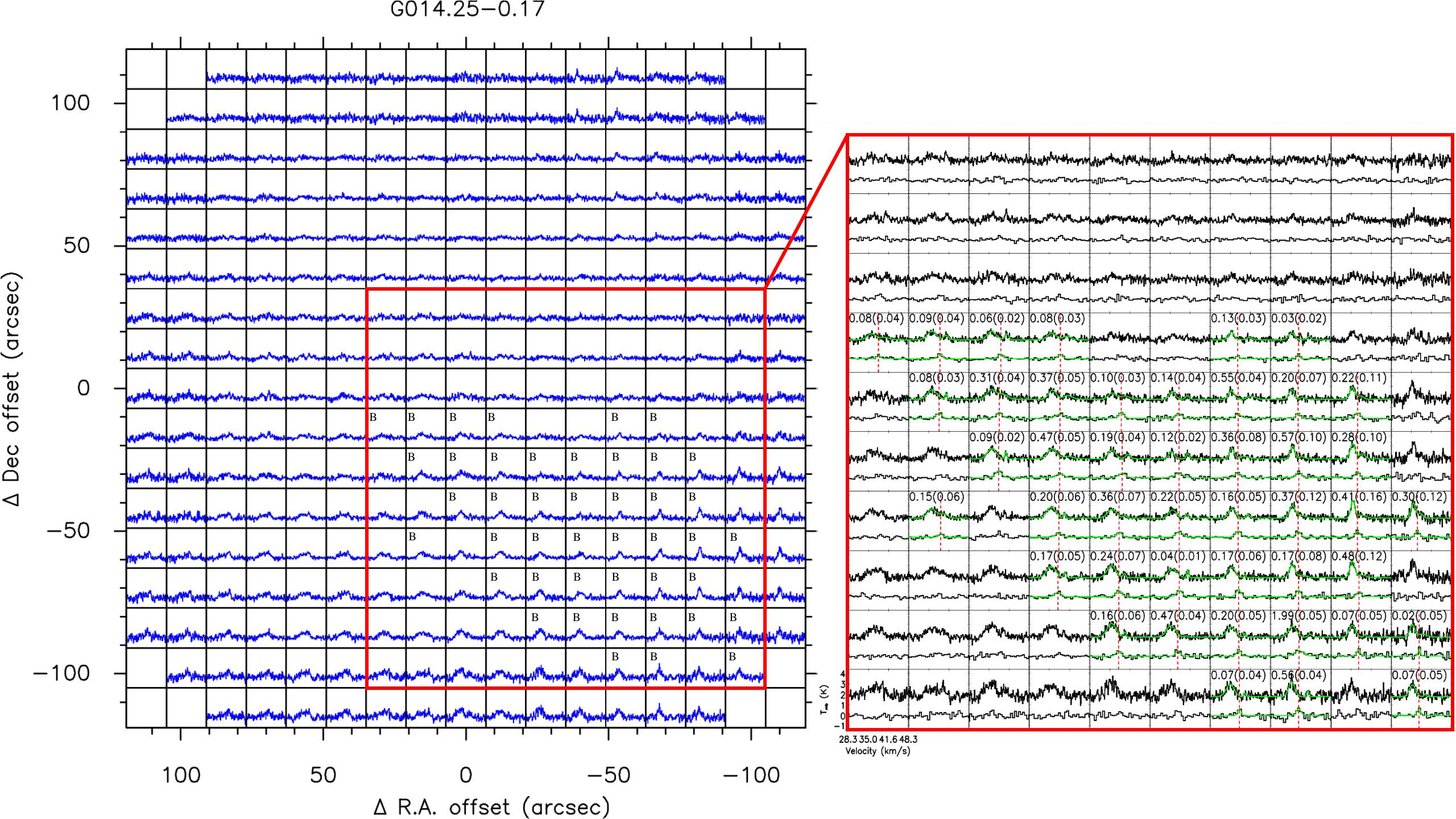}
\figsetgrpnote{HCO$^+$ (1-0) map grid (gridded to half of beam size). The axes plot the offsets $\Delta$ R.A. and $\Delta$ Dec relative to the coordinates from Table \ref{Tab:src-catalog}.}
\figsetgrpend

\figsetgrpstart
\figsetgrpnum{B2.6}
\figsetgrptitle{}
\figsetplot{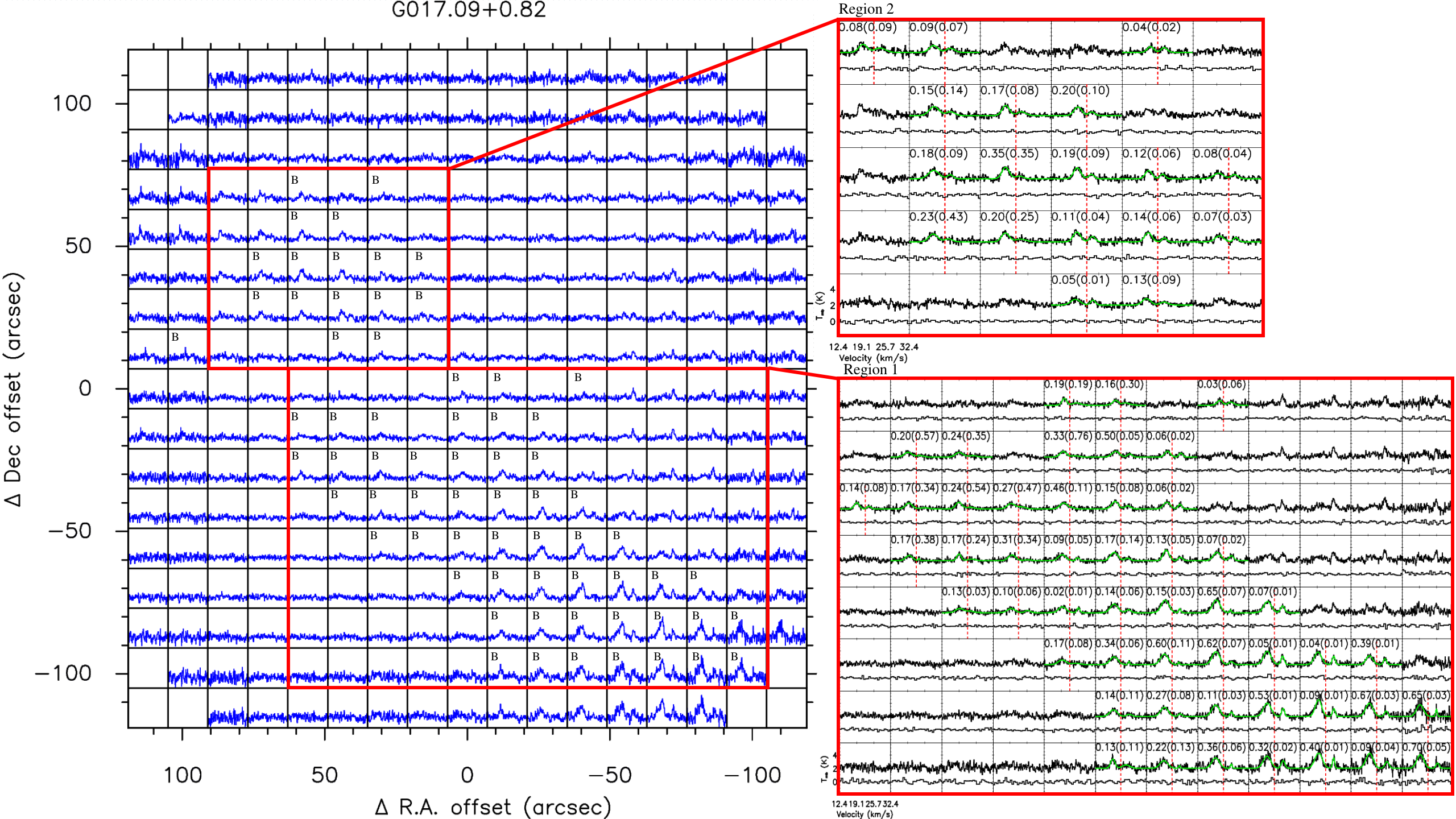}
\figsetgrpnote{HCO$^+$ (1-0) map grid (gridded to half of beam size). The axes plot the offsets $\Delta$ R.A. and $\Delta$ Dec relative to the coordinates from Table \ref{Tab:src-catalog}.}
\figsetgrpend

\figsetgrpstart
\figsetgrpnum{B2.7}
\figsetgrptitle{}
\figsetplot{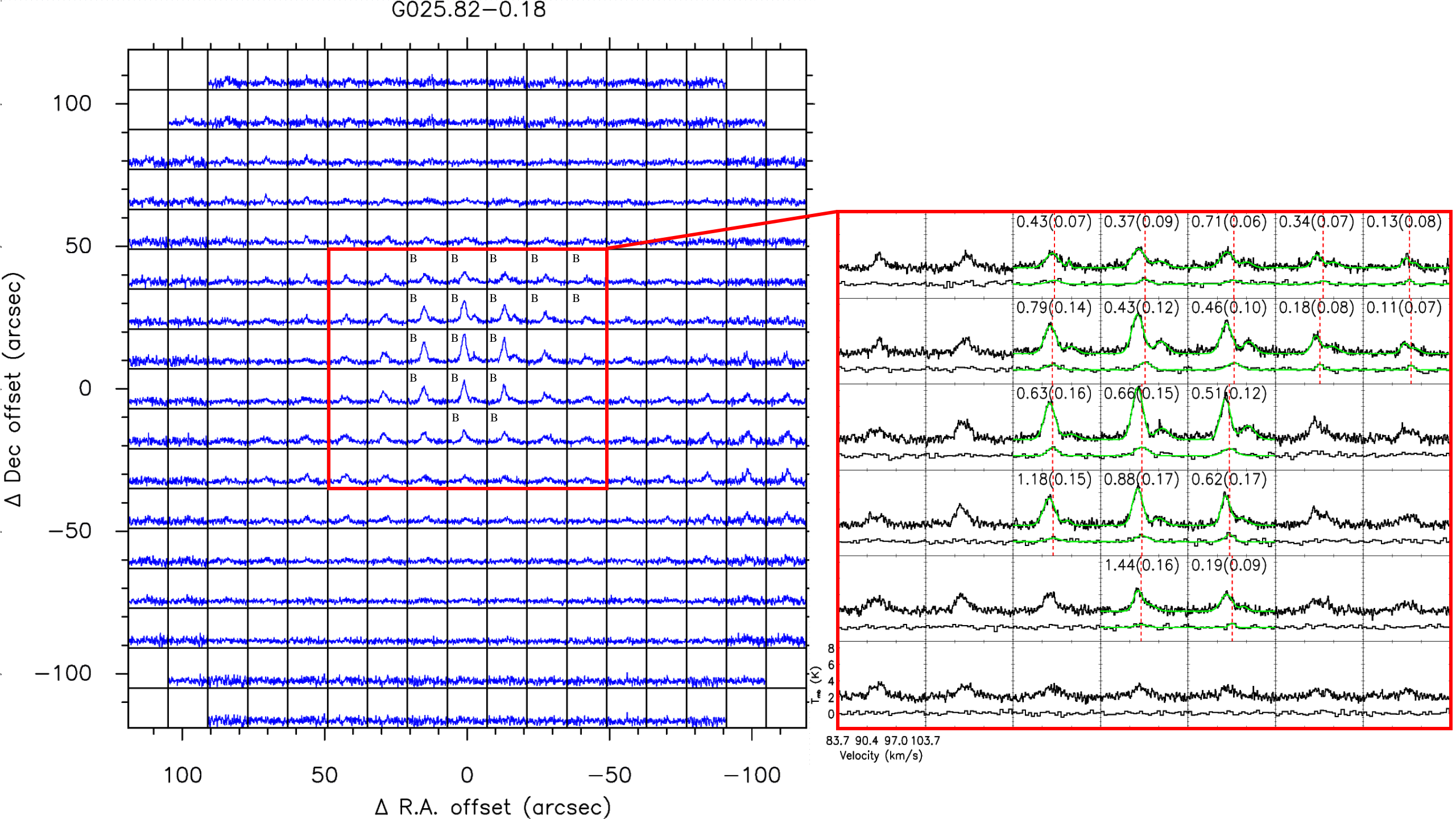}
\figsetgrpnote{HCO$^+$ (1-0) map grid (gridded to half of beam size). The axes plot the offsets $\Delta$ R.A. and $\Delta$ Dec relative to the coordinates from Table \ref{Tab:src-catalog}.}
\figsetgrpend

\figsetgrpstart
\figsetgrpnum{B2.8}
\figsetgrptitle{}
\figsetplot{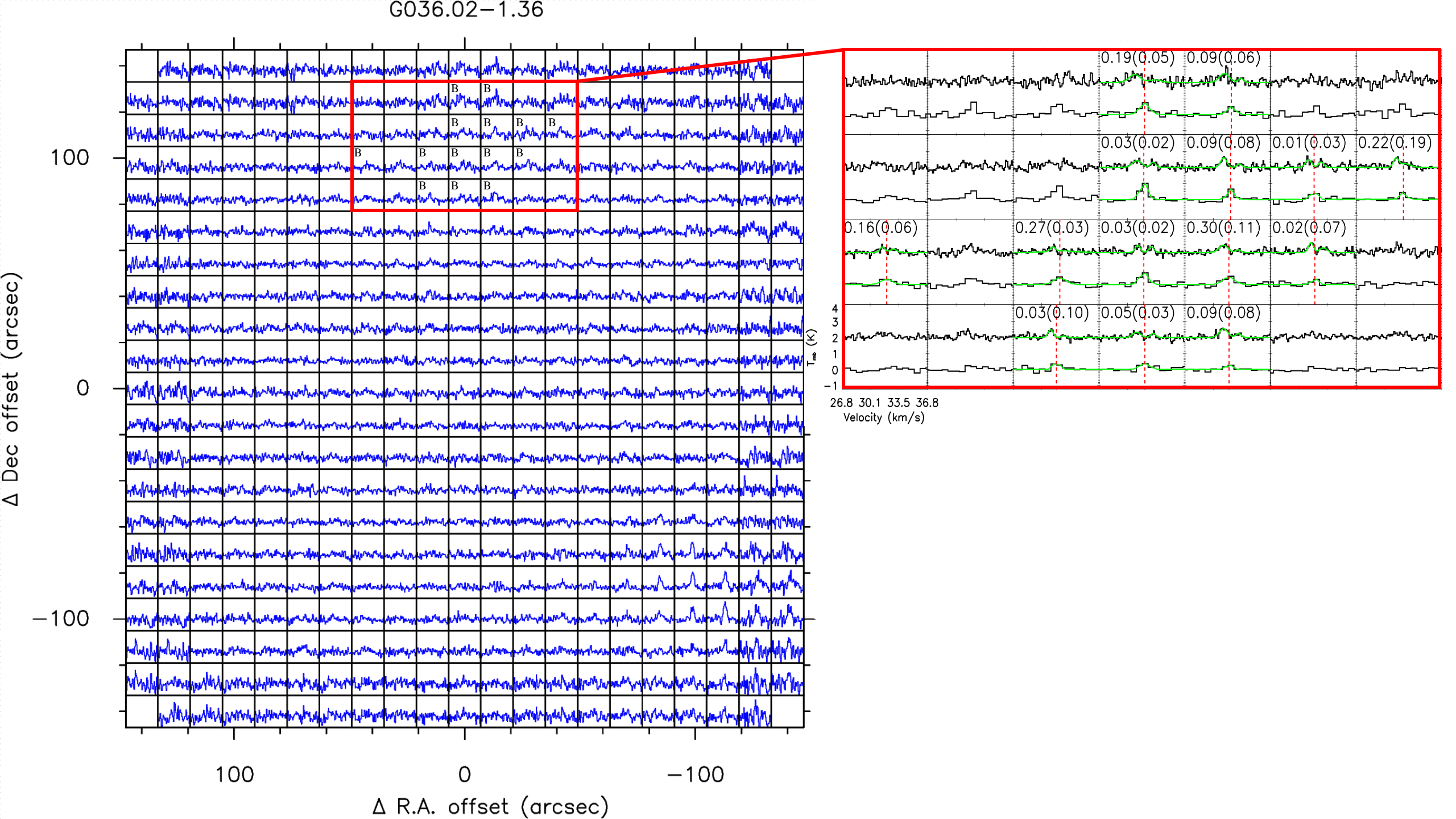}
\figsetgrpnote{HCO$^+$ (1-0) map grid (gridded to half of beam size). The axes plot the offsets $\Delta$ R.A. and $\Delta$ Dec relative to the coordinates from Table \ref{Tab:src-catalog}.}
\figsetgrpend

\figsetgrpstart
\figsetgrpnum{B2.9}
\figsetgrptitle{}
\figsetplot{fig_b1_09.png}
\figsetgrpnote{HCO$^+$ (1-0) map grid (gridded to half of beam size). The axes plot the offsets $\Delta$ R.A. and $\Delta$ Dec relative to the coordinates from Table \ref{Tab:src-catalog}.}
\figsetgrpend

\figsetgrpstart
\figsetgrpnum{B2.10}
\figsetgrptitle{}
\figsetplot{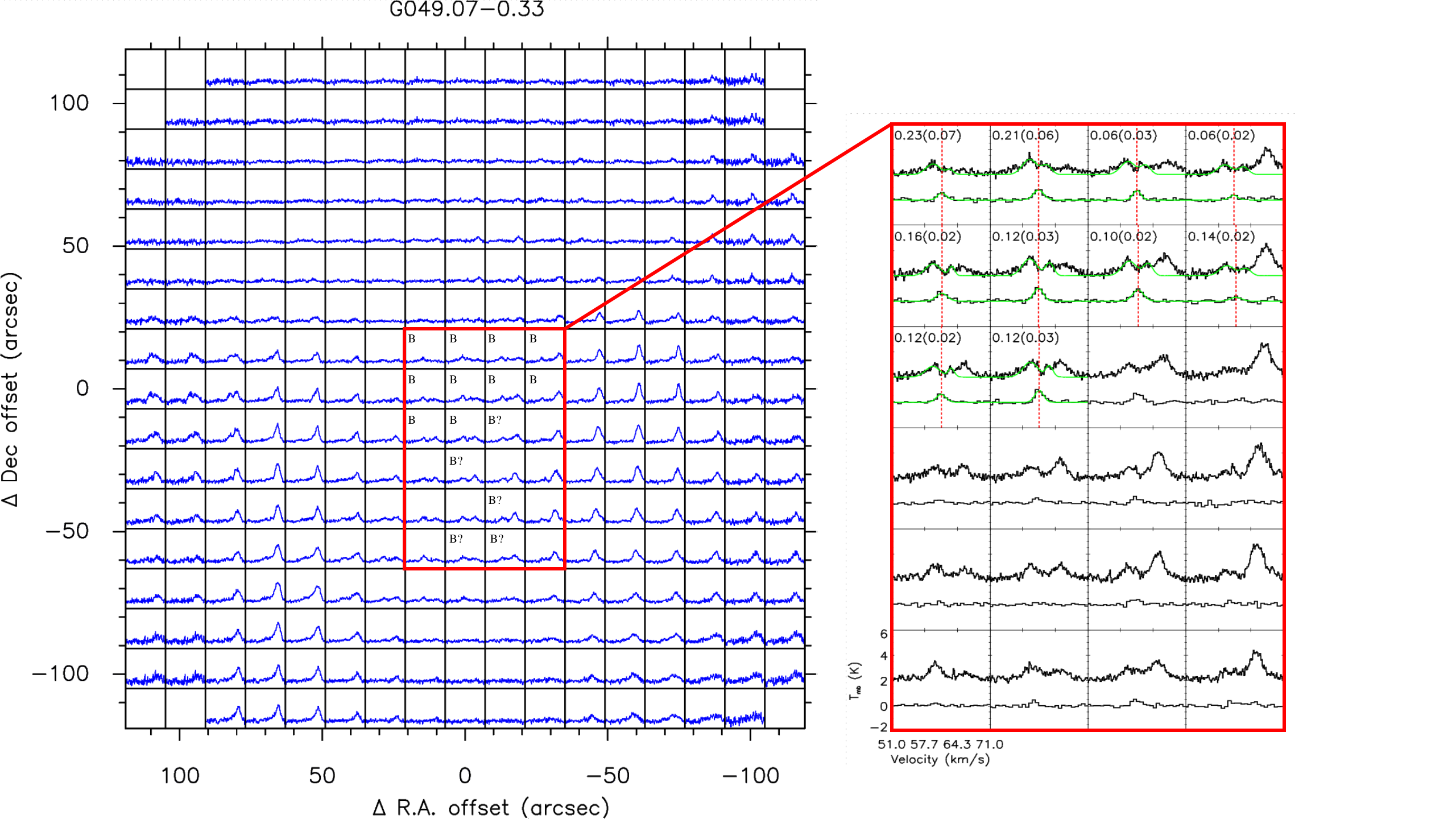}
\figsetgrpnote{HCO$^+$ (1-0) map grid (gridded to half of beam size). The axes plot the offsets $\Delta$ R.A. and $\Delta$ Dec relative to the coordinates from Table \ref{Tab:src-catalog}.}
\figsetgrpend

\figsetgrpstart
\figsetgrpnum{B2.11}
\figsetgrptitle{}
\figsetplot{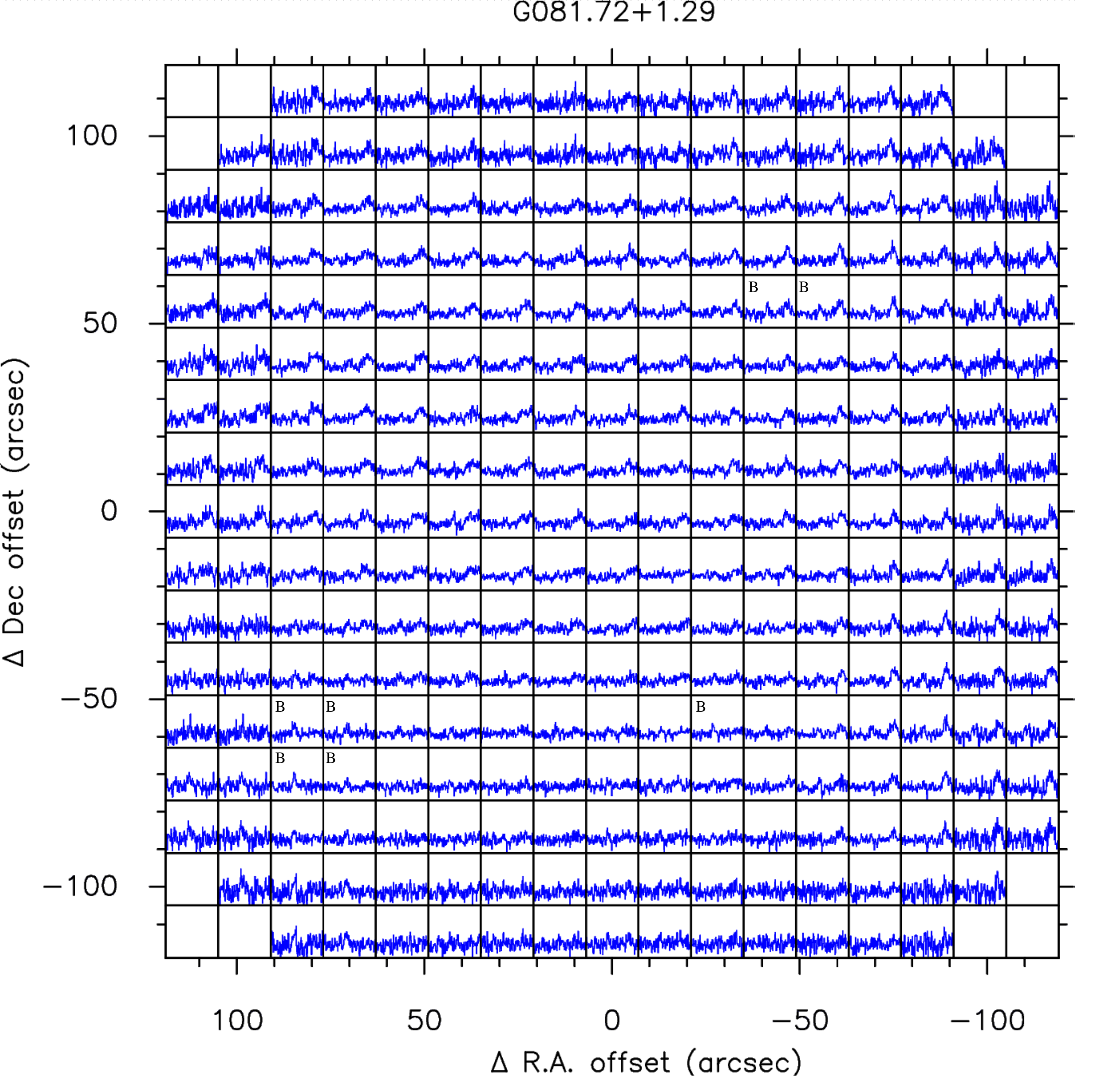}
\figsetgrpnote{HCO$^+$ (1-0) map grid (gridded to half of beam size). The axes plot the offsets $\Delta$ R.A. and $\Delta$ Dec relative to the coordinates from Table \ref{Tab:src-catalog}.}
\figsetgrpend

\figsetgrpstart
\figsetgrpnum{B2.12}
\figsetgrptitle{}
\figsetplot{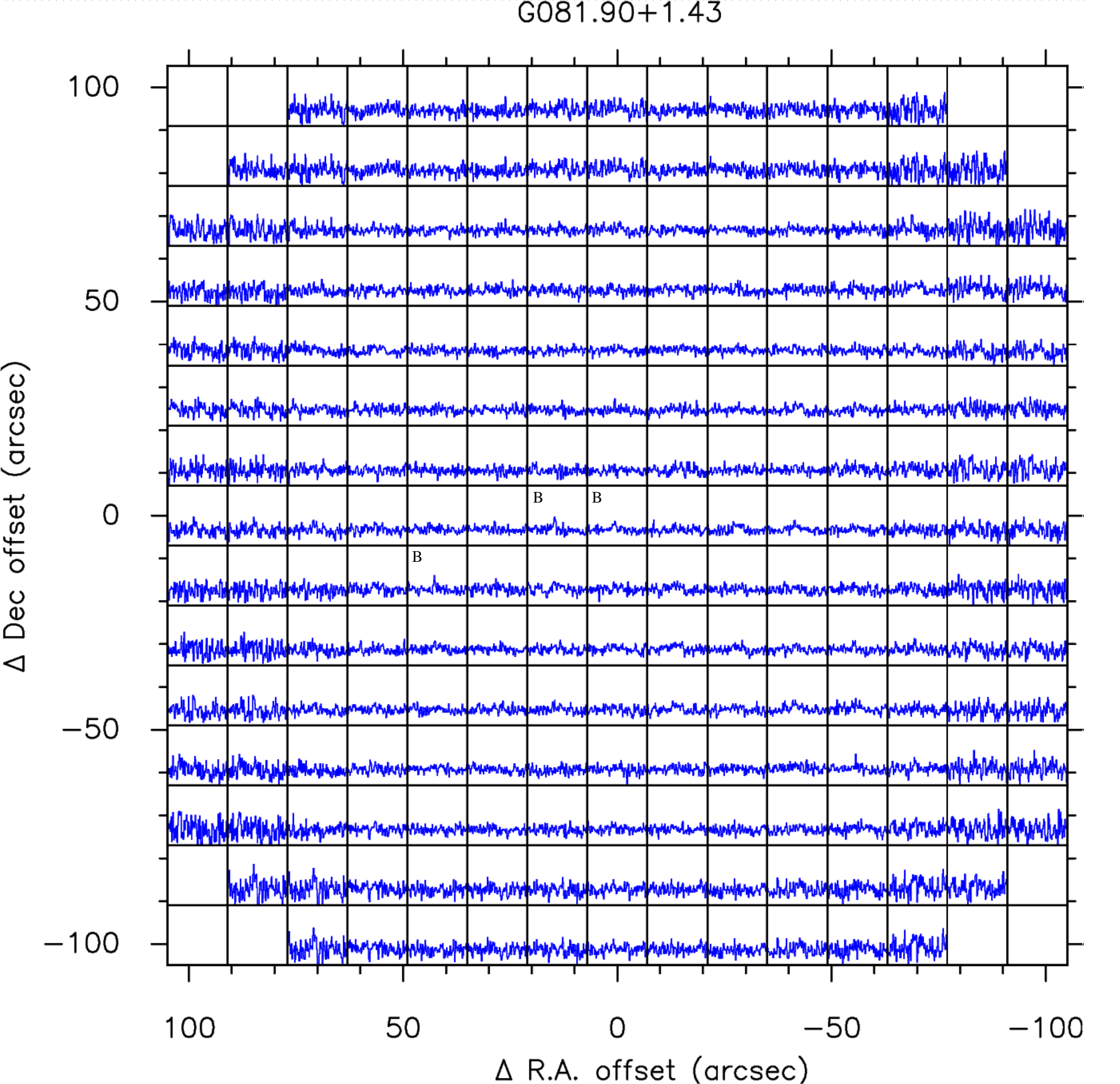}
\figsetgrpnote{HCO$^+$ (1-0) map grid (gridded to half of beam size). The axes plot the offsets $\Delta$ R.A. and $\Delta$ Dec relative to the coordinates from Table \ref{Tab:src-catalog}.}
\figsetgrpend

\figsetgrpstart
\figsetgrpnum{B2.13}
\figsetgrptitle{}
\figsetplot{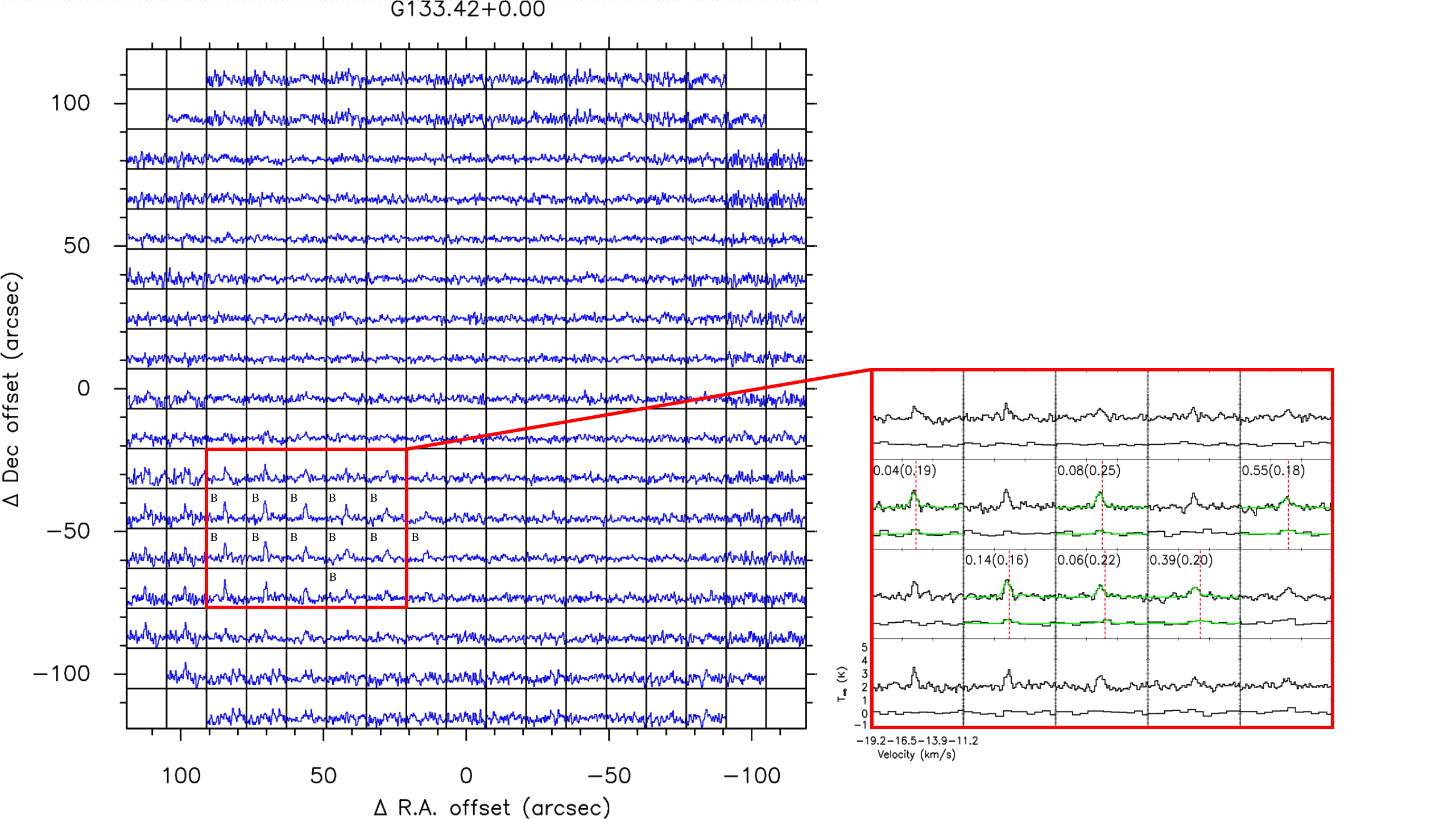}
\figsetgrpnote{HCO$^+$ (1-0) map grid (gridded to half of beam size). The axes plot the offsets $\Delta$ R.A. and $\Delta$ Dec relative to the coordinates from Table \ref{Tab:src-catalog}.}
\figsetgrpend

\figsetend

\clearpage

\bibliography{Infall_IV}{}
\bibliographystyle{aasjournal}



\end{document}